\documentclass[twoside]{article}

\usepackage{amsmath}
\usepackage{amsthm}
\usepackage{graphicx}

\renewcommand{\theequation}{\arabic{section}.\arabic{equation}}
\renewcommand{\thesection}{\arabic{section}.}

\begin{document}

\baselineskip .508cm

\newcommand{\dovd}[2]{\frac{\partial #1}{\partial #2}}
\newcommand{\rmd}{\mathrm{d}}
\newcommand{\rmi}{\mathrm{i}}
\newcommand{\rmiss}[1]{\mathrm{i}_{\scriptscriptstyle #1}}
\newcommand{\wt}[1]{\widetilde{#1}}
\newcommand{\ip}{\hspace{1pt} \mbox{\vrule depth-0.1pt height0.53pt width 6.7pt
                 \vrule depth-0.1pt height5.5pt} \hspace{3.5pt}}
\newcommand{\bp}{{\bf p }}
\newcommand{\dsplbl}[1]{}
\newcommand{\hi}{\frac{\hbar}{i}\,}
\newcommand{\ih}{\frac{i}{\hbar}\,}
\newcommand{\Sch}{Schr\"{o}dinger }
\newcommand{\Schs}{Schr\"{o}dinger's }
\newcommand{\br}{{\bf r }}
\newcommand{\bk}{{\bf k }}
\newcommand{\bl}{{\bf l }}
\newcommand{\bq}{{\bf q }}
\newcommand{\bn}{{\bf n }}
\newcommand{\bx}{{\bf x }}
\newcommand{\bj}{{\bf j }}
\newcommand{\bi}{{\bf i }}
\newcommand{\by}{{\bf y }}
\newcommand{\bz}{{\bf z }}
\newcommand{\bu}{{\bf u }}
\newcommand{\bv}{{\bf v }}
\newcommand{\bs}{{\bf s }}
\newcommand{\bS}{{\bf S }}
\newcommand{\bR}{{\bf R }}
\newcommand{\bA}{{\bf A }}
\newcommand{\bB}{{\bf B }}
\newcommand{\bsigma}{\boldsymbol{\sigma}}
\newcommand{\jahqt}{$\mbox{J}^a\mbox{HQT}$}
\newcommand{\ja}{$\mbox{J}^a$}
\newcommand{\tp}{\tilde{p}}
\newcommand{\tx}{\tilde{x}}
\newcommand{\wty}{\widetilde{y}}
\newcommand{\wtp}{\widetilde{\varphi}}
\newcommand{\wtP}{\widetilde{\Psi}}
\newcommand{\0}{\mbox{\o}}
\newcommand{\cP}{$\cal P$}
\newcommand{\cL}{{\cal L}}
\newcommand{\cN}{{\cal N}}
\newcommand{\cR}{{\cal R}}
\newcommand{\cH}{{\cal H}}
\newcommand{\cO}{{\cal O}}
\newcommand{\paqd}{$\cal P$AQD}
\newcommand{\rs}{^r_\sigma}
\newcommand{\rsi}{^r_{\sigma i}}
\newcommand{\rsj}{^r_{\sigma j}}
\newcommand{\dsp}{\displaystyle}
\newcommand{\veps}{\varepsilon}
\newcommand{\vphi}{\varphi}
\newcommand{\half}{\frac{1}{2}}
\newcommand{\ubar}{\bar{u}}
\newcommand{\vbar}{\bar{v}}
\newcommand{\dubar}{\partial_{\ubar}}
\newcommand{\dvbar}{\partial_{\vbar}}

\title{Analytical Quantum Dynamics in Infinite Phase Space}

\author{Maxim Raykin\footnote{E-mail: maxraykin@hotmail.com} \\
{\small \em{Millbury, Massachusetts}}}

\date{July 10, 2012}

\maketitle

\vspace{0.5cm}

\begin{abstract}

We develop a dynamical theory, based on a system of ordinary differential equations describing the
motion of particles which reproduces the results of quantum mechanics. The system generalizes
the Hamilton equations of classical mechanics to the quantum domain,
and turns into them in the classical
limit $\hbar\rightarrow 0$. The particles' motions are completely determined by the initial
conditions. In this theory, the wave function $\psi$ of quantum mechanics is equal to the
exponent of an action function, obtained by integrating some Lagrangian function along particle
trajectories, described by equations of motion. Consequently, the equation for the logarithm of a
wave function is related to the equations of motion in the same way as the Hamilton-Jacobi equation
is related to the Hamilton equations in classical mechanics.
We demonstrate that the probability
density of particles, moving according to these equations, should be given by a standard
quantum-mechanical relation, $\rho=|\psi|^2$. The theory of quantum measurements is presented, and
the mechanism of nonlocal correlations between results of distant measurements with entangled
particles is revealed. In the last section, we extend the theory to particles with nonzero spin.


\end{abstract}

\newpage
\tableofcontents
\newpage

\section{Introduction}

According to standard quantum mechanics\footnote{In this article we will use, in order of their
appearance, the following
abbreviations: QM --- quantum mechanics, ODE --- ordinary differential equation, PDE --- partial
differential equation, QHJE --- quantum Hamilton-Jacobi equation, \paqd\ --- analytical quantum 
dynamics in infinite phase space, OSFI --- one-step Feynman integral, CD --- Cartan distribution, 
HC --- Hamiltonian conditions, DBBT --- \mbox{de Broglie\,-\,Bohm} theory, FDS --- full
description space, RDS --- reduced description space.}
(QM), the state of every physical system is described by a wave function, whose time evolution is
determined by the \Sch equation. In this paper we will consider only closed systems, for which the
description by a wave function is sufficient. We know how to set up experiments with a known
initial wave function, and then using the \Sch equation we can calculate it at any later moment.
However, contrary to, say, an electric field in an electromagnetic wave, or a field of pressure in
a sound wave, the wave function is not an object of observation and measurement. Consequently,
besides \Schs equation, the theory additionally includes a set of rules, specifying the results of
experiments with quantum systems in terms of their wave functions. These rules were developed in
late 1920-s and collectively named the (statistical) interpretation of QM. Thus, the theory has two
parts: \Schs equation and interpretation.

Such structure of the theory may be viewed in various ways. The standard attitude consists of the
faith that the described construction constitutes the desired complete and fundamental law of
nature. However, there are a number of objections that may be raised against this point of view:
\newline -- It seems natural to expect from a fundamental theory that it reflects all observable
elements of physical reality and gives the law of evolution for them. Thus the very fact that
QM is formulated in terms of wave functions, which cannot be directly observed, and requires an
interpretation that establishes a connection between wave functions and results of experiments
creates doubts in its fundamental character.
\newline -- This interpretation is a separate and independent part of the theory's foundation,
whereas it seems desirable for a fundamental theory to allow the derivation of all its experimental
consequences by pure math from the dynamical laws of evolution alone.
\newline -- By necessity the interpretation, which describes the response of an approximately
classical apparatus to its interaction with a quantum system, is expressed in classical terms.
However the behavior of any apparatus, which is just a physical object built up of atoms, should
be derivable from QM. Consequently, QM contains an unacceptable for a fundamental theory logical
vicious circle: in the words of a classic textbook \cite{ll}: ``...~quantum mechanics occupies a
very unusual place among physical theories: it contains classical mechanics as a limiting case, yet
at the same time it requires this limiting case for its own formulation."
\newline -- The interpretation happens to be probabilistic, thus employing a series of similar
experiments, possibly performed in different places and at different times,
to establish the meaning
of a wave function in the experiment at hand. The wave function in this particular experiment,
however, certainly appears relevant. It therefore seems desirable for the theory to define the
meaning of a wave function in every individual experiment, without reference to its repetitions
(especially when such repetitions are clearly impossible, such as when discussing the wave
function of the universe) which QM fails to do. The fact that a wave function must have a
nonstatistical interpretation in ``internal" terms also clearly follows from the utility of a
concept of a wave function of quarks confined inside hadrons.
\newline -- According to this interpretation, during a measurement the wave function abandons
the unitary law of evolution, which it normally follows, and suffers a collapse. However, the
conditions under which this change of a character of evolution happens, are not specified in QM,
and attempts to formulate such conditions have not been convincingly successful.
\newline -- The collapse occurs randomly into different possible states, but QM does not explain
the reason for this randomness. Consequently, the values of corresponding probabilities (which one
would expect to see among the {\em results} of a fundamental theory) in QM are
not derived, but postulated, or, in other words, are taken from experiment.
\newline -- According to QM, unless a system is in an eigenstate of a measured quantity, the
result of its measurement does not exist before the measurement is done, but is rather created
during the measurement. In some cases this is completely obvious --- see a discussion of spin
measurement for a particle with spin 3/2 in \cite{peres}.
This means that a measurement is not a fundamental unanalyzable primitive,
but a nontrivial physical process for which QM fails to give an adequate description.
\newline -- Moreover, this process of measurement produces nonlocally correlated
results for measurements performed with space-like separated entangled particles, but QM does not
describe any mechanism which causes these correlations.

Thus it appears that the standard combination of \Schs equation and statistical interpretation is
too complicated, artificial and, in the words of John Bell ``unprofessionally vague and ambiguous"
\cite{jsb}, while the nature obviously prefers simple, natural, and clear fundamental
laws. Consequently, we suggest in this paper to regard the situation in the following alternative
way: The statistical interpretation does
not, of course, follow from the \Sch equation (simply because the latter only describes the
behavior of a wave function), but is a generalization of results of observations and experiments.
Our trust in statistical interpretation is based on its agreement with experiment, and only on this
agreement. Therefore, the interpretation is a phenomenological part of quantum theory, and so the
whole existing theory is semi-phenomenological. Then to this semi-phenomenological theory
the above objections are inapplicable, while at the same time there remains a possibility that
the nature is ruled by the other, ``simple, natural, and clear" fundamental theory, from which
statistical interpretation (and maybe \Schs equation as well) follows.

The conclusion about the phenomenological (or, as it is often called, ``pragmatic" \cite{stapp})
nature of existing QM may also be drawn from the works devoted to
its foundations. We read, for example, in Bohr \cite{bohr}: ``Strictly speaking, the mathematical
formalism of quantum mechanics and electrodynamics merely offers rules of calculation for the
deduction of expectations about observations obtained under well-defined experimental conditions
specified by classical physical concepts", or, in a frequently quoted more recent paper
\cite{fuchsperes}: ``...quantum theory does not describe physical reality. What it does
is provide an algorithm for computing probabilities for the macroscopic events", ``...the time
dependence of the wavefunction does not represent the evolution of a physical system. It only gives
the evolution of our probabilities for the outcomes of potential experiments on that system." Thus
according to these works, QM describes the results of our {\em observations} of electrons, atoms,
etc. Which theory, then, describes these particles, which we observe, {\em themselves}?

Of course, neither these quotations, nor the arguments presented above, can {\em prove}
that a better theory is needed. They can, however, {\em motivate} a search for such
a theory. Indeed, it is hard to help
feeling that peculiar features of QM are the consequences of a fact that it misses some important
part of a complete theory, a part which is substituted by a phenomenological description of the way
it works. In this searched-for complete theory, the wave function must have a definite meaning in
every individual experiment, and the theory must explain the nature of randomness and derive the
standard quantum-mechanical expression for probability. The expression for probability
will thus become just a property of a wave
function, rather than a basis for its interpretation. This theory should also explain the
properties of quantum measurements, describe the mechanism which creates nonlocal correlations, and
fill all other gaps listed above; in particular, it should contain an image of every observable
element of reality and predict its behavior directly from the theory's dynamical laws,
without the need for any special interpretation. Compared to QM, such a theory would be much less
vulnerable to suspicions of being a mere semi-phenomenology, and it is a goal of this paper to
present a theory which appears to satisfy these demands.

Before discussing this theory, we recall the Hamilton-Jacobi equation
\begin{equation}\label{fe}
    \dovd{p}{t} + \frac{1}{2m} \, (\nabla p)^2 + U = 0       \dsplbl{fe}
\end{equation}
for an action function $p(\bx,t)$ in classical mechanics. This function does not
describe any individual trajectory and motion of a particle along it; rather, it describes a
family of such trajectories, of which none can be singled out given an action function alone.
Individual trajectories and particles' motions along them are described in classical mechanics by
Hamilton ordinary differential equations (ODEs). Given these trajectories, the action function,
which solves the Hamilton-Jacobi equation, may be obtained by integrating a Lagrangian function
along them. On the other hand, the trajectories of a family, described by an action function
$p(\bx,t)$, may be reconstructed from this function using the equality $\bp(\bx,t) = \nabla
p(\bx,t)$, which says that the momentum $\bp$ at point $(\bx,t)$ of any trajectory is equal to the
gradient of an action function at this point. Thus the Hamilton and Hamilton-Jacobi equations
represent two parts of the same theory --- classical mechanics --- the former describing particle
motion along classical trajectories, and the latter, properties of the families or ensembles of
these trajectories.

Now we formulate the basic idea of the present approach. Its initial step is purely mathematical.
Namely, it is shown in a second section of the paper, that similar to the case of first order
partial differential equations (PDEs) with one unknown function, such as just discussed
Hamilton-Jacobi equation, for a large
class of PDEs of second and higher orders the solution of equation may be represented as an action
function, i.e. the value of the function $p(\bx,t)$ that solves the equation may in every point be
obtained as an integral from some ``Lagrangian" function along the curve that leads to this point
and is completely and uniquely determined by some system of ODEs. As is well known
\cite{olvr,zhar,szsumf,ibragim}, for PDEs of higher than first order the system of ODEs with such
properties neither exists in the usual phase space with coordinates
$t,x^i, p_i$, where $p_i = \partial p/\partial x^i$ are first derivatives of an unknown function,
nor even in the same space extended by adding to its coordinates the derivatives of an unknown
function up to any finite order. Such a system, however, exists in an infinite phase space, the
coordinates of which include all possible partial derivatives of an unknown function, and a
corresponding mathematical theory is developed in section~2. Although the very possibility of
solving higher-order PDEs in this way was known for quite some time \cite{klv}, the specific form
of solution presented in section~2 seems to be new. In spite of the presence of an infinite
number of variables and equations in the theory, it happens to be quite transparent and manageable;
in fact, the theory is remarkably similar to the Hamiltonian formalism in classical mechanics and
reproduces all its essential features. The theory of first order PDEs also can be formulated in an
infinite phase space and turns out to be a special case of our theory, but in this case ODEs for
$x^i$ and $p_i$, i.e. for coordinates in the usual phase space, decouple from other equations and 
can be considered independently. Thus, we obtain a general Hamiltonian formalism that covers a
large class of PDEs of first as well as higher orders on equal grounds.

\begin{figure}
\centerline{\includegraphics[bb=18 110 348 180,clip,scale=1.3]{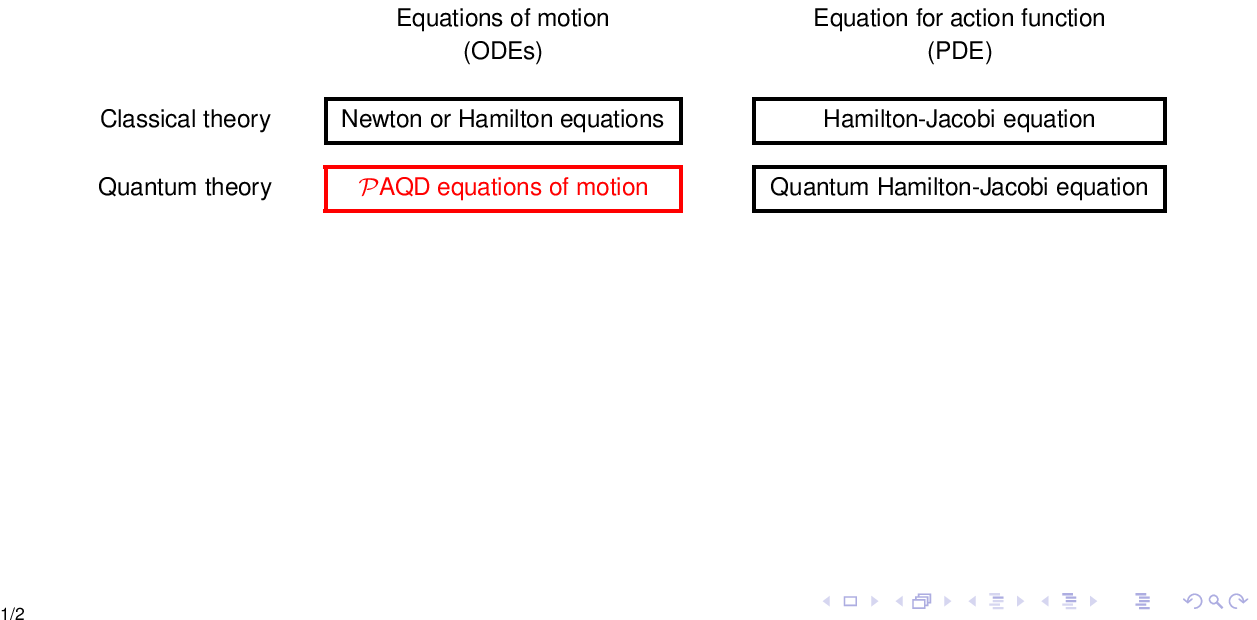}}
\caption{The structure of classical mechanics and \paqd. The relations between
equations of motion and equations for action functions are the same for both theories, and in the
limit $\hbar\rightarrow 0$ quantum equations turn into classical ones in both columns. In standard
QM the lower left rectangle is absent and substituted by the statistical interpretation.}
\label{fig1}
\end{figure}

Returning now to physics consider, along with a wave function $\psi$, a function
$p = (\hbar/i)\ln\psi$. Clearly, this function contains the same information as $\psi$, and may be
used instead of it in all discussions. For a spinless particle of mass $m$ in external
potential~$U$ we have from the \Sch equation the following PDE for $p(\bx,t)$:
\begin{equation}\label{qhjeint}
    \dovd{p}{t} + \frac{1}{2m} \, (\nabla p)^2 + U + \frac{\hbar}{2im} \Delta p = 0 \, .
    \dsplbl{qhjeint}
\end{equation}
Except for the last term, proportional to $\hbar$, this is the Hamilton-Jacobi equation for an
action function in classical mechanics. On the other hand, this equation happens to belong to the
class of PDEs considered in section~2, which have a solution in the form of an action function.
In this situation the following main idea of the present approach emerges with an absolute
inevitability: consider Eq.~(\ref{qhjeint}) as an equation for the action function in a new,
quantum, theory, the wave function --- as an exponent of the new action function (multiplied by
$i/\hbar$), the curves along which the Lagrangian function should be integrated to produce the
action function --- as particle's trajectories in the new theory, and ODEs that determine these
curves --- as new equations of motion, which correct Hamilton's equations. In exact analogy with
classical mechanics, the resulting theory will have two sides: ODE side, represented by the
equations of motion of particles along their trajectories, and PDE one, represented by
Eq.~(\ref{qhjeint}) for the action function that describes, along with the wave function,
ensembles of trajectories. In the following, Eq.~(\ref{qhjeint}) will be called ``quantum 
Hamilton-Jacobi equation" (QHJE). The motion of particles takes place in an infinite phase space,
\cP, defined in section 2, and the theory will be called analytical quantum dynamics in
infinite phase space, or \paqd. The general structure of \paqd\ and classical mechanics is 
presented in Fig.~\ref{fig1}.

In view of described above new mathematical possibility of dealing with Eq.~(\ref{qhjeint}), it
seems difficult to dispute that  this approach certainly appears quite natural; one could say that
by the very form of Eq.~(\ref{qhjeint}) the nature pushes us in this direction. Had the possibility
of solving second order PDEs in the way described above been known in 1926-1927, it is hard to
doubt that this work would be done right then! Further, being based on equations of motion and only
on them, the theory should be considered simple; for the same reason it promises to be clear and
unambiguous. Thus, it seems worth the efforts to investigate the possibilities which may open in
this direction; in doing so we will also finish the job
left unfinished eighty five years ago due to such
historical accident as an absence of a proper mathematical formalism at that time. Last but not
least, we note that the theory is completely fixed by Eq.~(\ref{qhjeint}) for its action function
(that is --- fixed by the \Sch equation) and doesn't contain any additional freedom to improve its
agreement with the second part of standard QM, the statistical interpretation. Therefore the fact
that, as we will soon see, such agreement is nevertheless achieved (or, in other words, that
statistical interpretation is deduced from the present theory) should be considered as a weighty
argument in the theory's favor.

The equations of motion of the theory are explicitly written down in the beginning of section 3,
and the rest of the paper is devoted to demonstrating that QM may be understood as a theory of
particles moving according to these equations, the difference between classical and quantum 
mechanics being the result of the different form and number of equations in these theories. 
Equations of motion are followed in section 3 by their general discussion. 
As in classical mechanics, these
equations are self-sufficient: given initial conditions, they define the particle's motion
unambiguously, without any need for using a wave function or the \Sch equation. In the classical
limit $\hbar \rightarrow 0$, the equations of motion turn directly into
the Hamilton equations of classical mechanics. The equations and particle trajectories live in an
infinite phase space. We show that
projections of these trajectories to physical space coincide with ``Bohmian trajectories,"
introduced by de Broglie and Bohm  \cite{dbb,dbb2} on very different grounds, and discuss the
relation between the present theory and that of de Broglie and Bohm.

The class of PDEs covered by the theory in section 2 is very large, and so the question arises:
what singles Eq.~(\ref{qhjeint}) out of this class? It is shown in section 4, that equations of
motion may be obtained from a ``one-step Feynman integral" (OSFI), combined with an appropriate
variational principle. OSFI provides, therefore, an alternative starting point of the theory,
bypassing the geometric theory of section 2. Moreover, OSFI is a functional of a Lagrangian
function, which is necessarily classical, i.e., depends on the position and velocity of a particle
only. Consequently, OSFI may be considered as a general source of quantum theories, obtained by
``quantization" of corresponding classical theories, represented by Lagrangian functions. As 
further discussed in section 4, the theories, obtained in this way, will automatically exhibit 
familiar features of QM: superposition principle, path-integral representation, and wave-particle
duality. We note that in a mathematical derivation of the latter feature, an infinite number of
variables and equations in our theory, which initially appears to be a theory's disadvantage, plays
a crucial role.

Another consequence of OSFI is that a corresponding PDE may be obtained from a variational
principle. This is shown in the beginning of section 5.1. By Noether's theorem, the
symmetries of such a PDE lead to conservation laws. We then use a fundamental invariance of all
PDEs, considered in section 2, with respect to a shift of the unknown function by a constant to
derive a continuity equation. In section 5.1 this is done for a standard Hamiltonian of the
\Sch equation, and in section 5.2 --- in a general case, without using an explicit form of
a Hamiltonian. In section 5.3, we use the current conservation to prove that a form
$|\psi|^2 dV$ is an integral invariant of our equations of motion, which replaces the canonical
integral invariant (Liouville measure) $d^3\!p\, dV$ of the Hamilton equations in classical 
mechanics.

Using the invariance of the form $|\psi|^2 dV$, section 6 demonstrates that a
pro\-ba\-bi\-li\-ty density in configuration space should be equal to $|\psi|^2$. We give two 
proofs, the second one using the maximization of a specially introduced functional of probability
density, analogous to the Gibbs entropy. We compare the situation in QM to the one in classical 
statistics. A brief review of equilibrium and nonequilibrium classical statistics is presented in 
the Appendix in a form convenient for such comparison. It is shown there that the repetition of 
steps which led to the expression of Gibbs entropy in QM leads in classical statistics (where for
invariant measure one uses the form $d^3\!p\, dV$, rather than $|\psi|^2 dV$) to its standard 
classical expression. The probability density $|\psi|^2$ that maximizes Gibbs entropy in QM has, 
therefore, a status identical to that of a microcanonical distribution, which maximizes Gibbs 
entropy in classical statistics. Note that the difference between these distributions results from
the difference between corresponding invariant measures, which, in turn, follows from the 
difference
in equations of motion. Regarding the claims \cite{dbb,bohm2} that the $|\psi|^2$ distribution in 
QM may arise in a way similar to relaxation to statistical equilibrium in classical statistics, the
Appendix also shows that this relaxation is related to the growth of Boltzmann, rather than Gibbs,
entropy, and is caused by the properties of macroscopic systems which cannot have any analogs in a
one-particle theory.

In section 7, the one-particle theory of the previous sections is generalized to multiparticle
systems. We also discuss how the standard physical picture of quantum particles in a potential
created by classical macroscopic objects emerges from our theory.

Section 8 considers the theory of quantum measurements. Von Neumann's measurements with
discrete and continuous spectra are considered in sections 8.1 and 8.2. The theory discussed
there is a \paqd-adaptation of the theory developed by Bohm \cite{dbb,dbb2}. In section 8.3, the
measurement of a particle's position by a photographic plate or in a bubble chamber, which is not a
von Neumann's measurement, is considered, and its properties are discussed. In the end of this
section we analyze the double-slit experiment discussed by Feynman~\cite{fnmn} and compare its
results with \paqd\ predictions.

Section 9 considers the mechanism of nonlocal correlations between the results of measurements,
performed with space-like separated, but entangled, particles. We argue that the
relativistic version of \paqd, although nonlocal, will be Lorentz invariant.

Section 10 considers particles with spin. We show that their theory, which
adequately generalizes the theory of spinless particles, may be developed based on extended
configuration space, which includes, besides the particle's space position, also its ``internal"
SU(2) coordinates.

Finally in Conclusion, we give a brief review of our theory, compare it with standard QM, and 
finish with several general remarks.

\section{Hamiltonian flow in infinite jet space}
\setcounter{equation}{0}

\subsection{Basic definitions and notation}

In this section we discuss the question of when the solution of a PDE
system may be obtained, as in the case of the Hamilton-Jacobi equation, via solving some related
system of ODEs. Consider evolutionary PDEs of the form
\begin{equation}\label{evpde}
    \dovd{p^r}{t} + H^r = 0\,, \,\,\,\, r = 1, \ldots, m \, , \dsplbl{evpde}
\end{equation}
where the $p^r$ are $m$ unknown functions (``dependent variables") of $n$ space variables
 $q^1, \ldots , q^n$ combined into a vector $\bq$, and time~$t$
(``independent variables"), and $H^r$ are functions of $t$, $\bq$, and partial
derivatives of unknown functions with respect to space variables up to some finite order.
Denote these derivatives by corresponding multi-indices, as in $p^r_{ij} = \partial^2 p^r
/ \partial q^i \partial q^j$, and include in the set of all possible multi-indices an
empty one, denoted as $\0$, which will correspond to the function $p^r$ itself. Use
$i,j,k$ for space indices, running from $1$ to~$n$, use $r$ and $s$
for function indices, running from 1 to~$m$, and use Greek letters for
multi-indices. The order of indices in a multi-index is arbitrary, and two
multi-indices which differ only by permutation are considered to be the same. Correspondingly,
only one such multi-index will be assumed to be included in a summation over all possible
multi-indices.
If $\sigma = i_1 i_2 \ldots i_k$, let $\sigma i$ or $i\sigma$ be the ``extended"
multi-index $i_1 i_2 \ldots i_k i$, and if $\mu=j_1\ldots j_l$, let $\sigma\mu$ or
$\mu\sigma$ be the multi-index $i_1\ldots i_kj_1\ldots j_l$. Let $\sigma_i$, $i = 1,\ldots,n$,
denote the number of times index~$i$ is found in the multi-index $\sigma$, so that $\sigma$ may be
represented as $\sigma_1$ ones, followed by $\sigma_2$ twos, etc. It is useful to think of the
multi-index $\sigma$ as an $n$-dimensional vector with nonnegative integer components $\sigma_i$.
Summation over all possible multi-indices $\sigma$ then reduces to summation over all
$\sigma_i$: $\sum_\sigma = \sum_{\sigma_1,\ldots,\sigma_n=0}^\infty\,$. Let $|\sigma|$
denote the total number of indices in multi-index~$\sigma$, so $|\sigma| = \sum_{i=1}^n \sigma_i$.
Let $\partial_0 = \partial_t = \partial / \partial t$. For every
multi-index~$\sigma$, let $\sigma!=\prod_{i=1}^n\sigma_i!$ and $\partial_\sigma =
\prod_{i=1}^n (\partial / \partial q^i)^{\sigma_i}$. For any
$n$-dimensional vector $\bx$, let $\bx^\sigma=\prod_{i=1}^n (x^i)^{\sigma_i}$.

Let $\{p\}$ denote the set of all unknown functions and all their derivatives.
By analogy with classical mechanics,
functions $p^r$ will be called action functions, or just actions, and their derivatives,
momentums. Denote the set of all momentums, i.e., all $p\rs$ with $\sigma\neq\0$,
by $\bp$, so $H^r = H^r(t,\bq,\bp)$. Denote the space of independent
variables~$\bq$ and~$t$ (``base space") by $M$, and the space of vectors~$\bq$
alone (``configuration space") by $Q$. Let~$J^\infty$
(``infinite jet space") be the space with coordinates
$t$, $\bq$, $\{p\}$, i.e., all independent as well as dependent
variables and all their space derivatives. Call the similar space
\cP\ with coordinates $t$, $\bq$, $\bp$ ``infinite phase space." We assume that $H^r$ depends
analytically on its arguments, and consider only analytic or real-analytic
solutions of Eq.~(\ref{evpde}). A mathematically rigorous treatment of geometry
of analytic jets in~$J^\infty$ may be found in~\cite{zhar-gaj}. Denote by $J^k$ a jet space of
$k$-times continuously differentiable functions, which includes derivatives only up to $k$-th order
(``k-jets space") \cite{olvr,zhar,szsumf}. We will also use the notation $J^k_m$ when
it is necessary to indicate the number $m$ of unknown functions explicitly, and
when $m=1$ we will drop the function number superscript in equations.
We will consider every solution of Eq.~(\ref{evpde}), with all
its space partial derivatives, as creating a graph in $J^\infty$. Denote such a graph by $\Gamma$
and note it is an $n+1$~--~dimensional surface in $J^\infty$. We want to find out when such graphs
can be usefully considered as formed by a congruence of curves described by a system of ODEs.

\subsection{Curves in the graph of a PDE solution}

\begin{figure}
\centerline{\includegraphics[bb=15 110 185 200,clip,scale=1.7]{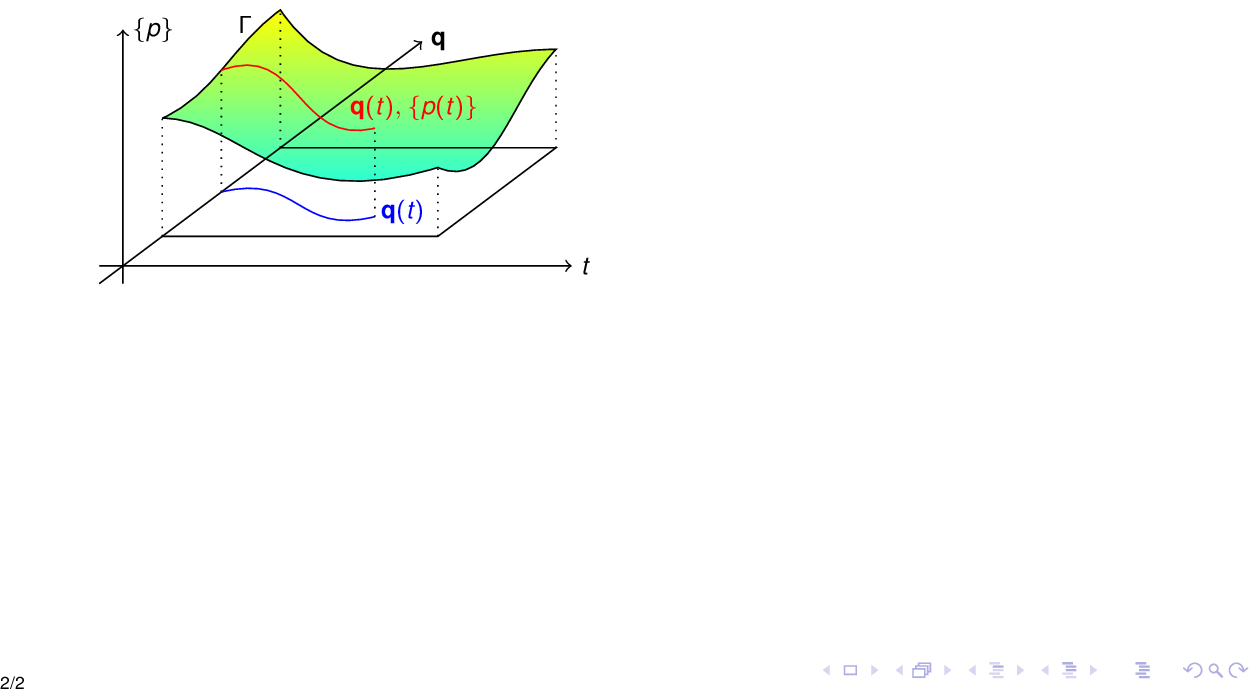}}
\caption{A graph $\Gamma$ of solution of PDE (\ref{evpde}) in the infinite jet space. When a 
point moves in the configuration space, the set of its coordinates $\bq$ and values $\{p\}$ of
solution and solutions's derivatives at $\bq$ moves in $\Gamma$ according to the system of 
ODEs~(\ref{tottd}).}
\label{fig2}
\end{figure}

It is easy to write an equation for an arbitrary curve which lies in the graph. Let $\bq(t)$ be the
projection of the curve to the base space $M$. Consider the operator of total differentiation
\begin{equation}\label{totdif}
    D_i = \dovd{}{q^i} + \sum_{r = 1}^m \sum_\sigma p\rsi \dovd{}{p\rs}\, ,  \dsplbl{totdif}
\end{equation}
where the second summation runs over all possible multi-indices $\sigma$. At every point of
$J^\infty$, the operator $D_i$ raises the partial derivative $\partial /\partial q^i$
to a graph of an analytic function of $\bq$, which  passes through the point
(see~\cite{olvr}). In other words, if $\Theta$ is such a graph,
$\{p_\Theta(\bq)\}$ is a set of values of unknown functions and their derivatives at a point of
$\Theta$ with a base coordinate $\bq$, and $F(t,\bq,\{p\})$ is some function in $J^\infty$, then
\begin{equation}\label{totder}
    \dovd{}{q^i}\, F\big(t,\bq,\{p_\Theta(\bq)\}\big) = D_i F\!\left.\big(t,\bq,\{p\}\big)
    \right|_{\{p\} = \{p_\Theta(\bq)\}}\, .     \dsplbl{totder}
\end{equation}
By consecutive differentiation of Eq.~(\ref{evpde}) with respect to space variables we now obtain
equations \big(``prolongations" of (\ref{evpde})\big) which describe the behavior
of space derivatives~$p\rs$ of the solution
\begin{equation}\label{prlngn}
    \dovd{p\rs}{t} + H\rs = 0 \, ,  \dsplbl{prlngn}
\end{equation}
where $H\rs = D_\sigma H^r$ and $D_\sigma$ is repeated total differentiation, i.e.,
$D_\sigma = \prod_{i=1}^n (D_i)^{\sigma_i}$.
Eq.~(\ref{prlngn}) gives the value of the partial derivative
of $p\rs$ with respect to $t$, while its partial derivative with respect to $q^i$ is,
by definition,~$p\rsi$. Consequently, the time dependence of the $\Gamma$-image of the
point $\bq(t)$ on the base is described by the system of ODEs\footnote{Summation over repeated
indices and multi-indices is assumed here and below unless stated otherwise. The summation will
\emph{not} be assumed when the expression with repeated indices or multi-indices stands on one
side of an equality if the other side of the equality also has the same (multi-)indices used
only once.}
\begin{equation}\label{tottd}
     \begin{array}{ccl}
       \dsp \dot{p}\rs & = & \dsp \dovd{p\rs}{q^i}\, \dot{q}^i + \dovd{p\rs}{t} \\[0.4cm]
                       & = & \dsp  p\rsi\, \dot{q}^i - H\rs \, ,    \dsplbl{tottd}
     \end{array}
\end{equation}
where by dot we denote the total derivative of a corresponding value with respect to $t$
along the curve $\bq(t)$ (see Fig.~\ref{fig2}).

Given the PDE (\ref{evpde}), we constructed the ODE system (\ref{tottd}), which
defines the evolution of values of unknown functions and their derivatives at a point $\bq(t)$
moving (in an arbitrary way) through the base.
It is easy to show that there also exists an inverse
correspondence: if there is a point $\bq(t)$ moving through the base, and a sequence of functions
of time $p^s_\sigma(t)$, which satisfies the system of ODEs (\ref{tottd}), then functions
$p^s({\bf r},t)$, which are the sums of Taylor series in a point $\bq(t)$ with coefficients
$p^s_\sigma(t)$, will satisfy PDE (\ref{evpde}). Indeed, these Taylor series are
\begin{equation}\label{us}
    p^s({\br},t)\, = \sum_\sigma \frac{1}{\sigma!} \, p^s_\sigma (t)
    \big(\br - \bq(t)\big)^\sigma .   \dsplbl{us}
\end{equation}
Taking the time derivative and using Eq.~(\ref{tottd}) for $\dot{p}^s_\sigma$, one obtains,
after some cancellations,
\begin{equation}\label{}
    \dovd{}{t}\, p^s(\br,t)\, = \,- \sum_\sigma \frac{1}{\sigma!} \, D_\sigma H^s \!\!
    \left. \big(t,\bq,\{p\}\big)\right|_{\{p\} = \{p^s_\sigma (t)\}}
    \big(\br - \bq(t) \big)^\sigma,    \dsplbl{}
\end{equation}
which by Eq.~(\ref{totder}) and analyticity of $H^s$ is equal to
$-H^s\big(t,\br,\{p_\Gamma(\br)\}\big)$, where $\Gamma$ is the graph of functions (\ref{us}) in
$J^\infty$, as required.

We see that there exists a simple and general relation between the evolution of
analytic functions,
described by PDE (\ref{evpde}), on one hand, and the evolution of coordinates $\bq(t)$ and
momentums~$p\rs(t)$, described by ODE system (\ref{tottd}), on the other. This relation is
just a straightforward consequence of the structure of Eqs.~(\ref{evpde}) and (\ref{tottd}),
and is satisfied for an arbitrary curve~$\bq(t)$. However, if one adds to system
(\ref{tottd}) an additional equation, expressing $\dot{q}^i$ through other variables, then this
extended system of ODEs will be closed with respect to the evolution of all variables involved,
and may be viewed as representing equations of motion that are generated by ``Hamiltonians" $H^r$
in the same way as Hamilton equations are generated by a Hamiltonian in classical mechanics.
Consequently, to obtain potential equations of motion, some plausible condition capable of
providing such an equation is needed.

\subsection{Generalized Hamiltonian fields and equations of motion} \label{ghfem}

To formulate the condition described above, we will use concepts and notation of the geometric
theory of PDEs\footnote{A reader unfamiliar with this theory may skip directly to
Eq.~(\ref{preheq}), which is a desired expression for velocity. An elementary justification of this
expression is presented in a footnote immediately after it.}
\cite{olvr,zhar,szsumf,ibragim,bcg3,amp}. The set $\left\{D_i, \, \partial /\partial
p\rs, \, \partial /\partial t \right\}$ is a basis for the tangent bundle $T(J^\infty)$. Define the
sequence of 1-forms
\begin{equation}\label{tcforms}
    \wt{\omega}\rs = \rmd p\rs - p\rsi \rmd q^i .  \dsplbl{tcforms}
\end{equation}
The set $\left\{\rmd q^i, \, \wt{\omega}\rs, \, \rmd t \right\}$ is a basis for the cotangent
bundle $T^*(J^\infty)$. The duality relations
\begin{equation}\label{ips}
    \begin{array}{lcllcllcl}
       D_i \ip \rmd q^j & = & \delta_{ij} \, ,\quad & D_i \ip \wt{\omega}\rs & = & 0 \, ,\quad
       & D_i \ip \rmd t & = & 0 \, , \\[0.3cm]
       \dsp \dovd{}{p\rs} \ip \rmd q^i & = & 0 \, ,\quad & \dsp \dovd{}{p\rs} \ip \wt{\omega}^s_\nu
       & = & \delta_{rs} \delta_{\sigma\nu} \, ,\quad & \dsp \dovd{}{p\rs} \ip \rmd t & = & 0 \, ,
         \dsplbl{ips} \\[0.5cm]
       \dsp \dovd{}{t} \ip  \rmd q^i & = & 0 \, ,\quad & \dsp \dovd{}{t} \ip \wt{\omega}\rs & = & 0
       \, ,\quad  & \dsp \dovd{}{t} \ip \rmd t & = & 1 \, ,
    \end{array}
\end{equation}
then follow at every point of $J^\infty$. In the following, let $T\rs$ be the distribution
generated by the fields $D_i$ and $\partial/\partial p\rsi$, $i = 1,\ldots ,n$. The curves
(\ref{tottd}) are now recognized as integral curves of a vector field
\begin{equation}\label{fldx}
     \begin{array}{ccl}
       X & = & \dsp \dovd{}{t} + \dot{q}^i \dovd{}{q^i} + \dot{p}\rs \dovd{}{p\rs} \\[0.4cm]
         & = & \dsp \dovd{}{t} + \dot{q}^i D_i - H\rs \dovd{}{p\rs} \, ,   \dsplbl{fldx}
     \end{array}
\end{equation}
which cancels the Pfaff system of differential forms (``Cartan forms")
\begin{equation}\label{cforms}
    \omega\rs = \rmd p\rs - p\rsi \, \rmd q^i + H\rs \, \rmd t \, ,   \dsplbl{cforms}
\end{equation}
i.e., satisfies equations
\begin{equation}\label{ixw}
    X \ip \omega\rs = 0, \quad \mbox{for all} \;\, r, \sigma \, .    \dsplbl{ixw}
\end{equation}
The system $\{\omega\rs\}$ defines in $J^\infty$ a distribution (``Cartan distribution" or CD), and
the graphs $\Gamma$ of solutions of (\ref{evpde}) are the integral manifolds of this distribution,
which is a necessary and sufficient condition for every surface in $J^\infty$ to satisfy two
requirements: First, to truly represent a graph of some function, i.e., to ensure agreement
between coordinates $p\rs$ in $J^\infty$ and the values of corresponding partial derivatives of a
function $p^r(\bq, t)$ represented by this graph. And second, to guarantee that this function
satisfies Eq.~(\ref{evpde}) \cite{olvr,zhar,szsumf}.

In the case of first order PDEs with $m=1$, considered in a space of one-jets $J^1_1$,
the trajectories, described by Hamilton
equations, are the characteristic curves of a corresponding exterior differential system. These
curves are uniquely defined in $J^1_1$, and so one and only one of them passes through every point
on the graph $\Gamma$ of the PDE solution. It is therefore natural to expect that the desired
trajectories in $J^\infty$ are the characteristic curves of the system $\{\omega\rs\}$. This
condition, however, happens to not be sufficient for the selection of $\dot{\bq}$. Indeed, by
direct calculation it is easy to check that the differential forms (\ref{cforms}) satisfy the
equation
\begin{equation}\label{dw}
    \rmd \omega\rs = \rmd q^i \wedge \omega\rsi - \dovd{H\rs}{p^s_\nu}\,\, \rmd t \wedge
    \omega^s_\nu \,.   \dsplbl{dw}
\end{equation}
From Eqs.~(\ref{dw}) and (\ref{ixw}) we immediately obtain that $X \ip \rmd \omega\rs$ is a
linear combination of forms (\ref{cforms}). Therefore,
for every $\dot{\bq}$ the vector field $X$ not only belongs to CD, but is also its characteristic
field, and so in sharp contrast with the case of first order PDEs,
there is a continuum of characteristic curves of CD passing through every point
of~$\Gamma$ in~$J^\infty$.

It is instructive to consider the source of this difference.
By analyticity, the graph of one and only one analytic function can pass in $J^\infty$ through
every point, which defines all spatial partial derivatives.
Therefore, $J^\infty$ is split into a foliation, each leaf of which is a graph $\Gamma$ of the
analytic solution of Eq.~(\ref{evpde}), so that every point of $J^\infty$ belongs to one and only
one leaf. These leaves are
the integral manifolds of CD, so CD in the infinite jet space of analytic functions
is completely integrable.\footnote{We note that this
complete integrability cannot be considered a consequence of the Frobenius theorem and
Eq.~(\ref{dw}), for the Frobenius theorem requires finite dimensionality of the space, and in
infinite-dimensional space is no longer true \cite{zhar}.} In contrast to~$J^\infty$,
in~$J^1_1$ CD is not completely integrable. Consequently, the dimension of the graphs $\Gamma$
there is less than the dimension of CD, and so through every point of $J^1_1$ different
integral manifolds of CD (i.e. graphs of solutions $\Gamma$) pass, namely,
with different values $p^{\scriptscriptstyle \Gamma}_{ik}$ of the second derivatives on $\Gamma$.
On the other hand, graphs $\Gamma$ are formed by characteristic curves, so at every
point these curves should belong to every graph that passes through this point. It turns out that
this condition alone is sufficient to uniquely specify the characteristic field $X$ at this point,
including the value of~$\dot{\bq}$. Indeed, the corresponding PDE has the form
\begin{equation}\label{j11pde}
    \dovd{p}{t} + H = 0\,.   \dsplbl{j11pde}
\end{equation}
On a graph $\Gamma$ of a given solution of (\ref{j11pde}) in $J^1_1$, the operator of the total
derivative is
\begin{equation}\label{DGamma}
    D^{\scriptscriptstyle \Gamma}_i = \dovd{}{q^i} + p_i \, \dovd{}{p} +
    p^{\scriptscriptstyle \Gamma}_{ik} \, \dovd{}{p_k} \, .   \dsplbl{DGamma}
\end{equation}
Similar to (\ref{tottd}), the curves on $\Gamma$ should satisfy a system of ODEs:
\begin{eqnarray}
    \dot{p} & = & \dsp p_i \, \dot{q}^i - H ,          \label{pdot}   \dsplbl{pdot} \\[0.1cm]
    \dot{p}_i & = & \dsp p^{\scriptscriptstyle \Gamma}_{ik}\,\dot{q}^k -
                    D^{\scriptscriptstyle \Gamma}_i H     \nonumber \\[0.1cm]
              & = & \dsp p^{\scriptscriptstyle \Gamma}_{ik} \left(
                    \dot{q}^k - \dovd{H}{p_k} \right) - \left( \dovd{}{q^i} +
                    p_i \dovd{}{p} \right) H  .   \label{pidot}   \dsplbl{pidot}
\end{eqnarray}
Now, as \textit{characteristic} curves should belong to every such graph, the dependence
on~$p^{\scriptscriptstyle \Gamma}_{ik}$ for them must disappear. Recalling also that $H$ does not
depend on $p$, we then obtain from Eq.~(\ref{pidot}) the standard Hamilton equations for
$\dot{q}^i$ and $\dot{p}_i$:
\begin{equation}\label{he}
    \dot{q}^i  =  \dovd{H}{p_i} \, , \quad\; \dot{p}_i = -\dovd{H}{q^i} \, .   \dsplbl{he}
\end{equation}

Thus, the fact that in $J^1_1$ characteristics of CD have a Hamiltonian form is based on the
specifics of~$J^1_1$, namely, on the lack of complete integrability of CD there, which
explains why this property cannot be generalized to $J^\infty$. Fortunately, however, the
requirement of being characteristics is not the only one which distinguishes Hamiltonian curves
from all other curves that lie on the graphs of solutions of Eq.~(\ref{j11pde}) in $J^1_1$.
Let $\omega = \rmd p - p_i \rmd q^i + H \rmd t$
be the (only) Cartan form in $J^1_1$. The Hamiltonian curves in $J^1_1$ may then be defined as
integral curves of a vector field $X$, which cancels the 2-form $\rmd \omega$, i.e., satisfies the
condition $X \ip \rmd \omega = 0$ \cite{arnld}. This condition may of course be rewritten as
$\xi \ip (X \ip \rmd \omega) = 0$, for all $\xi \in T\big(J^1_1\big)$.
Now, in $J^\infty$, CD is defined by a sequence of forms $\{ \omega\rs \}$, and a fruitful
generalization of the above condition in~$J^1_1$ is to require
\begin{equation}\label{hamcond}
    \xi \ip (X \ip \rmd \omega\rs) = 0, \quad \mbox{for all} \;\, r,\sigma, \xi \in T\rs,
    \dsplbl{hamcond}
\end{equation}
in $J^\infty$, which is a condition satisfied by the usual Hamiltonian curves of first-order PDEs
with $m=1$ after their prolongation from $J^1_1$ to $J^\infty_1$. In the following, we will call
fields $X$ that satisfy conditions (\ref{hamcond})
generalized Hamiltonian or, for short, simply Hamiltonian as they are called in $J^1_1$.
To find their form we observe that the internal product of~$X$, given by the first line of
Eq.~(\ref{fldx}), with $\rmd \omega\rs$, is
\begin{equation}\label{xipdrho}
    X \ip \rmd \omega\rs = \sum_i \left( \dot{q}^i - \dovd{H\rs}{p\rsi} \right) \omega\rsi \,
    - \, \sum_i \left( \dot{p}\rsi - l\rsi \right) \rmd q^i \, - \,
    {\sum_{s,\, \nu}}' \dovd{H\rs}{p^s_\nu} \, \omega^s_\nu \, + \, \sum_{s,\, \nu}
    \left( \dot{p}^s_\nu - l^s_\nu \right) \dovd{H\rs}{p^s_\nu} \, \rmd t \, ,   \dsplbl{xipdrho}
\end{equation}
where all summations are explicit, $\sum'_{s,\, \nu}$ omits terms with $(s,\nu) =
(r, \sigma i)$ for all $i$, and where we introduced functions
$l\rs = l\rs(t,\bq,\bp,\dot{\bq})$ by the relations
\begin{equation}\label{lrs}
    l\rs = p\rsi \dot{q}^i - H\rs \, .    \dsplbl{lrs}
\end{equation}
Now if the field $\xi$ belongs to $T\rs$, i.e. is a linear combination of $D_i$ and $\partial
/ \partial p\rsi$, then only the first two terms in (\ref{xipdrho}) contribute to
$\xi \ip (X \ip \rmd \omega\rs)$, and the condition (\ref{hamcond}) gives
\begin{eqnarray}
  \dot{q}^i & = & \dsp \dovd{H\rs}{p\rsi} \, , \label{preheq} \dsplbl{preheq} \\[0.2cm]
  \dot{p}\rsi & = & l\rsi , \label{prehep}  \dsplbl{prehep}
\end{eqnarray}
where in Eq.~(\ref{preheq}) summation over $r$ and $\sigma$ is not assumed.
Equations (\ref{prehep}) (with arbitrary $\sigma$) just reproduce Eqs.~(\ref{tottd}) (with
$\sigma\neq\0$), while Eq.~(\ref{preheq}) gives the desired expression for velocity.\footnote{Note
that for first order PDEs with one unknown function this expression does not depend on $\sigma$ and
gives the same value of $\dot{q}^i$ as in Eq.~(\ref{he}). Consequently, one can bypass the
geometric consideration above by simply postulating (\ref{preheq}) in a general case. For all PDEs
such that the right hand side of (\ref{preheq}) does not depend on $r$ and $\sigma$, the resulting
theory will be a generalization of the usual Hamiltonian formalism for first order PDEs with one
unknown function.} For this expression to make sense, we must additionally require that the
derivative there be independent of $r$ and $\sigma$. We will prove the following:
$\partial H\rs /\partial p\rsi$ does not depend on $\sigma$ if $H^r$ satisfies the condition
\begin{equation}\label{hcond}
    D_k \, \dovd{H^r}{p^r_{\nu k}} = 0, \quad \mbox{for all} \;\, k, \nu\neq\0   \dsplbl{hcond}
\end{equation}
(no summation over $r$ and $k$ here!), as, for example, if $H^r$ is linear in second and higher
order derivatives of $p^r$ with constant coefficients. First we show that if any function
$F(t,\bq,\{p\})$ depends on derivatives $p\rs$ only up to some finite order, and satisfies
the condition $D_k F = 0$, then it is a function of $t$ and $q^j$, $j\neq k$, only.
Indeed, let $r, \nu$ be such that $\partial F / \partial p\rs = 0$ for all~$\sigma$ such that
$|\sigma| > |\nu|$. Then applying the obvious identity
\begin{equation}\label{}
    \dovd{}{p^r_\nu} = \dovd{}{p^r_{\nu k}} D_k - D_k \dovd{}{p^r_{\nu k}} \, ,
\end{equation}
where summation over $k$ is not assumed, to the function $F$, we obtain immediately that
$\partial F / \partial p^r_\nu = 0$ also. Consequently, $F$ cannot depend on any~$p^r_\nu$ at
all, and is a function of~$t$ and~$\bq$ only, so that the equalities $0 = D_k F = \partial F /
\partial q^k$ prove our statement. Now for a function~$H^r$
which satisfies condition (\ref{hcond}),
this means that for any nonempty multi-index~$\nu$ and any~$k$, $\partial H^r / \partial
p^r_{\nu k}$ may only be a function of $q^j$, $j\neq k$, and $t$. Consequently, $H^r$ has the form
$H^r = H^r_1 + H^r_2$, where $H^r_1 = H^r_1\big(t,\bq,p^r_i,\{p^s_\sigma, s \neq r\}\big)$, and
$H^r_2=\sum_{|\nu|>1} a_\nu p^r_\nu$, where $a_\nu$ depends on~$t$ and $q^j$, $j \notin \nu$, only.
It can then be easily seen that $H^r_{2\sigma}$ does not contain $p\rsi$ at all, while for any
$\sigma\neq\0$ the only term with $p\rsi$ in $H^r_{1\sigma}$ is an additive term equal to
$p\rsi \,\partial H^r_1 / \partial p^r_i$, and so all $\partial H\rs /
\partial p\rsi$ are equal to $\partial H^r_1 / \partial p^r_i$.

We conclude that functions $H^r$ produce a Hamiltonian field if they
satisfy the following ``Hamiltonian conditions" (HC): First, for all
$r$, the $H^r$ satisfy Eq.~(\ref{hcond}) (HC1). Second, given $i$, $\partial H^r
/ \partial p^r_i$ is independent of $r$ (HC2). Equation~(\ref{preheq}) then gives for $\dot{q}^i$
the values that are the same for all $r$ and $\sigma$, and, along with Eq.~(\ref{tottd}) for
$\dot{p}\rs$, constitutes the desired system of equations of motion.
We will now discuss the structure of the resulting theory in more detail.

First note that the right hand sides of Eqs.~(\ref{preheq}) and (\ref{tottd}) depend only
on~$p\rs$ with $\sigma\neq\0$, and not on $p^r$.
Consequently, this system of equations splits into two subsystems. The first subsystem consists of
Eq.~(\ref{preheq}) and Eq.~(\ref{tottd}) with $\sigma\neq\0$, which determine time
histories (trajectories) $\bq(t)$, $\bp(t)$. The second subsystem, consisting of equations
(\ref{tottd}) with $\sigma=\0$, shows how actions $p^r$ vary along these trajectories. The
equations of the first subsystem, which is equivalent
to the system~(\ref{preheq}) and (\ref{prehep}), constitute the desired set of equations of
motion. After the equations of motion are solved, equations
of the second subsystem allow us to obtain expressions for the actions $p^r$ along corresponding
trajectories by quadrature. As will be discussed later, if one
makes equations of motion a starting point of a theory,
then these expressions may be considered as {\em defining} actions for given trajectories.

We will now show that Eq.~(\ref{prehep}) may be presented in a ``Hamiltonian" form, similar to the
second Eq.~(\ref{he}). Indeed, using Eq.~(\ref{preheq}) and making all summations explicit, we have
\begin{equation}\label{lrsi}
    \begin{array}{ccl}
      l\rsi & = & \dsp \sum_j p^r_{\sigma i j} \dot{q}^j - D_i H\rs \\[0.2cm]
                     & = & \dsp \sum_j p^r_{\sigma i j} \dot{q}^j - \left(\sum_j p^r_{\sigma j i}
                           \dovd{}{p^r_{\sigma j}} + \dovd{}{q^i} + {\sum_{s,\, \nu}}'
                           p^s_{\nu i} \dovd{}{p^s_\nu} \right) H\rs \\[0.4cm]
                     & = & \dsp - \left(\dovd{}{q^i} + {\sum_{s,\, \nu}}'
                           p^s_{\nu i} \dovd{}{p^s_\nu} \right) H\rs \, ,
    \end{array}\dsplbl{lrsi}
\end{equation}
where $\sum'_{s,\, \nu}$ omits terms with $(s,\nu) = (r, \sigma j)$ for all $j$. Now consider
$H\rs$ as a function of~$p\rsj$, $j = 1,\ldots,n$, and all its other arguments: $H\rs =
H\rs(t,\bq,\bp',p\rsj)$, where $\bp'$ is the set of all $p^s_\nu$ such that $\nu\neq\0$ and
for any~$j$, $(s,\nu) \neq (r, \sigma j)$.
Let $\Theta$ be the graph of a $t$-independent analytic function of $\bq$ passing through a point
with \cP-coordinates $(t, \bq, \bp)$ in~$J^\infty$. Let $\bp'_\Theta(\bq)$ be the set of values of
$\bp'$ at a point of~$\Theta$ with a base coordinate~$\bq$.
Then using (\ref{lrsi}), Eq.~(\ref{prehep}) may be written as
\begin{equation}\label{prsidot}
    \dot{p}\rsi = - \dovd{}{q^i} H\rs\big(t,\bq, \bp'_\Theta(\bq), p\rsj\big) \, . \dsplbl{prsidot}
\end{equation}
Although the right hand side of this equation is nothing but the last line of Eq.~(\ref{lrsi})
rewritten less explicitly, we prefer this form because of its obvious analogy with the standard
Hamilton equation for momentum. Making the reference to graph $\Theta$ implicit, the system of
equations of motion (\ref{preheq}) and (\ref{prehep}) can now be presented in the form
\begin{equation}\label{em}
    \dot{q}^i = \dsp\dovd{}{p\rsi}\, H\rs\big(t,\bq, \bp'(\bq), p\rsj\big)\,, \quad\; \dot{p}\rsi =
                - \dovd{}{q^i}\, H\rs\big(t,\bq, \bp'(\bq), p\rsj\big) \, .  \dsplbl{em}
\end{equation}
We will informally refer to variables $q^i$, $p\rs$, and $p\rsi$ as forming an $r$-$\sigma$ sector
of the theory. Evidently, in every $r$-$\sigma$ sector the theory looks like a standard theory of
characteristics for the Hamilton-Jacobi equation with the Hamiltonian
$H\rs$ in $J^1_1$ --- the same
conclusion that may be drawn by comparison of Eqs.~(\ref{prlngn}) and (\ref{j11pde}), or
(\ref{tottd}) and (\ref{pdot}). We will next explore other aspects of this similarity.

\subsection{Variational principles}  \label{varprinc}

If one introduces ``action forms" $\rho = p_i\, \rmd q^i - H\, \rmd t$ in $J^1_1$ and
$\rho\rs = p\rsi\, \rmd q^i - H\rs \,\rmd t$ in $J^\infty$, then the corresponding
Cartan forms may be written as $\omega = \rmd p - \rho$ and $\omega\rs = \rmd p\rs - \rho\rs$. For
an arbitrary curve $C$
in $J^1_1$, the difference of $p$ at its ends is $\Delta p = \int_C \rmd p$. Now if curve $C$ lies
on the graph $\Gamma$ of the solution, then the tangent vector $X$ at an arbitrary point of $C$
cancels the 1-form $\omega$,
and so $X\ip \rmd p = X\ip \rho$. For such curves, therefore, the difference of $p$ at their ends
is given by the invariant Hilbert integral $\Delta p = \int_C \rho$ \cite{gf,rnd}. Similarly, in
$J^\infty$, the difference of $p\rs$ at the ends of any curve $C$ that lies on the graph of
solution is
\begin{equation}\label{dprs}
    \Delta p\rs = \int_C \rho\rs \, .   \dsplbl{dprs}
\end{equation}
In $J^1_1$, the condition that the integral $\int_C \rho$ remains
stationary with respect to any variations of curve $C$ that do not change the space and time
coordinates of its ends may be used to select the curves (trajectories) which satisfy the Hamilton
equations of motion \cite{arnld}. Similarly, in $J^\infty$, the trajectories that satisfy
Eqs.~(\ref{em})
may be selected by the condition that for every $r$ and~$\sigma$, the integral $\int_C \rho\rs$ is
stationary with respect to variations of $C$ that do not change the base coordinates of its ends
and are generated by any vector field which belongs to $T\rs$. Indeed, let $C$ go
from point $A$ to point $B$ in~$J^\infty$, and let an infinitesimal $\veps$-variation
transform $C$ into the curve $C'$, which goes from $A'$ to $B'$. By the condition stated above,
the base coordinates of $A'$ and $B'$ are the same as of
$A$ and $B$, and so $\int_A^{A'} \rho\rs = \int_{B'}^B \rho\rs = 0$, where the integrals are taken
along straight segments connecting $A$ with $A'$ and $B$ with $B'$.
Consequently, the variation of the integral is equal to the integral over the closed loop
$AA'B'BA$, which in turn, using Stokes' theorem,
may be represented as an integral over an area $D$ inside the loop:
\begin{equation}\label{}
   \delta \int_C \rho\rs \,\,=\,\, \int_{C'} \rho\rs - \int_C \rho\rs \,\,
   =\,\, \oint_{AA'B'BA} \rho\rs \,\,=\,\, \int_D \rmd \rho\rs \, .
\end{equation}
If $C$ is an integral curve of a vector field $X$, parameterized by a base coordinate $t$, and the
variation is generated by a vector field $V$, then the above integral is equal to $\veps
\int_{t_A}^{t_B} V \ip (X \ip \rmd \rho\rs) \rmd t$. Since $\rmd \rho\rs =
- \rmd \omega\rs$, the integrand here has the same form as the left hand side of
Eq.~(\ref{hamcond}). Consequently, repeating the derivation that follows (\ref{hamcond}) we
conclude that this integral vanishes for arbitrary field $V \in T\rs$ if and only if
Eqs.~(\ref{preheq}) and (\ref{prehep}) are satisfied (or, equivalently, Eq.~(\ref{em}) is
satisfied). Clearly, this variational principle is completely equivalent to our original condition
(\ref{hamcond}), and may be considered to be its restatement.

As in $J^1_1$, our trajectories may be also obtained from a different variational
principle, a Lagrangian one. For that, we introduce the Legendre transformations of Hamiltonians
$H\rs$, i.e. the functions
\begin{equation}\label{lt}
    \begin{array}{rcl}
      L\rs & = & \dsp l\rs \left(t,\bq,\bp,\dovd{H\rs}{p\rsi}\right) \\[0.4cm]
           & = & \dsp p\rsi \, \dovd{H\rs}{p\rsi} - H\rs \, .   \dsplbl{lt}
    \end{array}
\end{equation}
For Hamiltonians satisfying condition (\ref{hcond}), when $\sigma\neq\0$, this transformation
is of a trivial nature: As was discussed above, in these cases $H\rs$ contains~$p\rsi$ only in an
additive term which is linear in it, so in Eq.~(\ref{lt}) this term gets canceled, and~$L\rs$ is
simply equal to the sum of the remaining terms with a minus sign. When $\sigma=\0$, however,
$H^r$ does not have to be linear in~$p^r_i$.
We assume that there are some values of $r$ such that $H^r$ is a nonlinear function of~$p^r_i$,
and will consider only those values of $r$ below. For those values of $r$,
velocities $v^i = \partial H^r /\partial p^r_i$ are nontrivial functions of momentums~$p^r_i$.
These relations between velocities $v^i$ and momentums $p^r_i$
are supposed to be resolved with respect to $p^r_i$, expressing them through $\bv$ as
$p^r_i = \varphi^r_i(t,\bq,\bp',\bv)$ with some functions $\varphi^r_i$, where, here and in what
follows, $\bp'$ is a set of all~$p^s_\nu$ such that $\nu\neq\0$ and $(s,\nu)\neq (r,i)$.
The obtained expression for $p^r_i$ should then be substituted into Eq.~(\ref{lt}) with
$\sigma=\0$, resulting in the definition
\begin{equation}\label{}
    L^r(t,\bq,\bp',\bv) = v^i \varphi^r_i(t,\bq,\bp',\bv) - H^r\big(t,\bq,\bp',
    \varphi^r_i(t,\bq,\bp',\bv)\big)\, .
\end{equation}
The above-defined Lagrangians $L^r$ can now be used to derive the principle of stationary action
in its Lagrangian form from the invariant Hilbert integral. The derivation closely follows the one
for the first order PDE \cite{rnd}; nevertheless, it is presented here for completeness and
because of the complicating presence of (absent in the first order case) higher momentums.

Let $A$ and $B$ be two points on the graph $\Gamma$ of the solution that both belong to the same
``true" \big(i.e., obtained by solution of Eqs.~(\ref{preheq}) and (\ref{tottd})\big) trajectory
$C_{AB}$. Let $\bq(t)$ be an arbitrary curve through the base, connecting the base projections of
$A$ and $B$. The difference between the values of $p^r$ in $B$ and $A$ may then be expressed, as in
Eq.~(\ref{dprs}), by the integral
\begin{equation}\label{dpab}
    \Delta p^r_{AB} = \dsp \int_{C_\bq} \big[p^r_i \dot{q}^i - H^r(t,\bq,\bp',p^r_i)
    \big] \rmd t \, ,     \dsplbl{dpab}
\end{equation}
where $C_\bq$ is the image of the curve $\bq(t)$ on the graph $\Gamma$. In this integral, $p^r_i$
and $H^r$ may be expressed \cite{zia} through $L^r$ as
\begin{equation}\label{}
    p^r_i = \dovd{}{v^i}\, L^r(t,\bq,\bp',\bv) \, , \quad \,
    H^r(t,\bq,\bp',p^r_i) = p^r_i v^i - L^r(t,\bq,\bp',\bv) \, ,
\end{equation}
where the value of $v^i$ in these formulas should be set to a known value
$\partial H^r(t,\bq,\bp',p^r_i) / \partial p^r_i$, i.e., to the velocity of a true trajectory,
passing through the corresponding point $\bq$. Substituting these expressions into
Eq.~(\ref{dpab}), obtain
\begin{equation}\label{}
    \Delta p^r_{AB} = \dsp \int_{C_\bq} \left[ L^r(t,\bq,\bp',\bv) + (\dot{q}^i - v^i)
    \dovd{}{v^i}\, L^r(t,\bq,\bp',\bv) \right] \rmd t \, .
\end{equation}
Now the integrand here contains the first two terms of a Taylor series expansion of
$L^r(t,\bq,\bp',\dot{\bq})$ in powers of~$\dot{q}^i - v^i$. When $\dot{q}^i$ is close to
$v^i$, i.e., when the curve $C_\bq$ is close to the true trajectory $C_{AB}$, we have
\begin{equation}\label{}
    \dsp \int_{C_\bq} L^r(t,\bq,\bp',\dot{\bq}) \rmd t - \Delta p^r_{AB} = \dsp \frac{1}{2}
    \int_{C_\bq}(\dot{q}^i - v^i)(\dot{q}^j - v^j) \frac{\partial^2}{\partial v^i \partial v^j}\,
    L^r(t,\bq,\bp',\bv) \rmd t + O \left( (\dot{q}^i - v^i)^3 \right) ,
\end{equation}
which is of second order with respect to $\dot{\bq}-\bv$ and, therefore, with respect to the
deviation of trajectory $C_\bq$ from the true one $C_{AB}$. Consequently, the ``action integral"
$\int_{C_\bq} L^r(t,\bq,\bp',\dot{\bq}) \rmd t$ is stationary with respect to variations of
$C_\bq$ around $C_{AB}$, which is the way the principle of stationary action
is formulated in $J^\infty$. The
stationarity of the action integral is, therefore, an alternative condition, which may be used to
select true trajectories, which satisfy Eqs.~(\ref{preheq}) and
(\ref{tottd}), from arbitrary curves on $\Gamma$, which satisfy Eq.~(\ref{tottd}) only.
Eq.~(\ref{prsidot}) with $\sigma=\0$ then becomes the Euler-Lagrange equation in the usual
way, and the difference between the values of the action function~$p^r$ in two points may be
expressed, as for first order PDE, as an integral from the Lagrangian along the true trajectory
that connects these points:
\begin{equation}\label{daction}
    \Delta p^r_{AB} = \dsp \int_{C_{AB}} L^r(t,\bq,\bp',\bv)\, \rmd t \, .   \dsplbl{daction}
\end{equation}
It is clear from Eqs.~(\ref{dprs}) and (\ref{lt}) that a similar expression may be written for
$\Delta p\rs$ with arbitrary $r$ and~$\sigma$, the only difference being that when $H\rs$ is
linear in $p\rsi$, the corresponding Lagrangian $L\rs$ does not depend on~$\bv$ and so is a
function of $t$, $\bq$, and $\{p^s_\nu\!: \nu\neq\0, (s,\nu) \neq (r,\sigma i)\}$ only. Therefore,
we always have
\begin{equation}\label{dactrs}
    \Delta p^r_{\sigma_{AB}} = \dsp \int_{C_{AB}} L\rs\, \rmd t \, . \dsplbl{dactrs}
\end{equation}

\subsection{From ODEs to PDE: infinite phase space formulation} \label{ode2pde}

So far, we started with the PDE (\ref{evpde}) and developed the system of ODEs~(\ref{em}). Now we
take system (\ref{em}) as the starting point and will construct a corresponding PDE from it.
The system lives in an infinite phase space \cP, which has a geometry almost
identical to that of~$J^\infty$:
The operator of total differentiation $D_i$ (\ref{totdif}), basis
forms $\wt{\omega}\rs$ (\ref{tcforms}), duality relations (\ref{ips}), and Cartan forms $\omega\rs$
(\ref{cforms}) in \cP\ will be the same as in $J^\infty$, with the only difference that in
\cP\ all multi-indices in these formulas and in all summations should be nonempty. Thus forms
$\omega\rs$ and $\wt{\omega}\rs$ exist only for $\sigma\neq\0$, and integral manifolds of CD,
defined by forms $\omega\rs$, are graphs of {\em derivatives} of analytic solutions of
Eq.~(\ref{evpde}). Also, vector field $X$, generated by Eq.~(\ref{em}), is tangent to these
graphs, has the same form (\ref{fldx}) with $q^i = \partial H^r / \partial p^r_i$ (no contribution
with $\sigma = \0$ there!), and satisfies Eq.~(\ref{xipdrho}). Equation (\ref{dactrs}) with
$\sigma=\0$ then may be considered as defining the actions $S^r_{AB}$, corresponding to arbitrary
curve $C_{AB}$, and for $r$ such that $H^r$ is not linear in $p^r_i$,
Eq.~(\ref{prsidot}) with $\sigma = \0$ will ensure that the Euler-Lagrange equations
for these actions are satisfied, and so the principle of stationary action for them holds.
The action functions $S^r(\bq,t)$ with
given initial condition $S^r_0(\bq)$ at $t=t_0$ are defined as in classical mechanics \cite{arnld}:
Namely, let ${S^r_0}_\sigma = \partial_\sigma S^r_0$, then for every space vector $\bq_0$ consider
a trajectory in \cP\ that is a solution of Eq.~(\ref{em}) with initial conditions
\begin{equation}\label{incond}
    \bq(t_0) = \bq_0\, , \quad\; p\rs(t_0) = {S^r_0}_\sigma(\bq_0) \, ,   \dsplbl{incond}
\end{equation}
and define functions $S^r(\bq,t)$ by
\begin{equation}\label{srdef}
    S^r(\bq,t) = \dsp S^r_0(\bq_0) + \int_{C_{AB}} L^r\, \rmd t \, ,   \dsplbl{srdef}
\end{equation}
where $C_{AB}$ is the trajectory that ends at time~$t$ in a point $B\in\cal P$ with a space
coordinate~$\bq$ and $\bq_0$ is a space coordinate of a starting point $A\in\cal P$ of this
trajectory. We will assume that trajectories don't intersect, and so this definition is
unambiguous. While equations of motion (\ref{em}) and their solutions describe individual
trajectories, action functions $S^r(\bq,t)$ describe a family of trajectories selected by
Eq.~(\ref{incond}). Consequently on trajectories that form the family, dependence of the initial
momentums ${p\rs}_0=p\rs(t_0)$ of the trajectory on its initial space coordinate~$\bq_0$ is given
by
\begin{equation}\label{inpofq}
    {p\rs}_0 (\bq_0) = {S^r_0}_\sigma(\bq_0) \, .    \dsplbl{inpofq}
\end{equation}
For the just-defined functions $S^r(\bq,t)$, the following generalizations of classical results
hold: First, for any $\sigma\neq\0$ the functions $S\rs = \partial_\sigma S^r$ satisfy
\begin{equation}\label{srsprs}
    S\rs(\bq,t) = p\rs(\bq,t) \, , \dsplbl{srsprs}
\end{equation}
where $p\rs(\bq,t)$ is a $p\rs$-coordinate of point $B$. Second, for any $\sigma$, including
$\sigma=\0$, the~$S\rs$ satisfy the Hamilton-Jacobi-type equation (\ref{prlngn})
\begin{equation}\label{shje}
    \dovd{}{t}\,S\rs(\bq,t) + H\rs(t,\bq,\bS) = 0 \, ,  \dsplbl{shje}
\end{equation}
where $\bS$ is a set of all partial derivatives $S\rs$, $\sigma\neq\0$, of functions $S^r$.
As in classical me\-cha\-nics, Eq.~(\ref{srsprs}) means that the values $p\rs$, which originally
were independent variables evolving according to Eq.~(\ref{em}), become also
partial derivatives of action functions~$S^r$.

As was the case with the principle of stationary action, the proof is similar to the standard one
\cite{arnld}, but we present it because there are some additional complications.
Let $C_{AB}$ and $C_{A'B'}$ be two close trajectories, with the base coordinates of $A$ and $A'$
being $(\bq_0,t_0)$ and $(\bq'_0,t_0)$ and of $B$ and $B'$ being $(\bq,t)$ and $(\bq',t')$. These
trajectories are integral curves of the vector field~$X$. Let action forms $\rho\rs$ for
any $\sigma$, including $\sigma=\0$, be defined as in $J^\infty$. Now connect point $A$ with
$A'$ and point $B$ with $B'$ by straight segments, and consider an integral of $\rho\rs$ along the
closed loop $AA'B'BA$. By Stokes' theorem we have
\begin{equation}\label{st}
     \oint_{AA'B'BA} \rho\rs \,\,=\,\, \int_D \rmd \rho\rs \, ,   \dsplbl{st}
\end{equation}
where $D$ is the region inside the loop. The difference $V(t)$ between points of $C_{AB}$ and
$C_{A'B'}$ with the same $t$ is given by the vector $\overrightarrow{AA'}$, dragged (with parameter
$t-t_0$) by a flow of the vector field~$X$. Now, the vector $\overrightarrow{AA'}$ is the vector
$\bq'_0 - \bq_0$ raised to the graph of the analytic function $S^r_0(\bq)$ in~\cP. Consequently,
if $\bq'_0 - \bq_0 = \veps^i \partial / \partial q^i$, then $\overrightarrow{AA'} =
\veps^i D_i$, so $\overrightarrow{AA'}$ is a linear combination of~$D_i$. It is easy to
calculate that the Lie derivative of $D_i$ in the direction of $X$ is
\begin{equation}\label{}
    [X,D_i] = - \,(D_i \dot{q}^j) D_j \, ,
\end{equation}
i.e., also a linear combination of $D_j$, and therefore, so is the difference $V(t)$. Now, the
integral on the right hand side of Eq.~(\ref{st}) is equal to $\int_{t_A}^{t_B} V(t) \ip (X \ip
\rmd \rho\rs)\, \rmd t$, but by Eq.~(\ref{xipdrho}) $X \ip \rmd \rho\rs$ is a linear combination of
Cartan forms that are canceled by any $D_i$ and, therefore, by $V(t)$. We have,
eventually, that for all $\sigma$ this integral vanishes, and with it the integral of~$\rho\rs$
along $AA'B'BA$, and so
\begin{equation}\label{ibb}
    \int_B^{B'} \rho\rs = \left(\int_B^A + \int_A^{A'} + \int_{A'}^{B'}\right) \rho\rs \, .
    \dsplbl{ibb}
\end{equation}
But for $\sigma\neq\0$, as in Eq.~(\ref{dprs}), $\int_B^A \rho\rs = p\rs(A) - p\rs(B)$, and
similarly for $\int_{A'}^{B'} \rho\rs$. On $AA'$, $t=t_0$ and so $p\rs = {S^r_0}_\sigma$,
$\overrightarrow{AA'}\ip\rmd t = 0$, and for all $\sigma$
\begin{equation}\label{iaa}
    \int_A^{A'}\rho\rs = \int_A^{A'} p\rsi \rmd q^i = \int_A^{A'} {S^r_0}_{\sigma i} \rmd q^i
                       = {S^r_0}_\sigma(A') - {S^r_0}_\sigma(A) \, ,  \dsplbl{iaa}
\end{equation}
which for $\sigma\neq\0$ is equal to $p\rs(A') - p\rs(A)$. Consequently, for $\sigma\neq\0$ Eq.
(\ref{ibb}) becomes
\begin{equation}\label{}
    \int_B^{B'} p\rsi \rmd q^i - H\rs \rmd t = p\rs(B') - p\rs(B) \, ,
\end{equation}
which in the limit $\veps^i \rightarrow 0$, $\Delta t \rightarrow 0$ gives
\begin{equation}\label{hhh1}
    \dovd{}{q^i}\, p\rs(\bq,t) = p\rsi(\bq,t)\, , \quad\; \dovd{}{t}\, p\rs(B) = - H\rs(B) \, .
    \dsplbl{hhh1}
\end{equation}
Now if for all $|\sigma| = N$, Eqs.~(\ref{srsprs}) are true, then for such $\sigma$ and all $i$,
\begin{equation}\label{}
    S\rsi(\bq,t) = \dovd{}{q^i} \, S\rs(\bq,t) = \dovd{}{q^i} \, p\rs(\bq,t) = p\rsi(\bq,t)\, ,
\end{equation}
and so (\ref{srsprs}) is also true for $|\sigma| = N+1$. Then for $\sigma=\0$, since trajectories
$C_{AB}$ and $C_{A'B'}$ satisfy Eq.~(\ref{em}), we have
\begin{equation}\label{}
    \int_B^A\rho^r = \int_B^A L^r \rmd t = S^r(A) - S^r(B) \, ,
\end{equation}
and similarly for $\int_{A'}^{B'} \rho^r$, while $\int_A^{A'}\rho^r$ is given by Eq.~(\ref{iaa})
with $\sigma=\0$. Eq.~(\ref{ibb}) now gives
\begin{equation}\label{}
    \int_B^{B'} p^r_i \rmd q^i - H^r \rmd t = S^r(B') - S^r(B),
\end{equation}
or in the limit $\veps^i \rightarrow 0$, $\Delta t \rightarrow 0$,
\begin{equation}\label{hhh2}
    \dovd{}{q^i} \, S^r(\bq,t) = p^r_i(\bq,t)\, , \quad\; \dovd{}{t}\, S^r(B) = - H^r(B) \, ,
    \dsplbl{hhh2}
\end{equation}
which completes the proof of Eq.~(\ref{srsprs}), and then the second relations in Eqs.~(\ref{hhh1})
and (\ref{hhh2}) prove Eq.~(\ref{shje}).

We see again in Eq.~(\ref{srdef}) that solution of the PDE (\ref{shje}) may be obtained from
solutions of the ODEs~(\ref{em}). Conversely, any sequence of functions $S\rs(\bq,t)$ which satisfy
\begin{equation}\label{sr0cond}
    \dovd{S\rs}{q^i} = S\rsi,  \quad \mbox{for all} \;\, \sigma,i,   \dsplbl{sr0cond}
\end{equation}
and which also satisfy Eq.~(\ref{shje})
with initial conditions corresponding to a family of trajectories with given initial distribution
of momentums (\ref{inpofq}), may be used for integration of the
equations of motion for trajectories of
this family: Equation~(\ref{srsprs}), read from right to left, gives for all $t$ the
distribution on the family's trajectories of momentums which satisfy Eq.~(\ref{em}). Indeed, we
have for these momentums $\partial p\rsi/\partial t = \partial_i\partial_t S\rs$. Using
Eq.~(\ref{shje}), it is then easy to show that if a point $\bq(t)$ moves with velocity $\dot{q}^k$
given by the first equation in~(\ref{em}), then the time derivative
\begin{equation}\label{}
    \dot{p}\rsi(\bq(t),t) = \dovd{p\rsi}{t} + \dovd{p\rsi}{q^k} \, \dot{q}^k
\end{equation}
of a function $p\rsi$ at a point $\bq(t)$ is given by the second equation in (\ref{em}).
For Hamiltonians of first order, this is the basis of a Jacobi method of integration of equations
of motion, and in the following we will call it the ``generalized Jacobi method" for arbitrary
Hamiltonians. Now, when this method is used, it is obviously desirable
to make it applicable to as large a class of trajectory families as possible. From this point of
view, the formulation we used above is unnecessarily restrictive and may be generalized. Indeed,
initial conditions that define the family's trajectories are given by Eq.~(\ref{inpofq}). In
this equation, ${S^r_0}_\sigma$ are derivatives of the functions $S^r_0$. However, we saw that the
sequence $S\rs$ that is used in the generalized Jacobi method does not contain $S^r$ and includes
only functions $S\rs$
with $\sigma\neq\0$. Therefore, all these functions are derivatives of $S^r_i$, $i=1,\ldots,n$,
while the functions $S^r_i$ themselves and, consequently,
their initial values $S^r_{0i}$, do not have
to be derivatives of any other functions. On the other hand, we have from Eq.~(\ref{sr0cond}) that
\begin{equation}\label{}
    \dovd{{S^r_0}_i}{q^j} = {S^r_0}_{ij} = \dovd{{S^r_0}_j}{q^i} \, .
\end{equation}
This means, that 1-forms ${S^r_0}_i \rmd q^i$ should be closed, $\rmd ({S^r_0}_i \rmd q^i) = 0$,
which will allow us to {\em define} the functions $S^r_0$ by
\begin{equation}\label{sr0}
    S^r_0(\bq) = \int_{\bq_0}^\bq {S^r_0}_i \rmd q^i + S^r_0(\bq_0) \, ,   \dsplbl{sr0}
\end{equation}
and so ${S^r_0}_i$ will be their derivatives. The integration in (\ref{sr0}) runs along
arbitrary curves in configuration space $Q$ that connect points $\bq_0$ and $\bq$, and
the value $S^r_0(\bq_0)$, as well
as the vector $\bq_0$ itself, are also arbitrary. Thus from the very beginning, the functions
$S^r_0$ are defined up to an arbitrary additive constant; moreover, they will be
usual, single-valued functions only if configuration space $Q$ is simply connected.
If the fundamental
group of $Q$ is nontrivial, then in general Eq.~(\ref{sr0}) defines functions $S^r_0$ as
multi-valued, or single-valued on the universal covering space of $Q$. The branches of
$S^r_0$ may differ only by a constant, and so they all have the same derivatives ${S^r_0}_\sigma$,
$\sigma\neq\0$. Therefore, these derivatives will be single-valued as they should be because
the family has one, and only one, trajectory starting at every point of
configuration space at $t=t_0$,
and the functions ${S^r_0}_\sigma$, $\sigma\neq\0$, define initial momentums of these trajectories.
Consequently, in the currently considered statement of the problem, which starts with equations of
motion in the infinite phase space \cP, the functions $S^r$ and
their initial values $S^r_0$ in the generalized Jacobi method are defined up to an additive
constant, and in cases where
the configuration space $Q$ is not simply connected, may be multi-valued.
Note that these conclusions are purely
topological, not dynamical --- they do not depend on the form of the
Hamiltonians $H\rs$ or on their
order. The simplest example is a family of trajectories on a circle that all have the same
initial velocity~$v$. The function $S_0$ is then equal to $vr\varphi+\mbox{const}$,
where $r$ is the radius of the
circle, and $\varphi$ is the angular coordinate on it. When $v\neq 0$, this function is
multi-valued on the circle, but single-valued on the universal covering space~${\rm R}^1$.

\subsection{The case of complex-valued solutions} \label{comunfun}

We now allow complex-valued solutions of the PDE~(\ref{evpde}).
We only consider the case of one complex function $p(\bq,t)$, the generalization to the
situation when there are several of them being obvious. Let $p^{1,2}$ be this function's real and
imaginary parts, so that $p = p^1 + ip^2$, and similarly $H = H^1 + iH^2$. The conjugated
values are $\bar{p} = p^1 - ip^2$ and $\bar{H} = H^1 - iH^2$.
By setting $m = 2$, the theory of the previous subsections
may be applied directly to the functions $p^1$ and $p^2$, treated as independent real functions
with Hamiltonians $H^1$ and~$H^2$. However, it is often more convenient to express the same
results via complex functions $p$ and $\bar{p}$, because in this representation they behave as if
they were independent and also because the equations for $\bar{p}$ are simply the conjugated
equations for $p$.

As is usually done, introduce vector fields
\begin{equation}\label{cderivs}
    \dovd{}{p} = \half \left(\dovd{}{p^1} - i \dovd{}{p^2}\right) \, , \quad \,
    \dovd{}{\bar{p}} = \half \left(\dovd{}{p^1} + i \dovd{}{p^2}\right)    \dsplbl{cderivs}
\end{equation}
and 1-forms
\begin{equation}\label{}
    \rmd p = \rmd p^1 + i \rmd p^2 \, , \quad \, \rmd \bar{p} = \rmd p^1 - i \rmd p^2 \, ,
\end{equation}
which satisfy duality relations
\begin{equation}\label{}
    \begin{array}{lcllcl}
       \dsp \dovd{}{p} \ip \rmd p & = & 1 \, ,\quad & \dsp \dovd{}{p} \ip \rmd \bar{p} & = & 0 \, ,
       \\[0.3cm]
       \dsp \dovd{}{\bar{p}} \ip \rmd p & = & 0 \, ,\quad & \dsp \dovd{}{\bar{p}} \ip \rmd
       \bar{p} & = & 1 \, .
    \end{array}
\end{equation}
We have
\begin{equation}\label{}
    \sum_{r = 1}^2 p\rsi \dovd{}{p\rs} = p_{\sigma i} \dovd{}{p_\sigma} + \bar{p}_{\sigma i}
    \dovd{}{\bar{p}_\sigma}
\end{equation}
and so the operator of total differentiation may be written as
\begin{equation}\label{ctotdif}
    D_i = \dovd{}{q^i} + p_{\sigma i} \dovd{}{p_\sigma} + \bar{p}_{\sigma i}
    \dovd{}{\bar{p}_\sigma} \, .   \dsplbl{ctotdif}
\end{equation}
Similarly, vector field $X$, Eq.~(\ref{fldx}), may be written as
\begin{equation}\label{}
    X = \dsp \dovd{}{t} + \dot{q}^i D_i - H_\sigma \dovd{}{p_\sigma} -
    \bar{H}_\sigma \dovd{}{\bar{p}_\sigma} \, .
\end{equation}

Since $H$ is an analytic function of $p_j\,$, it satisfies the Cauchy-Riemann equations
\begin{equation}\label{}
    \dovd{H^1}{p^1_j} = \dovd{H^2}{p^2_j}\, ,\quad \, \dovd{H^1}{p^2_j} = -\,\dovd{H^2}{p^1_j} \, .
\end{equation}
The first of these equations means that if $p^{1,2}$ are considered as independent real functions
with Hamiltonians $H^1$ and $H^2$, then HC2 are automatically satisfied with the corresponding
velocity
\begin{equation}\label{dotqj}
    \dot{q}^j \, =  \,\dovd{H^1}{p^1_j} \, = \, \dovd{H^2}{p^2_j} \, .
\end{equation}
Expressing here $\partial / \partial p^{1,2}$ through $\partial / \partial p$ and $\partial
/ \partial \bar{p}$, and similarly $H^{1,2}$ through $H$ and $\bar{H}$, and taking into account
that, being an analytic function, $H$ depends on $p^{1,2}$ only through the combination
$p=p^1+ip^2$, and $\bar{H}$ depends on $p^{1,2}$ only through $\bar{p} = p^1-ip^2$, obtain
\begin{equation}\label{cqdot}
    \dot{q}^j  \,= \, \dovd{H^1}{p^1_j}  \,= \, \left(\dovd{}{p_j} + \dovd{}{\bar{p}_j}\right)
    \frac{H + \bar{H}}{2}  \,= \, \half \left(\dovd{H}{p_j} + \dovd{\bar{H}}{\bar{p}_j}\right) \, .
    \dsplbl{cqdot}
\end{equation}
It is easy to see that $\partial H^2 / \partial p^2_j$ gives the same expression for the velocity.

For application to a theory of particles with spin, we also need to consider the case of a complex
analytic function $p(w,t)$ of complex coordinate $w = w^1 + i w^2$ with a Hamiltonian $H(p_w)$,
where $p_w = \partial p/\partial w$. If $w$ and $p$ were real, the $w$'s velocity would be equal to
\begin{equation}\label{vlctu}
    \dot{w} = \dovd{H}{p_w} \, .    \dsplbl{vlctu}
\end{equation}
It is easy to see that due to analyticity of all the functions involved and the corresponding
Cauchy-Riemann equations, the same expression for $\dot{w}$ remains true in a complex case. Indeed,
$p_w$ is given by the standard expressions
\begin{equation}\label{pu}
    p_w = p^1_{w^1} + i p^2_{w^1} = p^2_{w^2} - i p^1_{w^2} \, .
\end{equation}
From that, and using the Cauchy-Riemann equations, we have for the real and imaginary parts
of~(\ref{vlctu})
\begin{equation}\label{reimdotu}
    \Re \left(\dovd{H}{p_w}\right) = \dovd{H^1}{p^1_{w^1}} = \dovd{H^2}{p^2_{w^1}} \, ,\quad \,
    \Im \left(\dovd{H}{p_w}\right) = \dovd{H^2}{p^2_{w^2}} = \dovd{H^1}{p^1_{w^2}} \, .
\end{equation}
On the other hand, considering $p^{1,2}$ as independent real functions of real variables $w^{1,2}$
with Hamiltonians~$H^{1,2}$, we have
\begin{equation}\label{}
    \dot{w}^1 = \dovd{H^1}{p^1_{w^1}} = \dovd{H^2}{p^2_{w^1}} \, ,\quad \,
    \dot{w}^2 = \dovd{H^1}{p^1_{w^2}} = \dovd{H^2}{p^2_{w^2}} \, ,
\end{equation}
where again we used the Cauchy-Riemann equations, and the first (resp. second) representation of
$p_w$ in (\ref{pu}) for calculation of $\dot{w}^1$ (resp. $\dot{w}^2$). Thus, as it was for
velocity $\dot{q}^j$, the HC2 for $\dot{w}^{1,2}$ are automatically satisfied due to the
Cauchy-Riemann equations, and the $\dot{w}^{1,2}$ are equal to $\Re (\partial H/\partial p_w)$ and
$\Im (\partial H/\partial p_w)$ in Eq.~(\ref{reimdotu}), which proves (\ref{vlctu}).

\subsection{Discussion} \label{dscssn2}

The following general picture emerges from the above development. As for the standard case of
first order equations, the PDE (\ref{evpde}) with $H^r$ satisfying HC allows the introduction of a
corresponding system of ODEs~(\ref{em}). The values whose dynamical evolution
is governed by this system are coordinates of a point, moving in a base, and partial derivatives
at this point of unknown functions. On the other hand, as was just discussed, Eq.~(\ref{em}) may
be considered on its own, as Hamilton equations are in classical mechanics, with the action
functions being introduced later. In general, the system (\ref{em}) is an infinite hierarchical
system of coupled equations. We will not attempt its solution in this work; what will be important
for us here is that this system has solutions whenever Eq.~(\ref{evpde}) does, and as for any
system of first order ODEs, this solution is unique.
The system then defines in $J^\infty$ and \cP\ some trajectories, which
lie in graphs $\Gamma$ of solutions of (\ref{evpde}) and are the characteristic curves of a
corresponding exterior differential system. Like any characteristic curves \cite{amp},
these trajectories
express solutions of Eq.~(\ref{evpde}) with given initial conditions and their derivatives as
integrals of $\rho\rs$ or $L\rs \rmd t$ along them. The direct proof of this statement,
which doesn't use the theory of characteristics, is presented in section 2.5. The theory
splits the whole jet space $J^\infty$ into $r$-$\sigma$ sectors with coupled dynamics, described by
Hamiltonians $H\rs$. The sectors share common coordinates~$q^i$, but there are no conflicts,
because, thanks to HC, the dynamic they all define for these coordinates is the same.
The structure of the theory in every
$r$-$\sigma$ sector is similar to the one in $J^1_1$, but the value $p\rsi$ has a dual meaning:
while on one hand, in an $r$-$\sigma i$ sector it plays the role of an action, obtainable from the
above integrals, on the other hand, in sector $r$-$\sigma$ it is an
\mbox{$i$-th} component of momentum, evolving according to the corresponding ``Hamilton
equation" with Hamiltonian $H\rs$. In our equations, this duality
may be seen especially clearly in the comparison of Eq.~(\ref{prehep}), which describes the
evolution of $p\rsi$ as an action in an $r$-$\sigma i$ sector, with the second equation of
(\ref{em}), where it evolves as the $i$-th component of momentum in sector $r$-$\sigma$. As the
second equation of (\ref{em}) is just a different form of (\ref{prehep}), sectors $r$-$\sigma$ and
$r$-$\sigma i$ obviously agree on the dynamics of a variable $p\rsi$ which
they share. A variable $p^r$ belongs to only one sector $r$-$\0$, and so for its time derivative
we have only the ``action form" representation, given by Eq.~(\ref{tottd}) with $\sigma=\0$.

First order evolutionary PDEs with $m=1$ satisfy HC
automatically, and so the whole theory is completely applicable to
them. They are different, however, from higher-order equations in the following important aspect:
For any $N \geq 1$, the resulting equations of motion for $\bq$ and $p_\sigma$, $|\sigma|\leq N$,
form a closed subsystem, and the corresponding geometric theory may be formulated in
a space of $N$-jets $J^N_1$. Indeed, if the order of the PDE (\ref{evpde})
is equal to $k$, then $H_\sigma$ contains
the derivatives of orders up to $|\sigma|+k$. Since the Cartan forms $\omega_\sigma$ are expressed
through~$p_{\sigma i}$, which is of order $|\sigma|+1$, and $H_\sigma$, for first
order equations, i.e., $k=1$, and any integer $N\geq 1$ the system of exterior equations
$\{\omega_\sigma = 0\, , |\sigma| < N\}$ is closed with respect to the set of derivatives of $p$
it includes. For this system, the space of $N$-jets $J^N_1$ is sufficient, and the space
$J^\infty_1$ is not necessary. On the contrary, for, say, second order
equations, $H_\sigma$ in the exterior equation $\omega_\sigma = 0$ contains the
variables $p_\nu$ with $|\nu| = |\sigma| + 2$. In order to ensure that these variables do indeed
describe corresponding derivatives of the solution, which is represented by a graph in a jet space,
we need to require that this graph also solves an exterior equation $\omega_{\sigma'} = 0$ with
$|\sigma'| = |\sigma| + 1$.
But then $H_{\sigma'}$ in $\omega_{\sigma'}$ will contain variables $p_{\nu'}$ with
$|\nu'| = |\sigma'|+2 = |\sigma|+3$, and so the process will never stop, and the use of the
infinite jet space $J^\infty_1$ becomes inevitable.\footnote{Another reason to use an infinite
jet space is B\"{a}cklund's theorem \cite{szsumf,ibragim}, from which it follows that CD, defined
by 1-forms (\ref{cforms}) with Hamiltonians $H^r$ of higher-than-first order, cannot have
characteristic fields in any finite jet space $J^k$, $k<\infty$.} Similar
considerations show that, while for a higher order PDE the expression for $\dot{p}_\sigma$
with $\sigma\neq\0$ contains $p_\nu$ with $|\nu| > |\sigma|$, for a first order PDE it doesn't,
and so for it a system of equations $\dot{p}_\sigma = l_\sigma$ with $|\sigma| \leq N$
is closed for any $N\geq 1$. The system corresponding to $N=1$ is the simplest
possible, but it still describes the evolution of the  most important variables:
$\bq, \, p$, and~$p_i$. A theory of this
system in $J^1_1$ is a usual Hamiltonian theory of first order evolutionary PDE, which is a
part of their ``full", i.e. including all derivatives, theory, while the latter is a special
case of our theory of satisfying HC evolutionary PDE of arbitrary order and with arbitrary~$m$.

\section{Hamiltonian flow of quantum Hamilton-Jacobi equation}  \label{hflow}
\setcounter{equation}{0}

Now we apply the technique developed above to non-relativistic quantum theory. We start with the
one-particle case and consider the multi-particle situation later. By expressing the wave function
as
\begin{equation}\label{psi}
    \psi(\bx,t) = \dsp \exp \left(\ih\,p(\bx,t) \right) , \quad \quad p(\bx,t) = S(\bx,t) +
    \hi\,R(\bx,t) \, , \dsplbl{psi}
\end{equation}
where $p$ is complex and $S$ and $R$ are real functions of position and time, the
one-particle \Sch equation
\begin{equation}\label{se}
    i \hbar\, \dovd{\psi}{t} = -\frac{\hbar^2}{2 m}\, \Delta \psi + U(\bx,t) \psi  \dsplbl{se}
\end{equation}
may be equivalently presented as an evolutionary PDE for $p$ as
\begin{equation}\label{qhje}
    \dovd{p}{t} + H = 0 \, ,   \dsplbl{qhje}
\end{equation}
or as a system of evolutionary PDE for $R$ and $S$ as
\begin{equation}\label{szetadot}
    \dovd{S}{t} + H^S = 0\, , \quad \quad \dovd{R}{t} + H^R = 0 \, ,
        \dsplbl{szetadot}
\end{equation}
The Hamiltonian functions $H$, $H^S$, and $H^R$ in the above equations are
\begin{eqnarray}
  H & = & \frac{1}{2 m}\, p_j^2 + U + \frac{\hbar}{2 i m}\, p_{jj} \, ,  \label{hp}
  \dsplbl{hp} \\[0.2cm]
  H^S & = & \frac{1}{2 m}\, S_j^2 + U - \frac{\hbar^2}{2 m}\, \left(R_j^2 +
             R_{jj}\right) , \label{hs}  \dsplbl{hs} \\[0.2cm]
  H^R & = & \frac{1}{m} \left(S_j\, R_j + \half \, S_{jj}\right), \label{hz}  \dsplbl{hz}
\end{eqnarray}
where the indices denote corresponding partial derivatives and we extend the summation rule to
expressions like $p_j^2 = p_j p_j$.

Along with (\ref{qhjeint}), Eq.~(\ref{qhje}) and system (\ref{szetadot}) will be also called 
``quantum Hamilton-Jacobi equation(s)" (QHJE). Obviously, QHJE is equivalent to the \Sch equation
(\ref{se}), and the action function $p$ carries the same information as the wave function $\psi$.
In the following, for convenience, we will often discuss only one of these
functions/equations, with the understanding that our conclusions may be applied,
with proper modifications, to the other. Also, since there is only one wave function,
we will always say ``action function," even when there are several (two) of them.

We can now see immediately that Eqs.~(\ref{qhje}) and (\ref{szetadot}) satisfy HC1. This
is obvious in Cartesian coordinates used in Eqs.~(\ref{hp})-(\ref{hz}), and is instructive to
verify in the cylindrical and spherical coordinate systems. As they should (see
section~2.6), Eqs.~(\ref{qhje}) and (\ref{szetadot}) also satisfy HC2 with corresponding
velocity
\begin{equation}\label{vlct}
    v^j = \dovd{H^S}{S_j} = \dovd{H^R}{R_j} = \frac{1}{m} \, S_j\, ,  \dsplbl{vlct}
\end{equation}
which, in agreement with Eq.~(\ref{cqdot}), may also be expressed as
\begin{equation}\label{vlctc}
    v^j = \frac{1}{2m}\,(p_j + \bar{p}_j)\, .    \dsplbl{vlctc}
\end{equation}
Consequently, the theory of the previous section may be used. It means that in a space of
ana\-ly\-tic jets, corresponding to PDE (\ref{qhje}) \big(or to a system of PDEs
(\ref{szetadot})\big), there exist
trajectories, described by the system of ODEs (\ref{em}), such that the solutions of the PDE and
their derivatives may be obtained from the initial conditions by integrating the corresponding
Lagrangians along these trajectories \big(see Eq.~(\ref{srdef})\big).
The very existence of such ODEs and
trajectories is just a mathematical fact, proven in the previous section. However, it raises an
inevitable physical question: do the particles indeed move along these trajectories? Or, more
practically: can peculiar features of quantum mechanics be understood, and its predictions
reproduced, by assuming so? There are more questions. As we discussed, if the solution
of the PDE is known, then the values of $p\rs$ and $\bv$ may be obtained from it by the
generalized Jacobi method.
Therefore, the ODE and PDE formulations should be considered as two faces of the same
theory, exactly like Hamilton equations and Hamilton-Jacobi equation in classical mechanics.
But in classical mechanics, the roles of these equations are very different: while Hamilton (or
Newton) equations provide the description of individual trajectories, the Hamilton-Jacobi equation
describes an evolution of
the (action) function which does not correspond to any particular trajectory, but is associated
with a family of them. As was discussed at length in section 2.5,
similar roles are played by the equations of motion (\ref{em}) and the
PDE~(\ref{evpde}) in the mathematical theory of higher order equations.
Now the other question is whether the situation
in quantum mechanics is the same, so that the action or wave functions describe families or
ensembles of trajectories, while the description of individual events/trajectories is provided by
the system of ordinary differential equations of motion (\ref{em}). In the rest of this work we
defend a positive answer to these questions. As we already mentioned in the Introduction,
we call this approach an analytical quantum dynamics in infinite phase space (\paqd). We now start
with its general description.

As is clear from the previous section, the state of a particle at some moment $t$ in \paqd\ is
defined by a triple $(\bx,\bS,\bR)$ or, equivalently, $(\bx,\bp,\bar{\bp})$.
Here $\bx$ is the position of a particle at time $t$, and $\bS$,
$\bR$, $\bp$ and $\bar{\bp}$ are sets of all the derivatives of the corresponding action functions
in~$\bx$ at this time, so that $\bS$ is a set of all $S_\sigma(\bx,t)$ with $\sigma\neq\0$ and
similarly for $\bR$,
$\bp$ and $\bar{\bp}$. However, in the framework of \paqd, they are just a set of
independent fundamental variables, identified by their multi-indices, which describe the state of a
particle at time $t$ exactly like components of momentum in classical mechanics.
The evolution of a state is described by the equations of motion (\ref{em}) \big(the second of
these equations is easier to use in the form~(\ref{prehep})\big), where the functions $p^r$ in that
equation are now $p^1=S$ and $p^2=R$ or $p^1=p$ and $p^2=\bar{p}$.
For future reference, we present here expressions for Hamiltonians and some equations of motion.
The first Eq.~(\ref{em}), i.e. the equation for $\dot{\bx}$, takes the form of Eq.~(\ref{vlct}) for
the $(S,R)$ formulation and Eq.~(\ref{vlctc}) for the $(p,\bar{p})$ formulation. The operator of
total differentiation for the $(S,R)$ formulation is
\begin{equation}\label{}
    D_i = \dovd{}{x^i} + S_{\sigma i} \dovd{}{S_\sigma} + R_{\sigma i} \dovd{}{R_\sigma}\,.
\end{equation}
Expressions for $H^S$ and $H^R$ are given in Eqs.~(\ref{hs}) and (\ref{hz}) above. We also have
\begin{eqnarray}
  H^S_i & = & \frac{1}{m}\, S_j S_{ji} + U_i - \frac{\hbar^2}{m} \left(R_j R_{ji}
              + \half \, R_{jji}\right) , \\[0.2cm]
  H^R_i & = & \frac{1}{m} \left(S_j R_{ji} + S_{ji} R_j + \half \, S_{jji}\right) .
\end{eqnarray}
The time derivatives of the actions $S$ and $R$ and their first momentums are
\begin{eqnarray}
  \dot{S} & = & \frac{1}{2 m}\, S_j^2 - U + \frac{\hbar^2}{2 m} \left(R_j^2 +
                 R_{jj}\right) ,\label{sdot}  \dsplbl{sdot} \\[0.2cm]
  \dot{S}_i & = & -\, U_i + \frac{\hbar^2}{m} \left(R_j R_{ji} + \half \,
                  R_{jji}\right) ,\label{sidot}  \dsplbl{sidot} \\[0.2cm]
  \dot{R} & = & - \frac{1}{2 m} \, S_{jj}\, , \label{rdot}  \dsplbl{rdot} \\[0.2cm]
  \dot{R}_i & = & - \frac{1}{m} \left(S_{ji} R_j + \half \, S_{jji}\right) .
\end{eqnarray}

For the $(p,\bar{p})$ formulation, we only need equations for $\dot{p}_\sigma$ and $H_\sigma$,
since the equations for $\dot{\bar{p}}_\sigma$ and $\bar{H}_\sigma$ are obtained from them by
conjugation in an obvious way. The Hamiltonian~$H$ is given by Eq.~(\ref{hp}), and the operator of
total differentiation by Eq.~(\ref{ctotdif}). Then for $H_i$ and $H_{ik}$ we have
\begin{eqnarray}
  H_i & = & \frac{1}{m}\, p_j p_{ji} + U_i + \frac{\hbar}{2 i m}\, p_{jji} \, , \\[0.2cm]
  H_{ik} & = & \frac{1}{m}\, (p_j p_{jik} + p_{jk} p_{ji}) + U_{ik} + \frac{\hbar}{2 i m}\,
               p_{jjik} \, ,
\end{eqnarray}
while the time derivatives of the action and first momentums are
\begin{eqnarray}
  \dot{p} & = & \frac{1}{2 m} \, p_j \bar{p}_j - U - \frac{\hbar}{2 i m}\, p_{jj}\, ,
                \label{pdot2}  \dsplbl{pdot2} \\[0.2cm]
  \dot{p}_i & = & \frac{1}{2 m} \,(\bar{p}_j - p_j) p_{ji} - U_i - \frac{\hbar}{2 i m}\,
                  p_{jji}\, ,\\[0.2cm]
  \dot{p}_{ik} & = & \frac{1}{2 m} \,(\bar{p}_j - p_j) p_{jik} - \frac{1}{m} \, p_{ji} p_{jk}
                     - U_{ik} - \frac{\hbar}{2 i m}\, p_{jjik}\, .
\end{eqnarray}
It is not difficult to derive a general expression for $H_\sigma$. We say that a
multi-index $\nu$ is a subindex of the multi-index $\sigma$, and write $\nu \subset \sigma$, if
there exists a multi-index $\mu$ such that $\sigma = \nu\mu$. This multi-index
$\mu$ will then be denoted as $\sigma \setminus \nu$. Every multi-index is its own
subindex, and the empty multi-index is a subindex of every multi-index. We also say that the
multi-index $\nu \subset \sigma$ is chosen from the multi-index $\sigma$ if $\nu$ is obtained from
$\sigma$ in the following way: write $\sigma$ as a sequence of indices
$i_1,\ldots,i_{|\sigma|}$, then with this sequence fixed select $|\nu|$ members of the sequence
to form $\nu$, and the others form $\sigma\setminus\nu$. Denote the summation over all
such choices from a fixed sequence by $\sum_{\nu\prec\sigma}$. With this
definition, it is easy to prove by induction that
\begin{equation}\label{hsigma}
    H_\sigma = \frac{1}{2m} \,\sum_{\nu\prec\sigma} p_{j\nu} p_{j\sigma \setminus \nu}
               + U_\sigma + \frac{\hbar}{2 i m}\, p_{jj\sigma} \, ,   \dsplbl{hsigma}
\end{equation}
where the factor $1/2$ accounts for the fact that in the sum over $\nu$ every term appears twice.
If not all indices in $\sigma$ are different, then the same subindex $\nu \subset \sigma$ may be
chosen from $\sigma$ in different ways. Consequently, there will be different choices that give
the same (i.e., with the same $\nu$) contribution to the sum in
(\ref{hsigma}). For example, this will {\em always} happen when configuration space is
one-dimensional, and the reader is encouraged to write formulas for $H_\sigma$ with
$|\sigma| = 1,2,3,\ldots$ in this case.
It may be useful to present the summation in (\ref{hsigma}) in a form that contains only
different contributions. Since the number of ways by which $\nu_i$ indices $i$ may be chosen from
$\sigma_i$ of them in a multi-index $\sigma$ is equal to $C_{\sigma_i}^{\nu_i} = \sigma_i!/\nu_i!
\,(\sigma_i - \nu_i)!$, the total number of ways by which a multi-index $\nu \subset \sigma$ may
be chosen from $\sigma$ is $C_\sigma^\nu=\prod_{i=1}^n C_{\sigma_i}^{\nu_i}$. Therefore,
Eq.~(\ref{hsigma}) may be rewritten as
\begin{equation}\label{hsigma2}
    H_\sigma = \frac{1}{2m} \,\sum_{\nu \subset \sigma} C_\sigma^\nu p_{j\nu}
    p_{j\sigma\setminus\nu} + U_\sigma + \frac{\hbar}{2 i m}\, p_{jj\sigma} \, ,  \dsplbl{hsigma2}
\end{equation}
where the summation now is over all {\em different} subindices $\nu$ of $\sigma$.
Correspondingly, the equations of motion for the $p_\sigma$, $\sigma\neq\0$, become
\begin{equation}\label{psigdot}
    \dot{p}_\sigma = \frac{1}{2 m} \,(\bar{p}_j - p_j) p_{j\sigma} - \frac{1}{2m} \,
                     {\sum_{\nu \subset \sigma}}' C_\sigma^\nu p_{j\nu} p_{j\sigma\setminus\nu}
                     - U_\sigma - \frac{\hbar}{2 i m}\, p_{jj\sigma}\, .   \dsplbl{psigdot}
\end{equation}
where the summation $\sum_{\nu \subset \sigma}'$ excludes terms with $\nu=\0$ and $\nu=\sigma$.

The emerging theory is in many respects similar to classical mechanics, but there are also
important differences. As in classical mechanics, the particles in \paqd\ move along well-defined
trajectories, with definite values of position, velocity, and all momentums at every
moment of time. The states of the particle belong to an infinite phase space \cP, and the
equations of motion (\ref{em}) describe the evolution of these states in terms of Hamiltonian flow
in \cP. For the $(p,\bar{p})$ formulation, these equations take the form of (\ref{vlctc}) and
(\ref{psigdot}), and the corresponding Hamiltonian flow is generated by the vector field
\begin{equation}\label{fldxpp}
     \begin{array}{ccl}
       X & = & \dsp \dovd{}{t} + v^i \dovd{}{x^i} + \dot{p}_\sigma \dovd{}{p_\sigma}
               + \dot{\bar{p}}_\sigma \dovd{}{\bar{p}_\sigma} \\[0.4cm]
         & = & \dsp \dovd{}{t} + v^i D_i - H_\sigma \dovd{}{p_\sigma}
               - \bar{H}_\sigma \dovd{}{\bar{p}_\sigma} \, ,   \dsplbl{fldxpp}
     \end{array}
\end{equation}
where $v^i$, $\dot{p}_\sigma$, $D_i$, and $H_\sigma$ are given by Eqs.~(\ref{vlctc}),
(\ref{psigdot}), (\ref{ctotdif}), and (\ref{hsigma2}) respectively, $\dot{\bar{p}}_\sigma$ and
$\bar{H}_\sigma$ are obtained by conjugation, and summation over $\sigma$ does not include
$\sigma=\0$. As in classical mechanics, initial value of the state uniquely determines its future
evolution. The action function and QHJE are not needed for solution of the equations of motion. The
action function is brought into use as an additional mathematical structure either by introducing a
Taylor series (\ref{us}) or via Eqs.~(\ref{incond}) and (\ref{srdef}) of the previous section. For
the $(p,\bar{p})$ formulation, Eqs.~(\ref{srdef}) with initial condition $p(\bx_0,t_0)=p_0(\bx_0)$
take the form
\begin{equation}\label{lp}
    p(\bx,t) = \dsp p_0(\bx_0) + \int_{C_{AB}} L\, \rmd t \, ,   \dsplbl{lp}
\end{equation}
and conjugated equation for $\bar{p}(\bx,t)$, and similarly for the $(S,R)$ formulation, they take
the form
\begin{equation}\label{lsandr}
    \begin{array}{ccl}
      S(\bx,t) & = & \dsp S_0(\bx_0) + \int_{C_{AB}} L^S\, \rmd t \, , \\[0.4cm]
      R(\bx,t) & = & \dsp R_0(\bx_0) + \int_{C_{AB}} L^R\, \rmd t \, ,   \dsplbl{lsandr}
    \end{array}
\end{equation}
where $L$, $L^S$, and $L^R$ are given by the right hand sides of Eqs.~(\ref{pdot2}), (\ref{sdot}),
and (\ref{rdot}) respectively, and $C_{AB}$ is the particle's trajectory, connecting points
$A=(\bx_0,t_0)$ and $B=(\bx,t)$.
As was discussed in section 2.5, in the spaces with a nontrivial fundamental group,
the action function may be multi-valued, and it is always defined up to an additive constant.
Consequently, only the derivatives of the action function are relevant, and so this function may be
represented by a graph in \cP. It then describes a family of trajectories, determined
by the given momentums at each position at some initial time. The same
is true in classical mechanics; the important difference, however, is that while in \paqd\
Eqs.~(\ref{incond}) fix all momentums/derivatives, the corresponding classical equations
\cite{arnld} fix only the first of them. As a result, in classical mechanics the action function
cannot be considered as characterizing the individual state/trajectory of a particle: a given
trajectory may belong to any of a continuum of different families of trajectories, with different
action functions. Contrary to that, in \paqd, if
the state of a particle belongs to some family, described by an action function, then by
Eq.~(\ref{incond}) it determines all derivatives of this function at a point where the particle is.
As an action function is analytic, it is equal, up to a constant, to the sum of a corresponding
Taylor series. Consequently, in \paqd\ the state of a particle determines the action/wave function
of a family, which includes it, and is, therefore, described or characterized by this function.
This description, however, is not complete:
since an analytic function can be expanded in a Taylor series
at any point of space, there are different (i.e., with different~$\bq$) members of a family that
all have the same action/wave function. Thus a complete description of particle's state may be
given either by a point in \cP\ or, equivalently, by a point in a base and an action function,
defined up to an additive constant (or wave function, defined up to a constant factor).

The action function introduced in this way satisfies QHJE, which expresses its time derivative
through this function itself, regardless of which particular trajectory is responsible for its
appearance, and so the action function obtains its own dynamics. Nevertheless, it
is clear that in the framework of \paqd, on the fundamental level of equations of
motion, the action function is a useful, but purely mathematical entity: for determination of
particle's trajectory, its use is neither necessary nor sufficient.
However, the action function, or rather the wave function, gains physical significance when a
{\em family} of trajectories described by it gains phy\-si\-cal significance. This will
be the case when one considers the preparation of an experiment. Namely, as we will see later,
using macroscopic control tools one can usually fix the wave function, but not the trajectory
(i.e., not the specific \paqd\ state) of a particle. This means, that with every macroscopically
identical repetition of an experiment, the wave function of a prepared particle
will be reproduced, but with a different specific
trajectory. These trajectories belong to the just-described family,
and make up an ensemble that the wave function is associated with. Thus the wave
function reflects the preparation procedure and describes the properties of an emergent ensemble,
but not individual events (trajectories) in it, in agreement with Einstein's views (see
corresponding discussion in \cite{blntn}). The statistical distribution of trajectories in this
ensemble will be discussed later.

When a particle moves in an infinite phase space, its position in configuration space moves
with the velocity given by
Eq.~(\ref{vlct}). This is the same velocity that is attributed to the particle in the
\mbox{de Broglie\,-\,Bohm} theory (DBBT) \cite{dbb,dbb2}, where the wave function and particle's
position are considered as fundamental ele\-ments of physical reality. It is then postulated that
the wave function evolves according to \Schs equation (\ref{se}) and guides the motion of a
particle according to Eq.~(\ref{vlct}). Alternatively, it is assumed that relation (\ref{vlct}) is
satisfied at some initial moment of time, and then the particle moves according to Newton's law,
but under the influence of an additional ``quantum potential", which is created by the wave
function and is given by the part of $H^S$, Eq.~(\ref{hs}), proportional to~$\hbar^2$. With an
additional assumption about initial statistical distribution of particles, DBBT is known to
reproduce experimental predictions of QM. From the \paqd\ point of view, the relation between
\paqd, DBBT, and standard QM is as follows: While \paqd\ develops both the ODE part of the theory,
describing the particle's motion, and the PDE part, which describes the evolution of the action
function, the standard QM restricts itself to the PDE part, thus being an
analog of the Hamilton-Jacobi part of classical mechanics without its Newton/Hamilton ODE part.
Consequently, to compensate for this missing part of the theory, QM employs the statistical
interpretation, which postulates the missing part's results. The progress achieved by DBBT is based
on the observation that the need for the interpretation disappears if one postulates just described
dynamical law of particle's motion, for all experimental predictions of QM can be deduced from
this law mathematically. However, in the absence of a full geometric picture and the theory of
equations of motion, developed in section 2, this modification of the theory required a promotion
of the wave function to the rank of a real physical field that guides
the particle or acts on it (but is not acted upon) with a quantum potential. As a result, DBBT drew
a picture of the world so alien to the generally accepted ideas about a possible structure of
physical theory, that the majority of the physical community found it too hard to accept, in spite
of the theory's success with some difficult issues of QM, such as the measurement problem. As was
discussed above, far from declaring the wave function a real physical field, \paqd\ may deal
without it at all. However, using the wave function may be convenient from the practical point of
view. Thus for \paqd, DBBT just
implements the generalized Jacobi method: rather than solve the ordinary differential equations of
motion, one can instead solve the \Sch equation or QHJE, and then get the particle's velocity from
Eq.~(\ref{vlct}), where momentum $S_j$ is obtained from the real part of the action function by a
simple differentiation. The same procedure works in classical mechanics, and so for \paqd\ the DBBT
program sounds exactly like a suggestion to consider classical mechanics as a theory of particles
and real physical ``action field" $S$ that evolves according to the Hamilton-Jacobi equation and
guides particles, forcing them to move with the velocity $\bv = \nabla S/m$. Besides different
physical picture, \paqd\ also differs from DBBT by an extra requirement of analyticity, which will
become increasingly important in what follows. Nevertheless, the particles in \paqd\ move along the
same ``Bohmian trajectories" with velocity (\ref{vlct}) as in DBBT, which will allow us to use, 
with proper modifications, some of its important results.

The classical limit of \paqd\ is best seen in the $(S,R)$ formulation. The Hamiltonian $H^S$ and
the equations of motion for the action function $S$ and its derivatives contain terms proportional
to~$\hbar^2$. When these terms are small compared to other, ``classical" ones,
they may be neglected. The equations for $S$ and $S_\sigma$ then decouple from the equations for
$R$ and $R_\sigma$, and directly turn into
the system of equations for the theory with a Hamiltonian, given by the first two terms
of~$H^S$, Eq.~(\ref{hs}). This is a first-order Hamiltonian of classical mechanics, and
the theory is classical mechanics, prolonged from the classical space of 1-jets $J^1_1$ to the
corresponding infinite jet space $J^\infty_1$ of the ``full" theory, which describes all
derivatives of the action function. As was discussed in section 2.7, in the infinite
system of equations of this theory the standard equations of classical mechanics form a closed
``classical"
subsystem, which provides full information about the evolution of the action function $S$, its
first derivatives, i.e., components of classical momentum, and, most importantly, the particle's
position~$\bq$. If the equations of the classical subsystem are solved, the higher derivatives
$S_\sigma$ can be obtained from the solution either by
quadrature~(\ref{dactrs}) or simply by direct differentiation of the action function~$S$.
Thus in a classical limit (or in a formal limit $\hbar \rightarrow 0$) \paqd\
dramatically simplifies, both conceptually and in terms of its complexity, and reduces to this
subsystem, i.e., to classical mechanics.

\section[The form of Hamiltonian, superposition principle, path integration, and
wave-particle duality]
{The form of Hamiltonian, superposition principle, path\newline integration, and wave-particle
duality} \label{fh}
\setcounter{equation}{0}

The evolution of the wave function $\psi(\bx,t)$ over an infinitesimal time interval $\veps$ may
be represented by a one-step Feynman integral as
\begin{equation}\label{1sfi}
    \psi(\bx,t+\veps) = \int \exp \left[ \ih \veps L\left( \frac{\bx-\by}{\veps},
    \by, t \right)\right] \psi(\by,t) \prod_{i=1}^n \frac{dy^i}{A}\,+\,\cO(\veps^2)\,,
    \dsplbl{1sfi}
\end{equation}
where $L(\bv,\by,t)=m \bv^2/2 - U(\by,t)$ is a classical Lagrangian, and $A=\sqrt{2\pi i\hbar
\veps/m}$ is a nor\-ma\-li\-za\-tion constant~\cite{fh}. Let $\bq(t)$ be some curve in the base
space, and $\{p(t)\}$ be the set of values of the action function $p$ and its derivatives at time
$t$ at the point $\bq(t)$: $\{p(t)\} = \big\{ \partial_\sigma (\hbar/i) \ln
\psi\big(\bq(t),t\big) \big\}$. At any time, we have then the wave function
\begin{equation}\label{psiqt}
    \psi(\bx,t) = \exp \left[ \ih \sum_\sigma \frac{1}{\sigma!}\, p_\sigma(t)
                  \big(\bx - \bq(t)\big)^\sigma \right] .   \dsplbl{psiqt}
\end{equation}
At time $t=0$, let the curve pass through a point $\bq=0$ with velocity $\bv$, so that $\bq(0)=0$
and $\bq(\veps)=\bv\veps$. Using Eq.~(\ref{psiqt}) and letting $\bz=\bx-\bv\veps$, $p_\sigma =
p_\sigma(0)$, and $p'_\sigma = p_\sigma(\veps)$, we have from Eq.~(\ref{1sfi}):
\begin{equation}\label{psigser}
    \sum_\sigma \frac{1}{\sigma!}\, p'_\sigma \, \bz^\sigma = \hi \ln \int \exp \left[ \ih
    \veps L\left( \frac{\bz-\by}{\veps}+\bv,
    \by, t \right) + \ih\sum_\sigma \frac{1}{\sigma!}\, p_\sigma \by^\sigma
     \right] \prod_{i=1}^n \frac{dy_i}{A} \,\,+\,\, \cO(\veps^2)\, .  \dsplbl{psigser}
\end{equation}

\Schs equation is a consequence of Eq.~(\ref{1sfi}), therefore, Eq.~(\ref{qhje}) with the
Hamiltonian function~(\ref{hp}), and then Eq.~(\ref{tottd}) follow from it as well. It is, however,
instructive to obtain that the evolution of momentums $p_\sigma$ along the curve $\bq(t)$
corresponds to Eq.~(\ref{tottd}), i.e., that
\begin{equation}\label{psigprime}
    p'_\sigma = p_\sigma + \veps ( p_{\sigma i} v^i - H_\sigma ) + \cO(\veps^2)
    \dsplbl{psigprime}
\end{equation}
with $H_\sigma$ given by Eq.~(\ref{hsigma2}), directly from Eq.~(\ref{psigser}). For that,
we need to find the coefficients of the expansion of the integral in (\ref{psigser}) in
powers of $z^i$. It is convenient to introduce one more variable $\bu = \by - \bz$ and, using
the expression for the Lagrangian, rewrite this integral as
\begin{equation}\label{psiglast}
    \hi \ln \int \exp \left[ \frac{im}{2\hbar\veps} \, u^2 - \frac{im}{\hbar} \, \bu \bv
    + \frac{im}{2\hbar} \, v^2 \veps - \frac{i\veps}{\hbar} \, \sum_\sigma
    \frac{1}{\sigma!} \, U_\sigma (\bz+\bu)^\sigma + \ih \sum_\sigma \frac{1}{\sigma!} \,
    p_\sigma (\bz+\bu)^\sigma \right] \prod_{i=1}^n \frac{du^i}{A} \, .   \dsplbl{psiglast}
\end{equation}
The integral here is of the kind that may be evaluated using standard rules of the diagram
technique \cite{vslv}. The logarithm in front of the integral means that
we should include only connected
diagrams. The first term in the exponent defines a contraction $\langle u^j u^k \rangle =
-(\veps\hbar/im)\,\delta_{jk}$. Since we are only interested in the zero-order
and first-order contributions of $\veps$, and the contraction is proportional to $\veps$, we
have to consider only diagrams with one contraction or with no contractions at all. Then the
expression in Eq.~(\ref{psiglast}) will become the sum of the following contributions: The
contraction of the term $-(im/\hbar)\bu\bv$ in the exponent with itself gives $-m v^2 \veps/2$,
where $1/2$ is a symmetry factor, and cancels the contribution of the third term in the
exponent. As the potential term in the exponent already has a coefficient $\veps$ in front
of it, we can write there
$\bz^\sigma$ instead of $(\bz+\bu)^\sigma$, and then the contribution of this term to
(\ref{psiglast}) will be equal to $- \veps \sum_\sigma (1/\sigma!) \, U_\sigma \bz^\sigma$.
In the last term in the exponent, it is sufficient to expand $(\bz+\bu)^\sigma$ up to
the second power of~$u^i$. This term then becomes equal to $(i/\hbar)\sum_\sigma
(\bz^\sigma/\sigma!)\,(p_\sigma + p_{\sigma i}u^i + p_{\sigma ij}u^iu^j/2)$. Now the
contribution to (\ref{psiglast}) of the term $p_\sigma$ here is equal to $\sum_\sigma
(\bz^\sigma/\sigma!)p_\sigma$, the contribution of the contraction of $p_{\sigma i}u^i$ with
$-(im/\hbar)\bu\bv$ is equal to $\veps \sum_\sigma(\bz^\sigma/\sigma!)p_{\sigma i}v^i$, the
contribution of the contraction of $p_{\sigma i}u^i$ with itself is equal to
$-(\veps/2m)\sum_{\nu\mu}(\bz^\nu\bz^\mu/\nu!\mu!)p_{\nu i}p_{\mu i}$, where $1/2$ is a
symmetry factor, and finally the contribution of the contraction of $u^i$ with $u^j$ in
$p_{\sigma ij}u^iu^j$ is equal to $-(\veps\hbar/2im)\sum_\sigma(\bz^\sigma/\sigma!)
p_{\sigma jj}$. Now, collecting all terms and comparing the coefficients for equal powers of $z^i$
in both parts of Eq.~(\ref{psigser}), we obtain Eq.~(\ref{psigprime}) with $H_\sigma$ given by
Eq.~(\ref{hsigma2}).

We have thus demonstrated that the whole system of equations (\ref{tottd}) with
Hamiltonian (\ref{hp}) may be compactly represented
by one equation (\ref{psigser}). On the other hand, starting from the ODEs~(\ref{tottd}) with
Hamiltonian~(\ref{hp}), and reversing the above arguments, we can derive Eq.~(\ref{psigser}) in the
framework of \paqd. Equation~(\ref{psigser}), therefore, is equivalent to the system (\ref{tottd})
with Hamiltonian (\ref{hp}), and, being augmented with the variational principles of
section~2.4 for determination of the velocity $\bv$,
may be taken as an alternative starting point of the theory. Equation~(\ref{psigser}) then will
fit in the general scheme of \paqd\ as a separate postulate, restricting the possible form of the
Hamiltonians in~(\ref{tottd}). For more general situations than the just-considered motion of a
particle in a flat space under the influence of a potential force, the form of the Lagrangian
function in (\ref{psigser}) will be different, for example, in magnetic field it will include a
linear in velocity term $(e/c)\bA\cdot(\bx-\by)/\veps$, where~$\bA$ is a vector potential evaluated
at the ``midpoint" $(\bx+\by)/2$ \cite{schlmn}. The Hamilton operator in \Schs equation and the
Hamiltonian function $H$ in Eq.~(\ref{tottd}) will then be determined by this Lagrangian function
in the same way as for the standard case above, and to be able to develop \paqd\ we have to require
that the function $H$ satisfies HC1, Eq.~(\ref{hcond}). In the development based on
Eq.~(\ref{psigser}), this additional condition appears completely arbitrary and artificial. We will
see in the next two sections, however, that it is also necessary for a derivation of the
probabilistic interpretation of the wave function.

As was just mentioned, the quantum Hamiltonian function in Eq.~(\ref{tottd}), which describes the
dynamics of a particle in \paqd, is determined by the Lagrangian function in~(\ref{psigser}),
or by the corresponding classical Hamiltonian, obtained from it in the limit $\veps\rightarrow 0$.
This last Hamiltonian will necessarily be of first order, i.e., it will depend only on position and
the usual momentums $p_i$, and not on any $p_\sigma$ with $|\sigma|>1$. Thus the approach that
starts from Eq.~(\ref{psigser}) automatically reduces the variety of possible quantum Hamiltonians
in~(\ref{tottd}), which were previously restricted by the Hamiltonian conditions only,
to those which are obtainable in the described
way from some classical Hamiltonian of the first order. As was discussed above, this classical
Hamiltonian will then describe the classical limit of the corresponding quantum theory, which in
turn will become its quantization. However, this quantization does not have to be unique. Indeed,
while the quantum Hamiltonian in (\ref{tottd}) is determined by a Lagrangian function in
(\ref{psigser}), which is written for finite $\veps$, the corresponding classical
Hamiltonian is obtained from this function in the limit $\veps\rightarrow 0$. Consequently,
there might be cases when different Lagrangian functions in (\ref{psigser}) define different
quantum Hamiltonians, but the same classical Hamiltonian in the $\veps\rightarrow 0$ limit.
For example, the quantum Hamiltonian in magnetic field would be different, if the vector potential
in the term $(e/c)\bA\cdot(\bx-\by)/\veps$ in Lagrangian was evaluated at other point than
$(\bx+\by)/2$; this ambiguity reflects operator ordering ambiguity in canonical quantization
\cite{schlmn}. In such cases, we will regard these different Lagrangians as defining physically
different quantum theories that nevertheless share a common classical limit. In other words, the
more fundamental quantum theory must uniquely define its classical approximation, but not the other
way around. In such situations, if competing theories are supposed to describe nature, then not
more than one of them can do it right, and it should be chosen based on its phenomenological
success.

The next observation regarding the approach based on Eq.~(\ref{psigser}) as a foundation of the
theory is that it automatically introduces the wave function, which in \paqd\ is
{\em defined} by Eq.~(\ref{psiqt}), as an object with a linear law of evolution~(\ref{1sfi}). The
superposition principle then follows immediately. As was discussed above, the particle's state in
\paqd\ may be described by its position and a wave function, and it is the wave
function part of this description that is the subject of the superposition principle: if at some
initial time $t=t_0$, the wave function~$\psi$, Eq.~(\ref{psiqt}), is equal to a linear
combination of other functions $\psi_k$, $k=1,\ldots,n_k$, of the form~(\ref{psiqt}) with the same
$q(t_0)$ as $\psi$, then it continues to be that combination
as time evolves. The position $q(t)$ of the particle then evolves according to the
equations of motion with the wave function $\psi$.
There is no such concept as superposition of a particle's position, and the time evolution
of this position in a state with the wave function~$\psi$ is not related in any simple way to
evolutions in states with wave functions $\psi_k$.

Equation (\ref{psigser}) is equivalent to Eq.~(\ref{1sfi}), from which \Schs equation (\ref{se})
immediately follows. An even more important property of the approach that selects Hamiltonians in
(\ref{tottd}) using Eq.~(\ref{psigser}) is that while the integral in~(\ref{psigser}) is
mathematically well defined, and does not suffer from any difficulties that are usually
associated with the path integration, the ite\-ration of Eq.~(\ref{1sfi}) leads to a Feynman
path-integral representation of wave and action functions. Thus the value of the action function at
some point of the base space may be obtained in two seemingly very different ways: either as action
integrals (\ref{lp}) and (\ref{lsandr}) along a particle's well-defined trajectory, coming to this
point, or as a logarithm of a sum over paths. Obviously, this remarkable duality is a consequence
of the fact that the quantum Hamiltonian (\ref{hp}), which determines the particle's dynamics in
\paqd, was obtained from the Lagrangian function of the path integral via Eq.~(\ref{psigser}).
Still, it is not immediately clear how the action
integrals (\ref{lp}) and (\ref{lsandr}), which operate only with the values defined directly on a
particle's trajectory, and not anywhere else, can reproduce the sum over paths, which is obviously
affected by the whole neighborhood of the trajectory. The answer is that the trajectory itself, and
therefore the action integrals, are determined by an infinite system of (ordinary differential)
equations, which depend on all derivatives
$p_\sigma$ of the action and $U_\sigma$ of the potential.
But the action function is analytic (this is one of the postulates of \paqd), and we also assume
that the potential function is analytic (and believe it always is analytic in nature).
Consequently the trajectory, using these derivatives, obtains the full knowledge of the action and
potential functions everywhere, and with it the ability to reproduce results obtained by path
integration. In other words, the action integrals (\ref{lp}) and (\ref{lsandr}) utilize the
information about analytic action and potential functions which is contained in their derivatives
at the points of the actual particle's trajectory,
while summation over paths uses the values of these functions on the whole space
directly. But path integration provides a purely wave description of a particle's behavior, which
naturally explains such characteristically wave phenomena as interference and diffraction.
Therefore, it is because of a special form of the Hamiltonian, obtained from Eq.~(\ref{psigser}),
and analyticity of the action/wave function, that the particle, which moves along a single
trajectory, exhibits at the same time
the characteristics of a wave, thus possessing the property of wave-particle duality. For
example, in agreement with conclusions of \cite{sonego}, in a two-slit experiment the motion of a
particle, passing through one slit, may depend
crucially on whether the other slit is open or closed, even when the difference between the
classical forces acting on the particle in these two cases is negligible.
This is of course a purely
quantum effect, completely impossible in classical theory, where a particle's trajectory is
determined by a finite system of equations that depend only on the first derivative of the
potential. Therefore, the wave-particle duality in the quantum domain receives a simple and natural
mathematical explanation in~\paqd.

\section[Variational principle, continuity equation, and invariant measure]
{Variational principle, continuity equation, and invariant\newline measure}
\setcounter{equation}{0}

Equations (\ref{szetadot}), (\ref{hz}) for $R(\bx,t)$ may be rewritten in the form of a continuity
equation
\begin{equation}\label{curcons}
    \dovd{j^0}{t} + \mbox{div} \bj = 0,    \dsplbl{curcons}
\end{equation}
where $j^0 = |\psi|^2 = e^{2R}$, $\bj = j^0 \bv$, and $\bv = \nabla S/m$ is the particle velocity.
The invariance of the measure, associated with conserved current $(j^0,\bj) = |\psi|^2(1,\bv)$,
is used in the next section to demonstrate that the particle's probability density is equal to 
$|\psi|^2$. Therefore, the conservation of this current is a very important element of the 
theory, and in this section we present several different proofs of it, which will allow us to 
better elucidate its origin. We will also derive an important expression for a corresponding 
invariant measure.

\subsection{Variational principle and current conservation} \label{vpcc}

Equation (\ref{1sfi}) implies that \Schs equation may be obtained from a stationary action
principle. Indeed, consider the value
\begin{equation}\label{hpsipsi}
    \mathbf{H}(\psi,\psi^*) = \frac{d}{d\veps} \int \psi^*(\bx) \exp \left[ \ih \veps L
    \left( \frac{\bx-\by}{\veps},\by, t \right)\right] \psi(\by) \prod_i \left.
    \frac{dx^idy^i}{A} \,\, \right|_{\veps=0} .  \dsplbl{hpsipsi}
\end{equation}
Since the integration with a factor $\exp(i\veps L/\hbar)$ propagates $\psi(\by)$ from time
$t$ to time $t+\veps$, and $\psi^*(\bx)$ from $t$ to $t-\veps$, we obtain, obviously,
\begin{equation}\label{psigam}
    \dot{\psi}(\bx) = \frac{\delta}{\delta\psi^*(\bx)}\, \mathbf{H}(\psi,\psi^*) \, , \quad\;
    \dot{\psi^*}(\bx) = -\frac{\delta}{\delta\psi(\bx)}\, \mathbf{H}(\psi,\psi^*) \, ,
    \dsplbl{psigam}
\end{equation}
so that $\mathbf{H}(\psi,\psi^*)$ is a Hamiltonian function with respect to the canonical field
coordinates
$\psi(\bx)$ and their conjugate momentums $\psi^*(\bx)$ (or coordinates $\psi^*(\bx)$ and
momentums $-\psi(\bx)$, which differs just by a canonical transformation). Being Hamilton
equations, Eqs.~(\ref{psigam}), which are equivalent to \Schs equation and its conjugate, follow,
after standard discretization, from a stationary action principle in a Hamiltonian form
$\delta\!\int\! \mathbf{L}(\psi,\psi^*)\,dt = 0$, where
\begin{equation}\label{}
    \mathbf{L}(\psi,\psi^*) = \hi\left[\int \psi^*(\bx)\dot{\psi}(\bx) \prod_i dx^i -
    \mathbf{H}(\psi,\psi^*)\right]
\end{equation}
(see \cite{arnld} and section 2.4) and the factor $\hbar/i$ is introduced for
convenience. To calculate $\mathbf{H}(\psi,\psi^*)$, integrate over $\prod_i dy^i$ in
Eq.~(\ref{hpsipsi}) to get
\begin{equation}\label{}
    \mathbf{H}(\psi,\psi^*) = -\ih \int \psi^*(\bx) \widehat{H} \psi(\bx) \prod_i dx^i \, ,
\end{equation}
where $\widehat{H}$ is a Hamilton operator, and so $\mathbf{L}(\psi,\psi^*) = \int\! \cL \prod_i
dx^i$, where the Lagrangian density~$\cL$ is
\begin{equation}\label{lpsipsi}
    \cL = \psi^*(\bx) \left(\hi\, \partial_t + \widehat{H} \right) \psi(\bx)\, .  \dsplbl{lpsipsi}
\end{equation}
For the standard Hamiltonian of \Schs equation (\ref{se}) we have then
\begin{equation}\label{}
    \cL = \psi^*(\bx) \left(\hi\, \partial_t - \frac{\hbar^2}{2m}\,\partial_j^2 + U \right)
          \psi(\bx) \, .
\end{equation}
As is well known \cite{olvr,bk}, both equations of motion and conservation laws, considered below,
remain invariant with respect to adding a total divergence to the Lagrangian density. By adding
appropriate terms to the above expression, we then obtain a familiar variational principle
$\delta\int\cL \prod_i dx^i \, dt = 0$ with symmetric Lagrangian density
\begin{equation}\label{ldens}
    \cL = \frac{\hbar}{2i} \left(\psi^*\partial_t\psi - \psi\partial_t\psi^*\right) +
    \frac{\hbar^2}{2m}\, \partial_i \psi^* \partial_i \psi + U \psi^* \psi \, , \dsplbl{ldens}
\end{equation}
and it is indeed easy to verify directly that the corresponding Euler-Lagrange equation
\begin{equation}\label{}
    \dovd{\cL}{\psi^*} - \sum_{i=0}^n \partial_i \dovd{\cL}{\,(\partial_i \psi^*)} = 0 \, ,
\end{equation}
is equivalent to the \Sch equation (\ref{se}).

By Noether's theorem \cite{olvr,ibragim,bk}, if a transformation $\psi \rightarrow \psi + \alpha
\Delta$, $\psi^* \rightarrow \psi^* + \alpha \Delta^*$ with infinitesimal real parameter $\alpha$
changes the Lagrangian density $\cL$ just by adding a total divergence to it,
\begin{equation}\label{chngL}
    \cL \rightarrow \cL + \alpha \sum_{i=0}^n \partial_i \Lambda^i \, ,   \dsplbl{chngL}
\end{equation}
then the solutions of the equations of motion \big(i.e., in our case
\Schs equation (\ref{se})\big) satisfy a local conservation law
\begin{equation}\label{lcl}
    \partial_0 j^0 + \partial_k j^k = 0     \dsplbl{lcl}
\end{equation}
with a current $j^i$ which is, for a first order Lagrangian $\cL$, equal to
\begin{equation}\label{crnt}
    j^i = \dovd{\cL}{\,(\partial_i \psi)} \,\Delta + \dovd{\cL}{\,(\partial_i \psi^*)} \,\Delta^* -
    \Lambda^i \, , \quad i = 0,\ldots,n.    \dsplbl{crnt}
\end{equation}
As was discussed at the end of section 2.5 and in section 3, by their
very construction the action function is defined
up to an additive constant and the wave function up to a constant factor. Therefore, we
should expect that the corresponding transformation does not change the equations of
motion, and so the original and transformed Lagrangian densities differ by a total divergence only.
Since the wave function is complex, we should consider two different transformations. Under the
scale transformation $\psi \rightarrow e^\alpha \psi$, $\psi^* \rightarrow e^\alpha \psi^*$,
when the wave function satisfies the \Sch equation, the Lagrangian density $\cL$ indeed changes as
in~(\ref{chngL}) with
\begin{equation}\label{}
    (\Lambda^0,{\bf \Lambda}) = \left(0,\, \frac{\hbar}{2m}\,(\psi^* \nabla \psi + \psi \nabla
    \psi^*)\right),
\end{equation}
but the sum of the first two terms in Eq.~(\ref{crnt}) in this case
is equal to $\Lambda^i$, and so the total current $j^i$ vanishes and the scale
invariance does not lead to any conservation law. The phase transformation $\psi\rightarrow
e^{i\alpha/\hbar}\psi$, $\psi^*\rightarrow e^{-i\alpha/\hbar}\psi^*$ is more useful. Lagrangian
density $\cL$ is invariant with respect to it, i.e., satisfies Eq.~(\ref{chngL}) with
$\Lambda^i=0$. Consequently, the corresponding current
\begin{equation}\label{concur}
    (j^0, \bj) = \left(\psi^*\psi,\, \frac{\hbar}{2im}\,(\psi^* \nabla \psi - \psi \nabla
    \psi^*)\right)      \dsplbl{concur}
\end{equation}
is conserved, i.e., satisfies Eq.~(\ref{lcl}), which in this case coincides with a continuity
equation~(\ref{curcons}). Therefore, the current $(j^0,\bj)$ is equal to $|\psi|^2(1,\bv)$, where
the velocity $\bv$ is given by Eqs.~(\ref{vlct}) or (\ref{vlctc}). To obtain this form of the
conserved current, we used an explicit form of a Hamiltonian here. However, in section 5.2
we will show that this result has a much more general character, namely, for a wide range of
possible Lagrangian functions $L$ in Eq.~(\ref{hpsipsi}), the current, which is conserved due to
the phase invariance, is equal to $|\psi|^2(1,\bv)$ with~$\bv$ given by Eq.~(\ref{cqdot})
\big(which for the standard Hamiltonian coincides with~(\ref{vlctc})\big).

\subsection{Current conservation for Hamiltonian operators of general form} \label{cchogf}

It is desirable to derive the conservation of the current $(j^0,\bj) = |\psi|^2(1,\bv)$
under more general assumptions than above where we used an explicit form
of a standard Hamiltonian. Here we will show that this conservation follows from the phase
invariance of the Lagrangian density~$\cL$, Eq.~(\ref{lpsipsi}), for an arbitrary quadratic in
velocity Lagrangian function $L$ in Eq.~(\ref{hpsipsi}), provided it satisfies some simple
conditions.

First substitute into (\ref{lpsipsi}) the representation~(\ref{psi}) to get
\begin{equation}\label{cl}
    \cL = |\psi(\bx)|^2 \left(p_t + H\right) ,    \dsplbl{cl}
\end{equation}
where the Hamiltonian $H$, which corresponds to the Lagrangian function $L$ in~(\ref{hpsipsi}), is
a function of the space derivatives of the action function $p$. We have from~(\ref{psi}) and
(\ref{hpsipsi})
\begin{equation}\label{w}
    H = - \left.\frac{dW}{d\veps}\,\right|_{\veps = 0} , \quad \quad W = \hi \ln \int \exp \left[
        \ih \veps L + \ih p(\bx + \bu) \right] \prod_i \frac{du^i}{A} \, ,   \dsplbl{w}
\end{equation}
where we introduced $\bu = \by - \bx$.
As in section~4, expand the exponent in (\ref{w}) in powers of $u^i$ and
consider $W$ as a generating function for connected diagrams. We will assume that similar to the
case of a standard Hamiltonian, the corresponding contraction $\langle u^j u^k \rangle$ is purely
imaginary and proportional to~$\veps$. To account for a possible presence of a magnetic field, we
allow the product $\veps L$ to have a vector potential term $\,-(e/c) A_k u^k$ inside it
\cite{schlmn}, but assume that there are no other $\veps$-independent and linear in $u^k$
terms there. The Lagrangian density~(\ref{lpsipsi}) will then be a function of the wave
functions~$\psi$ and $\psi^*$ and derivatives of~$\psi$ up to a second order. \mbox{Using} the
corresponding formulas for second-order Lagrangian functions
\cite{olvr,ibragim,bk}, the current, which conserves due to invariance of the Lagrangian density
$\cL$, Eq.~(\ref{lpsipsi}), with respect to the phase transformation $\psi\rightarrow
e^{i\alpha/\hbar}\psi$, $\psi^*\rightarrow e^{-i\alpha/\hbar}\psi^*$, will then be equal to
$(j^0,\bj)$, where
\begin{equation}\label{current}
    \begin{array}{ccl}
      j^0 & = & \dsp \ih \psi \dovd{\cL}{\psi_t} \, , \\[0.4cm]
      j^k & = & \dsp \ih \psi \left(\dovd{\cL}{\psi_k} - D_l \dovd{\cL}{\psi_{kl}}\right) +
                \ih \psi_l \dovd{\cL}{\psi_{kl}} \, .   \dsplbl{current}
    \end{array}
\end{equation}
We now want to rewrite these expressions in terms of derivatives of $H$ over $p_k$ and $p_{kl}$.

We have from Eq.~(\ref{psi}) the following relations between partial derivatives of $p$ and $\psi$:
\begin{equation}\label{pkpkl}
    \begin{array}{ccl}
      p_k & = & \dsp \hi \frac{\psi_k}{\psi} \, , \\[0.3cm]
      p_{kl} & = & \dsp \hi \left(\frac{\psi_{kl}}{\psi} - \frac{\psi_k \psi_l}{\psi^2}\right) .
      \dsplbl{pkpkl}
    \end{array}
\end{equation}
Let $p_{kl}$ and $p_{lk}$ enter the expression for $\cL$ symmetrically. We have then
from~(\ref{pkpkl})
\begin{equation}\label{}
    \begin{array}{ccl}
      \dsp\dovd{\cL}{\psi_k} & = & \dsp  \hi \,\frac{1}{\psi}\left(\dovd{\cL}{p_k} - 2 \ih p_l
      \dovd{\cL}{p_{kl}}\right) , \\[0.4cm]
      \dsp\dovd{\cL}{\psi_{kl}} & = & \dsp \hi \,\frac{1}{\psi}\, \dovd{\cL}{p_{kl}} \, ,
    \end{array}
\end{equation}
where the factor of 2 in the first equation compensates for the dropped contribution of $\partial
\cL/\partial p_{lk}$. Substituting these expressions into Eq~(\ref{current}) and using
Eq.~(\ref{cl}) and condition~(\ref{hcond}), we obtain for the current
\begin{equation}\label{currnt}
    \begin{array}{ccl}
      j^0 & = & |\psi|^2 , \\[0.1cm]
      j^k & = & \dsp |\psi|^2 \left[\dovd{H}{p_k} + \ih \left(\bar{p}_l - p_l\right)
                \dovd{H}{p_{kl}} \right] ,    \dsplbl{currnt}
    \end{array}
\end{equation}
so that $j^0$ has the right form, and we need to evaluate the derivatives of $H$ in an expression
for $j^k$. For every function $B$ of coordinates~$u^i$, we denote by $\langle B\rangle$ the
corresponding sum of connected diagrams produced by the generating function~$W$:
\begin{equation}\label{}
    \langle B\rangle = \frac{\dsp\int \exp \left[\ih \veps L + \ih \left(p + p_k u^k + \half\,
    p_{kl} u^k u^l + \cdots \right) \right] B \prod_i \frac{du^i}{A}}
    {\dsp\int \exp \left[\ih \veps L + \ih \left(p + p_k u^k + \half\, p_{kl} u^k u^l + \cdots
    \right) \right] \prod_i \frac{du^i}{A}}\, ,
\end{equation}
where the derivatives $p_\sigma$ are taken at the point $\bx$ where the Lagrangian
density~(\ref{cl}) is evaluated. We have then from Eq.~(\ref{w})
\begin{equation}\label{}
    \dovd{H}{p_k} = - \left.\frac{d}{d\veps}\, \langle u^k \rangle\,\right|_{\veps = 0} =
    - \left.\frac{d}{d\veps}\, \ih \left(p_l - \frac{e}{c}\, A_l\right) \langle u^k u^l \rangle \,
    \right|_{\veps = 0} \, ,
\end{equation}
where we used the fact that the only nonzero contribution to $d \langle u^k \rangle /d \veps$ for
$\veps = 0$ comes from contraction of $u^k$ with $\veps$-independent terms in the exponent. On the
other hand, we have
\begin{equation}\label{}
    2\, \dovd{H}{p_{kl}} = -\left.\frac{d}{d\veps}\,\langle u^k u^l \rangle\,\right|_{\veps = 0}\,,
\end{equation}
and so
\begin{equation}\label{}
    \begin{array}{ccc}
      \dsp \dovd{H}{p_k} & = & \dsp 2\, \ih \left(p_l - \frac{e}{c}\,A_l\right)
      \dovd{H}{p_{kl}} \, ,\\[0.4cm]
      \dsp \dovd{\bar{H}}{\bar{p}_k} & = & \dsp 2\, \ih \left(\bar{p}_l - \frac{e}{c}\,A_l\right)
      \dovd{H}{p_{kl}}\, ,
    \end{array}
\end{equation}
where to obtain the second equation we conjugated the first one and used that $\partial H/\partial
p_{kl}$ is purely imaginary. Now using these equalities in Eq.~(\ref{currnt}) for $j^k$, we obtain
\begin{equation}\label{jk}
    j^k = \dsp \half \,|\psi|^2 \left(\dovd{H}{p_k} + \dovd{\bar{H}}{\bar{p}_k}\right)
        = \dsp |\psi|^2 v^k ,    \dsplbl{jk}
\end{equation}
where $\bv$ is the particle's velocity (\ref{cqdot}), as was required.

\subsection{Invariant measure}  \label{intmeas}

For an arbitrary current $(i^0,\bi)$ in the base space $M$ that satisfies a continuity equation
$\partial_t i^0 + \partial_k i^k = 0$, a form $\nu = i^0\Omega$,
$\Omega = \rmd x^1 \wedge \cdots \wedge \rmd x^n$, integrated over any subspace of $M$
corresponding to a fixed time $t$, is invariant with respect to a vector field
$Y = \partial_t + u^k \partial_k$ with $\bu = \bi/i^0$, so that this form defines on
$M$ a measure that is invariant with respect to the flow of~$Y$. Indeed, let $D$ be an
arbitrary cell in configuration space, every point of which moves with velocity~$\bu$. If
$ds$ is an element of the boundary of $D$, orthogonal to a unit vector $\bn$ pointing outside,
then by the continuity equation over a time interval $dt$ the integral $\int_D \nu$ will reduce by
$\bi\bn\, dsdt$ due to the current~$\bi$ through $ds$.
On the other hand, since the element~$ds$ moves
with velocity~$\bu$, during time~$dt$ a volume $\bu\bn\,dsdt$ will be added to $D$, and with it
a value $i^0\bu\bn\,dsdt$ added to $\int_D \nu$. Therefore, this integral will not change, and
so $\nu$ is invariant with respect to $Y$.

The same result may also be obtained by direct calculation. Indeed, using equalities $Y(\nu) =
\rmd(Y\ip\nu) + Y\ip\rmd\nu$ and $\rmd\Omega = 0$, we obtain
\begin{equation}\label{}
    \begin{array}{ccl}
      Y(\nu) & = & \rmd(i^0Y\ip\Omega) + Y\ip\rmd(i^0\Omega) \\[0.2cm]
      & = & \rmd i^0\wedge(Y\ip\Omega) + i^0\rmd(Y\ip\Omega) + Y\ip\rmd i^0\wedge\Omega \, .
    \end{array}
\end{equation}
In the last term, substitute $Y\ip\rmd i^0\wedge\Omega = Y(i^0)\,\Omega - \rmd
i^0\wedge(Y\ip\Omega)$ to get
\begin{equation}\label{xofomega}
    Y(\nu) = i^0\rmd(Y\ip\Omega) + Y(i^0)\,\Omega \, .   \dsplbl{xofomega}
\end{equation}
In the second term of this equation, we have $Y(i^0) = \partial_t i^0 + u^k\partial_k i^0$, and in
the first term
\begin{equation}\label{}
    Y\ip\Omega = \sum_k (-1)^{k-1}u^k(\rmd x^1\wedge\cdots\wedge\rmd x^n)',
\end{equation}
where $(\,)'$ means that the factor $\rmd x^k$ in the product is dropped. From this, we have
\begin{equation}\label{}
    \rmd(Y\ip\Omega) = \Omega\,\partial_k u^k + \rmd t \sum_k(-1)^{k-1}\partial_t u^k (\rmd x^1
    \wedge \cdots \wedge \rmd x^n)'.
\end{equation}
We will integrate $Y(\nu)$ over surfaces with
fixed $t$ in the base space $M$, so we are interested in a pullback $\pi_t^*Y(\nu)$, where
$\pi_t\!:\,Q\rightarrow M$ is the natural embedding that maps configuration space $Q$ into such
surfaces. Then obviously the term with $\rmd t$ does not contribute to such integrals,
$\pi_t^*\rmd t= 0$, and collecting remaining terms we have $\pi_t^*Y(\nu) = \Omega\,(\partial_t
i^0 + u^k\partial_k i^0+i^0 \partial_k u^k)$. But the last two terms sum to $\partial_k(u^k i^0) =
\partial_k i^k$, and so from the continuity equation we obtain $\pi_t^*Y(\nu) = 0$.

The vector flow $X$ in \paqd\ is defined by the first line of Eq.~(\ref{fldxpp}). Its first two
terms have the form of the just-considered vector field $Y$ with
respect to a current (\ref{concur}),
and so conserve the form $\omega = j^0\Omega$, while the last two terms, where $\sigma\neq\0$, when
acting on this form give zero. However, the flow $X$ is defined in the infinite phase space \cP,
rather than in the base space~$M$. Consequently, to formulate the invariance condition, we use the
map $\pi_t\!:\,Q\rightarrow \cal{P}$, which projects configuration space $Q$ into a part of the
graph of a solution of \Schs equation with given~$t$. With so-defined $\pi_t$, we have then the
desired identity $\pi_t^*X(\omega) = 0$, which expresses the invariance of the form $\omega$ and
corresponding measure with respect to the Hamiltonian flow~(\ref{fldxpp}).

\section{Probability density}
\setcounter{equation}{0}

In the previous sections we studied the mathematical structure of \paqd. Here we start considering
its physical implications, i.e., experimental consequences of the assumption that particles move
along the trajectories that we discussed. It is then natural to think, and we will confirm it
later, that in \paqd\ the experimentally measured particle's position should be equal to its
position in \cP. In this case, what can \paqd\ say about the distribution of this measurement's
results?

We believe that in every repetition of an experiment, in which the particle is described by a wave
function~$\psi$, its position coordinate in \cP\ assumes a random value, determined by a specific
history of this particular repetition. In the mathematical limit of an infinite number of such
repetitions, the results form an ensemble that determines the particle's probability density
$\rho$: the probability of finding the particle in any volume element is equal to the relative
number of
ensemble members with the particle inside that element. It seems natural to assume, in agreement
with experiment, that this probability density is determined by the wave function only, and not by
the way in which the ensemble with this wave function was created. Then once created at some time
$t$, the ensemble remains representative for all future time, for one of the ways to create an
ensemble at any time $t' > t$ is to create it at time $t$ and let it evolve till time $t'$. But
during such evolution, every volume element $dV$, propagating with \paqd\ Hamiltonian flow,
continues to contain the same ensemble members, and so the probability $\rho dV$ to find the
particle in this element remains constant, which means that the change in the probability density
$\rho$ in the
element is inversely proportional to $dV$. On the other hand, as was demonstrated in the
previous section, the product $|\psi|^2 dV$ in this element also remains constant. Consequently,
along any given trajectory, the probability density $\rho$ should be proportional to~$|\psi|^2$.
The coefficient of proportionality can, by this reasoning, depend on trajectory. However, we note
that once selected, this coefficient should remain fixed in the presence of any external fields
that may be applied to the particle in the future. Since such fields can shuffle trajectories in an
arbitrary way, but the coefficient should remain a continuous function of trajectory, it is clear
that for all trajectories it must be the same. Moreover, even if for some reason configuration
space $Q$ splits into two subspaces $Q_1$ and $Q_2$ such that trajectories never cross from one of
them to the other, according to the way the wave function is brought into \paqd, that will only
mean that rather than being defined up to one constant factor in $Q$, the wave function is now
defined up to two independent constant factors in $Q_1$ and $Q_2$. Obviously, these factors can be
chosen in such a way as to make the coefficients of proportionality between $\rho$ and $|\psi|^2$
in $Q_1$ and $Q_2$ equal. We conclude that the probability density $\rho$ should
be proportional to $|\psi|^2$ with a constant coefficient, or equal to it if $\int |\psi|^2 dV =
1$, i.e., if the wave function is normalized. Thus in \paqd\ this classic relation between the wave
function and probability density becomes just a property of the wave function, rather than its main
physical meaning. It is worth recalling that this property is a consequence of such fundamental
elements of the theory as the possibility of obtaining the \Sch equation from a variational
principle (which in turn follows from the quantization via one-step Feynman integral --- see the
corresponding discussion at the beginning of section 5) and the definition of an action function,
which leaves the freedom to add an arbitrary constant to it and which, therefore, for the theory
following from a variational principle, results in a corresponding current conservation. Also, in
the process of proving that the space part of the current has the desired form (\ref{jk}) for a
general Hamiltonian function $H$ obtained from the one-step Feynman integral, we had to require
that this function $H$ satisfies HC1~(\ref{hcond}). Thus for any theory based on the one-step
Feynman integral, the condition (\ref{hcond}) is needed for the probabilistic interpretation of the
wave function as well as for the very possibility to develop \paqd\ in the first place.

The same expression for the probability density may be also obtained in a different manner, if we
count the number of possible ways by
which a given ensemble can be created. This means the following: We break configuration space~$Q$
into small cells, so that in every cell the probability density $\rho$ can be considered constant.
The ensemble of $N$ points, representing the particle in a state with a wave function~$\psi$, is
described by the numbers $N_i$ of points in every cell, $\sum_i N_i = N$. The series of $N$
experiments, which form an ensemble, is then characterized by the sequence $i_1,\ldots,i_N$ of
cells the particle was found in in each experiment, so that such ensemble can be created in
$\cN = N!/\prod_i N_i!$ different ways. These numbers $N_i$ and consequently $\cN$, depend on the
specific way the space $Q$ is split into cells, and so some reasonable prescription for
a way the splitting is done should be made. We will demand, as we did above, that once created,
the ensemble should remain representative for all future time. The splitting of configuration space
$Q$ into cells, therefore, should be such that once it is done and fixed, the number $\cN$ for
every ensemble, i.e., the number of ways this ensemble can be created, remains constant with time.
But the image of every cell $i$, corresponding to the particle's flow over arbitrary time $t$, has
the same invariant measure $|\psi|^2 dV$ and contains the same $N_i$ ensemble members as the cell
itself. It is then clear that the demand will be satisfied, if in the limit of vanishing cell
volumes, they will all have the same invariant measure, which we denote as $\Delta \Gamma$. For our
ensemble, described by the distribution of points in configuration space, we need now a
characteristic of the number of its realizations that remains finite as $N\rightarrow\infty$ and
$\Delta\Gamma\rightarrow 0$ (in this order). The number of cells in a small area of configuration
space $Q$ of a particle is proportional to $1/\Delta\Gamma$, and in a small area of configuration
space $Q^N$ of the ensemble --- to $1/(\Delta\Gamma)^N$. As $\cN$ is proportional to this number
we need to factor it out, i.e., to consider $(\Delta\Gamma)^N\cN$. The function of this last number
that stays finite as $N\rightarrow\infty$ is the $N$th root of it, and so we come to
considering~$\Delta\Gamma\cN^{1/N}$. The value
\begin{equation}\label{ge}
    \begin{array}{ccl}
      S_G(\rho) & = & \dsp \lim_{\Delta\Gamma\rightarrow 0} \lim_{N\rightarrow\infty} \ln
      \left(\Delta\Gamma\,\cN^{1/N}\right) \\[0.3cm]
      & = & \dsp \lim_{\Delta\Gamma\rightarrow 0} \left(\ln\Delta\Gamma + \lim_{N\rightarrow\infty}
      \frac{1}{N}\,\ln\cN\right)   \dsplbl{ge}
    \end{array}
\end{equation}
will be called the Gibbs entropy of the ensemble, representing probability density $\rho$ in
configuration space $Q$. We have then, using Stirling's formula,
\begin{equation}\label{}
    \lim_{N\rightarrow\infty}\frac{1}{N}\ln\cN = \lim_{N\rightarrow\infty}\frac{1}{N}\ln
    \frac{N!}{\prod_i N_i!} = -\lim_{N\rightarrow\infty} \sum_i \frac{N_i}{N}\ln \frac{N_i}{N}\, .
\end{equation}
Further, $\lim_{N\rightarrow\infty} N_i/N = \rho_i \Delta V_i = (\rho_i/|\psi_i|^2) \Delta
\Gamma$ and $\sum_i \rho_i \Delta V_i = \lim_{N\rightarrow\infty} \sum_i N_i/N = 1$, therefore
\begin{equation}\label{sgr}
    \begin{array}{ccl}
      S_G(\rho) & = & \dsp \lim_{\Delta\Gamma\rightarrow 0} \left[\ln\Delta\Gamma - \sum_i \Delta
      V_i \rho_i \ln \left(\frac{\rho_i}{|\psi_i|^2}\Delta\Gamma\right)\right] \\[0.4cm]
        & = & \dsp \lim_{\Delta V_i\rightarrow 0} \sum_i \Delta V_i \rho_i \ln
              \frac{|\psi_i|^2}{\rho_i} \\[0.5cm]
        & = & \dsp \int \rho \ln \frac{|\psi|^2}{\rho}\, dV \, .   \dsplbl{sgr}
    \end{array}
\end{equation}
In the absence of circumstances that make some cells preferable compared to others, every sequence
$i_1,\ldots,i_N$ should be assigned equal probability. In the limit $N\rightarrow\infty$, the
emerging ensemble (``Gibbs ensemble") should then maximize $\cN$ or $S_G(\rho)$. Indeed, in the
limit $N\rightarrow\infty$, the relative frequency of emergence of two ensembles with $S_G^{(2)} <
S_G^{(1)}$ is
\begin{equation}\label{rf0}
    \lim_{N\rightarrow\infty} \frac{\cN_2}{\cN_1} = \lim_{N\rightarrow\infty} \exp \left[N \left(
    S_G^{(2)} - S_G^{(1)} \right) \right] = 0 \, .     \dsplbl{rf0}
\end{equation}
Now, since $\ln x \leq x - 1$, we have from Eq.~(\ref{sgr}), when the wave function is normalized,
\begin{equation}\label{}
    S_G(\rho) \leq \int \rho \left(\frac{|\psi|^2}{\rho} - 1\right) dV = 0 \, ,
\end{equation}
$S_G(\rho)$ achieving its maximum possible value of $0$ for $\rho = |\psi|^2$, which will,
therefore, be the observed probability density.

As was discussed in section 3, the particles in \paqd\ move with the same velocity
(\ref{vlct}) as in the theory of \mbox{de Broglie\,-\,Bohm}. It was shown by Bohm \cite{dbb} that
this law of motion preserves the standard quantum form of the probability density: if $\rho$ is
equal to $|\psi|^2$ at some initial time~$t_0$, then it will stay equal to it for all $t>t_0$.
It was hypothesized
\cite{dbb,bohm2} that an arbitrary initial distribution would converge to the stable density
$|\psi|^2$ for $t$ of the order of some ``relaxation time," in the same way as macroscopic systems
converge to thermal equilibrium. The derivation, presented above, shows that this hypothesis is
unnecessary. Nevertheless, especially because we are using the concept of entropy, and looking for
a distribution which maximizes it, it is instructive to compare the situation in \paqd\ with that
in classical statistics.
We present a brief sketch of statistical distribution and entropy growth in classical
statistics, based mostly on the works \cite{jayns,lbvtz}, in a form convenient for
such comparison in the Appendix. From the discussion there, the following conclusions may be
drawn:
\newline -- The convergence to thermal equilibrium in classical statistics is related to the
growth of Boltzmann entropy~$S_B$, rather than Gibbs entropy $S_G$, which is maximized in \paqd.
\newline -- The growth of Boltzmann entropy is related to such properties of macroscopic systems
as possibility of their crude, but adequate, description; as ty\-picality (i.e., practical equality
of observable magnitudes of additive physical values to their averages over microcanonical
ensembles); and as possibility of replacement of one ensemble by the other in the process of these
systems' time evolution (see details in the Appendix). These properties exist only in
macroscopic systems that consist of enormous number of particles, and don't have any analogs in
one-particle dynamics, classical or quantum.
\newline -- The nature of the quantum distribution $|\psi|^2$ is identical to that of the
microcanonical distribution in classical statistics. Both distributions maximize the corresponding
Gibbs entropies, and emerge not because of the large number of particles in a system, but because
of the infinite number of systems,
be they one- or multi-particle, in the Gibbs ensemble. According to Eq.~(\ref{rf0}), the ensemble
with less than maximum Gibbs entropy has zero probability to arise. Consequently, all observed
distributions automatically have the maximum possible values of their Gibbs entropy, in contrast to
macroscopic systems' Boltzmann entropy, which grows due to the physical process of
thermalization.

\section{Multiparticle systems and quantum particles in a macroscopic classical environment}
\setcounter{equation}{0}

We now extend our approach to multiparticle systems. For a system of $n_p$ particles in
\mbox{$n$-dimensional} space, we do it by directly combining $n_p$ one-particle $n$-dimensional
configuration spaces into $n_p n$-dimensional configuration space $Q$ of a system. The theory of
previous sections will be generalized in a straightforward way to look like a one-particle
theory with corresponding Hamiltonian in an $n_p n$-dimensional space $Q$. In particular, the wave
function of a system is related to the corresponding action function as in Eq.~(\ref{psi}):
\begin{equation}\label{psimp}
    \begin{array}{ccl}
      \psi(\bx_1,\ldots,\bx_{n_p},t) & = & \dsp \exp \left(\ih\,p(\bx_1,\ldots,\bx_{n_p},t)\right),
      \\[0.4cm]
      p(\bx_1,\ldots,\bx_{n_p},t) & = & \dsp S(\bx_1,\ldots,\bx_{n_p},t)
                                        + \hi\,R(\bx_1,\ldots,\bx_{n_p},t) \, , \dsplbl{psimp}
    \end{array}
\end{equation}
and momentums in the infinite phase space are partial derivatives of the action function with
respect to the components of $\bx_1,\ldots,\bx_{n_p}$. The Hamiltonian function \big(the
multiparticle analog of the one-particle Hamiltonian~(\ref{hp})\big)
\begin{equation}\label{hmulti}
%    H = \half \,\sum_{k=1}^{n_p} \frac{p^2_{j_k}}{m_k} + U + \frac{\hbar}{2i} \sum_{k=1}^{n_p}
%        \frac{p_{j_k j_k}}{m_k} \, ,
    H = \sum_{k=1}^{n_p} \frac{p^2_{j_k}}{2 m_k} + U + \hi \sum_{k=1}^{n_p}
        \frac{p_{j_k j_k}}{2 m_k} \, ,    \dsplbl{hmulti}
\end{equation}
where summation over repeating indices $j_k$ is from $1$ to $n$, is obtained from the multiparticle
\Sch equation in the same way as in section 3, and defines the evolution of momentums and
particle velocities by the equations of motion (\ref{em}), so that in particular the velocity of
the $k$-th particle is $\bv_k = \nabla_k S(\bx_1,\ldots,\bx_{n_p},t)/m_k$.

For macroscopic systems, the part of the action function related to their directly observable
macroscopic degrees of freedom is much larger than Planck's constant $\hbar$. As was discussed at
the end of section 3, in the corresponding equations of motion
the terms with $\hbar$ may be dropped,
and then these equations reduce to those of classical mechanics, so that these degrees of freedom
will exhibit a classical behavior. The wave function describing these classical degrees of
freedom is given by Eq.~(\ref{psimp}) with the action function $S$ that solves the multiparticle
analog of the classical part of Eqs.~(\ref{szetadot}), (\ref{hs}), i.e., the classical
Hamilton-Jacobi equation
\begin{equation}\label{hje}
    \dovd{S}{t} + H_c\left(\bq,\, \dovd{S}{\bq}\right) = 0\,,   \dsplbl{hje}
\end{equation}
where the macroscopic degrees of freedom are combined into the vector $\bq$, $H_c(\bq,\bp)$ is a
cor\-res\-pon\-ding classical Hamiltonian, and where by derivative with respect to a vector we
understand a vector made from derivatives over the corresponding components.
As was discussed in section~5, the wave function
amplitude $A = e^R$ always satisfies a continuity equation, which in this case has the form
\begin{equation}\label{ce}
    \dovd{A^2}{t} + \sum_i \dovd{}{q_i} \left[ A^2 \left. \dovd{H_c(\bq,\, \bp)}{p_i} \,
    \right|_{\bp = \partial S/\partial \bq} \right] = 0 \,,   \dsplbl{ce}
\end{equation}
The standard quantum-mechanical derivation of equations~(\ref{hje}), (\ref{ce}) for the action
function and amplitude in the quasiclassical case may be found, for example, in \cite{dirac}.

We now want to consider a combined system, consisting of macroscopic objects interacting with
quantum particles. The same consideration applies to the interaction of macroscopic objects with
their own internal (like electrons' or phonons') microscopic degrees of freedom, which should be
described quantum-mechanically. In fact, it will be sufficient for our analysis to consider an
extremely simplified situation where a macroscopic object is represented by one particle with a
large (macroscopic) mass $M$ in the limit $M \rightarrow \infty$ interacting with a quantum
particle with a fixed (microscopic) mass $m$. Let the Hamiltonian of this system be
\begin{equation}\label{}
    H = \frac{p_x^2}{2M} + \frac{p_y^2}{2m} + U(x) + V(x,y) \, ,
\end{equation}
where $x$ and $y$ are the particle coordinates (their dimensionality will be irrelevant for us, so
we may consider them one-dimensional), $U(x)$ is the potential energy of the heavy particle that
scales proportionally to $M$ as $M \rightarrow \infty$, and $V(x,y)$ is the potential of the
particle interaction and of the light particle alone and is independent of $M$. The wave function
$\Psi(x,y,t)$ of the system satisfies the \Sch equation
\begin{equation}\label{se2p}
    \hi \dovd{\Psi}{t} \,-\, \frac{\hbar^2}{2M}\, \frac{\partial^2 \Psi}{\partial x^2}
    \,-\, \frac{\hbar^2}{2m}\, \frac{\partial^2 \Psi}{\partial y^2} \,+\, \big[U(x) + V(x,y)
    \big] \Psi \,=\, 0 \, .    \dsplbl{se2p}
\end{equation}
We take the point in $(x,y)$-configuration space where the system is at initial time $t=0$, as
a coordinate system's origin. Then the initial action function $p(x,y)$ is a power series in $x$
and $y$, and it may be presented as a sum $p(x,y) = p_M(x) + p_m(x,y)$, where $p_M$ collects all
the terms of the series with the powers of $x$ alone, and $p_m$ the remaining terms, which contain
nonzero powers of $y$. The wave function $\Psi(x,y,t)$ may always be represented as a product
$A(x,t)e^{(i/\hbar)S(x,t)}\phi(x,y,t)$ with real functions $A(x,t)$ and $S(x,t)$. We have then from
Eq.~(\ref{se2p})
\begin{equation}\label{se2f}
    \begin{array}{l}
      \dsp \phi(x,y,t) \left(\hi \, \dovd{}{t} \,-\, \frac{\hbar^2}{2M}\, \frac{\partial^2}
      {\partial x^2} \,+\, U\right) A e^{\ih S} \\[0.4cm]
      \dsp \quad + \, A e^{\ih S} \left[ \left(\hi \, \dovd{}{t} \,-\, \frac{\hbar^2}{2m}\,
      \frac{\partial^2}{\partial y^2} \,+\, V\right)\phi \,+\, \left(\hi\frac{1}{M} \,
      \dovd{S}{x} \,-\, \frac{\hbar^2}{M}\, \frac{1}{A}\, \dovd{A}{x} \right) \dovd{\phi}{x} \,-\,
      \frac{\hbar^2}{2M}\, \frac{\partial^2 \phi}{\partial x^2} \right] \,=\, 0\,.  \dsplbl{se2f}
    \end{array}
\end{equation}
Let now the function $A(x,t)e^{(i/\hbar)S(x,t)}$ cancel the first term in (\ref{se2f}), i.e., it
satisfies the equation
\begin{equation}\label{se3}
    \dsp \left(\hi \, \dovd{}{t} \,-\, \frac{\hbar^2}{2M}\, \frac{\partial^2}
      {\partial x^2} \,+\, U\right) A e^{\ih S} \,=\, 0 \, ,  \dsplbl{se3}
\end{equation}
with initial condition $A(x,0)e^{(i/\hbar)S(x,0)} = e^{(i/\hbar)p_M(x)}$. The function
$\phi(x,y,t)$ must then cancel the second term in (\ref{se2f}), i.e., satisfy an equation
\begin{equation}\label{se4}
    \dsp \left(\hi \, \dovd{}{t} \,-\, \frac{\hbar^2}{2m}\,
      \frac{\partial^2}{\partial y^2} \,+\, V\right)\phi \,+\, \left(\hi\frac{1}{M} \,
      \dovd{S}{x} \,-\, \frac{\hbar^2}{M}\, \frac{1}{A}\, \dovd{A}{x} \right) \dovd{\phi}{x} \,-\,
      \frac{\hbar^2}{2M}\, \frac{\partial^2 \phi}{\partial x^2} \,=\, 0\,.  \dsplbl{se4}
\end{equation}
To investigate the $M \rightarrow \infty$ limit, expand the functions $S$, $A$, and $\phi$ in
powers of $1/M$ as
\begin{equation}\label{}
    S = S^{(c)} + \sum_{k=0}^\infty \frac{S^{(k)}}{M^k}\,, \quad \quad A = \sum_{k=0}^\infty
    \frac{A^{(k)}}{M^k}\,, \quad\quad \phi = \sum_{k=0}^\infty \frac{\phi^{(k)}}{M^k}\,,
\end{equation}
where $S^{(c)}(x,t)$ is proportional to $M$ while coefficients $S^{(k)}(x,t)$, $A^{(k)}(x,t)$, and
$\phi^{(k)}(x,y,t)$ are $M$-independent, and neglect all contributions with positive powers of
$1/M$. For $S^{(c)}(x,t)$, we have then the classical Hamilton-Jacobi equation
\begin{equation}\label{clhje}
    \dovd{S^{(c)}}{t} + \frac{1}{2M} \left(\dovd{S^{(c)}}{x}\right)^2 + U = 0\,,  \dsplbl{clhje}
\end{equation}
and for $A^{(0)}(x,t)$, a continuity equation
\begin{equation}\label{}
    \dovd{{A^{(0)}}^2}{t}+\frac{1}{M}\,\dovd{}{x}\left({A^{(0)}}^2\,\dovd{S^{(c)}}{x}\right)=0\,.
\end{equation}
The solution of Eq.~(\ref{clhje}) is given by integrals of the Lagrangian function along
classical trajectories in the potential $U(x)$, and so $S^{(c)}(x,t)$ will scale proportionally to
$M$ as $M \rightarrow \infty$, as expected. The velocity of the heavy particle will converge for
$M\rightarrow\infty$ to an $M$-independent limit $v(x,t)=(1/M)\,\partial S^{(c)}(x,t)/\partial x$,
and since the action $S^{(c)}(x,t)$ satisfies the Hamilton-Jacobi equation, this particle will
exhibit a classical motion in the potential~$U(x)$. Let now $x(t)$ be the trajectory of the heavy
particle. Since it represents a macroscopic object, this trajectory is directly observable and, as
such, known. The behavior of the light particle is described by momentums $p_\sigma$ with
multi-indices $\sigma$ that include $y$ at least once. These momentums are derivatives of
$(\hbar/i) \ln \Psi(x,y,t)$ taken at $x=x(t)$. That means that the light particle is described by
the wave function $\phi(x,y,t)$ at a point $x(t)$, i.e., in the $M \rightarrow \infty$ limit, by
the function $\psi(y,t) = \phi^{(0)}\big(x(t),y,t\big)$. From (\ref{se4}), the function
$\phi^{(0)}$ satisfies the equation
\begin{equation}\label{se5}
    \dsp \left(\hi \, \dovd{}{t} \,-\, \frac{\hbar^2}{2m}\, \frac{\partial^2}{\partial y^2}
    \,+\, V\right)\phi^{(0)} \,+\, \hi\, v(x,t)\, \dovd{\phi^{(0)}}{x} \,=\, 0\, .   \dsplbl{se5}
\end{equation}
Combining the last term in (\ref{se5}) with the first one, and letting $W(y,t) =
V\big(x(t),y,t\big)$, we then obtain the equation for $\psi(y,t)$:
\begin{equation}\label{}
    \hi \, \dovd{\psi}{t} \,-\, \frac{\hbar^2}{2m}\, \frac{\partial^2 \psi}{\partial y^2}
    \,+\, W(y,t)\, \psi \,=\, 0 \, ,
\end{equation}
which is the \Sch equation for a light particle in the potential $W(y,t)$ created by a heavy
particle moving along the classical trajectory $x(t)$. Thus in \paqd, the experimentally observed
separation of reality into a macroscopic world that behaves classically and a microscopic one that
exhibits quantum behavior in a classical macroscopic environment is not postulated as
in standard quantum mechanics, but obtained as a direct consequence of its equations of motion.

\section{The theory of quantum measurements}
\setcounter{equation}{0}

Besides different equations of motion, the difference in the measurement procedure is pro\-bably
the most important difference between classical and quantum theory. For every physical quantity,
quantum mechanics specifies a corresponding linear hermitian operator~$O$.
In \paqd, $O\psi(\bx)/\psi(\bx)$ may be interpreted as a numerical
value, which this quantity has if a particle with wave function $\psi$ happens to be at a point
$\bx$. If $\psi$ is an eigenstate of~$O$, then this value is the same for all $\bx$ (i.e., for all
possible trajectories of the particle) and is a corresponding real eigenvalue of $O$. If, on the
other hand, $\psi$ is not an eigenstate, then this value will be different for different $\bx$, and
for a given point~$\bx$ will in general be an arbitrary complex number that would have been
the result of a measurement of $O$, if this measurement had its classical meaning.
In quantum theory, however, the situation is more complicated. Indeed, in contrast
to classical theory, which deals with macroscopic objects, quantum theory describes microscopic
ones, whose properties are usually not directly observable. In order to find the value of any
physical quantity that such objects possess, one has to produce the interaction of this quantity
with another one that {\em is\,} observable, and to infer the value of the quantity of interest
from the reaction of that observable quantity. The observable quantity is a characteristic of the
``apparatus", and may have a macroscopic character,
like the position of a pointer, or a microscopic
one, as in a Stern-Gerlach experiment, where the measured quantity is a particle's spin and the
observable (or rather, in this case, detectable) quantity is this particle's position, and the role
of an apparatus is played by the particle itself. Thus the measurement procedure in the quantum
domain is highly indirect, which causes its peculiar properties. To analyze them, we will apply
our theory to the combination of a particle and an apparatus. We will identify several different
kinds of quantum measurements, and consider them in turn.

\subsection{von Neumann's measurements with discrete spectrum}

The measurements of the first kind were originally investigated by von Neumann \cite{vNn}, and so
we will call them von Neumann's measurements. In this subsection we will consider the case where
the spectrum of a measured observable $O$ is discrete.
According to von Neumann, if the apparatus performs a
measurement of this observable, and the particle's state is its eigenstate~$\psi_i$ (which is
assumed to be normalized, $\int |\psi_i|^2 dx = 1$), corresponding to an eigenvalue~$O_i$ (so
that, in \paqd, the quantity $O$ has the value of $O_i$ for arbitrary position $\bx$ of the
particle) then the reading of the apparatus should have the corresponding $i$-th value, clearly
distinguishable from others. In more detail this means the following: Before the measurement, at
initial time $t=0$, the apparatus is
set into the state $\varphi(y)$, where $y$ is the apparatus coordinate, which is assumed to be
directly observable. We also assume that $\varphi(y)$ is centered at $y=0$ and has width
$\Delta y$. Since before the measurement a particle and an apparatus are independent, if a particle
is in a state $\psi_i$, then an initial wave function of the combined particle-apparatus system is
$\varphi(y)\psi_i(\bx)$. If $\Psi^{(i)}(y,\bx,t)$ is the result of an evolution of this state
during the measurement, then it is required that for $t$ larger than the duration of
measurement~$\Delta t$, $\Psi^{(i)}(y,\bx,t)$ should be centered around some $y_i(t)$ and have such
a width that the overlap of $\Psi^{(i)}(y,\bx,t)$ and $\Psi^{(j)}(y,\bx,t)$ in $y$-space
could be neglected for all
$j \neq i$ (in \paqd, $\Psi^{(i)}$ and $\Psi^{(j)}$ are analytic functions, and so they always
overlap, but we can require each of them to be negligibly small in the area where the other one is
centered). In other words, over the measurement time $\Delta t$ different packets $\Psi^{(i)}$
should diverge in $y$-space far enough to make their overlap negligible.
By observing the value of $y$ after the measurement, we can then infer the value of $O$
before it. In particular, if the measurement time $\Delta t$ is so short, and particle-apparatus
interaction Hamiltonian $H_{int}$ is so strong, that during the measurement all other terms in the
total Hamiltonian of the combined particle-apparatus system may be neglected compare to $H_{int}$,
and if $H_{int}$ is proportional to~$O$, then the wave function of the system
$\Psi^{(i)}(y,\bx,\Delta t)$ immediately after the measurement will have the form $\varphi^{(i)}
(y) \psi_i(\bx)$, i.e., the particle after the measurement will remain in an eigenstate $\psi_i$ of
$O$. But this is not necessary. Explicit models of such a measurement are considered in \cite{vNn}
and, in great detail, in \cite{bohmqt}.

If such an apparatus is built, then an interesting situation occurs when, before the measurement, a
particle in {\em not\,} in an eigenstate of $O$, i.e., if its wave function is $\psi(\bx) = \sum_i
c_i \psi_i(\bx)$ with more than one nonzero coefficient $c_i$. We assume that this wave
function is normalized, so that $\sum_i |c_i|^2 = 1$. By the linearity of \Schs equation, in
this case the initial wave function of the combined system $\varphi(y)\psi(\bx)$ evolves during the
measurement into $\sum_i c_i \Psi^{(i)}(y,\bx,t)$, in direct contradiction with experiment, from
which we know that in fact the combined system will end up in one of the states
$\Psi^{(i)}(y,\bx,t)$. To save the theory, von Neumann postulated, besides the unitary evolution
described by the \Sch equation, the second law of evolution, which acts only during the
measurements:
a random, unpredictable, and unanalyzable collapse of the linear combination $\sum_i c_i
\Psi^{(i)}(y,\bx,t)$ into one of $\Psi^{(i)}(y,\bx,t)$ with experimentally observed probability~
$|c_i|^2$. Nobody, however, was able to formulate convincingly when the unitary evolution should
be replaced by the collapse (or, in other words, what exactly allows us to qualify an experiment as
being a measurement). Similar issues arise in other orthodox approaches to the interpretation of
quantum theory. This is the essence of the quantum measurement problem, which found a
simple and natural resolution in the framework of DBBT \cite{dbb,dbb2}. We now reproduce Bohm's
solution of the problem using the language of \paqd.

In \paqd, the state of the combined system is characterized by its position in configuration space
and all its momentums, all of which evolve according to the corresponding equations of motion. As a
consequence of this evolution, the action function, which is just a corresponding Taylor series,
evolves according to the quantum Hamilton-Jacobi equation, while the wave function evolves
according to the \Sch equation as was described above. During this process, the combined evolution
of the system's position and wave function
is such that the system normally stays in the areas of configuration space where the wave function
is not small. Consequently, when the packets $\Psi^{(i)}(y,\bx,t)$ start to diverge, the apparatus
position $y(t)$ will end up in the area where one of them, say the $k$-th, is not small, i.e., near
$y_k(t)$. Now, the momentums are derivatives of the system's action function, i.e., $(\hbar/i) \ln
\sum_i c_i \Psi^{(i)}(y,\bx,t)$, at the point $\big(y(t),\bx(t)\big)$ (where $\bx(t)$ is the
particle's position) in configuration space, and the
further the packets move away from each other the closer are these derivatives to the ones of
$(\hbar/i) \ln \Psi^{(k)}(y,\bx,t)$. The measurement ends when the overlap of the packets becomes
negligible, and with it the difference between the exact momentums and the derivatives
of $(\hbar/i) \ln \Psi^{(k)}(y,\bx,t)$ becomes negligible also. Consequently, although the wave
function is still equal to $\sum_i c_i \Psi^{(i)}(y,\bx,t)$, the motion of the ``physical"
variables, i.e., the system's position in
configuration space and momentums, will be the same as if the wave function was equal to
$\Psi^{(k)}(y,\bx,t)$, in agreement with experiment. This explains the apparent wave function
collapse. The probability of observing the $k$-th result of the measurement is calculated according
to the general rules of section 6 as an integral from $\big|\sum_i c_i \Psi^{(i)}(y,\bx,t)\big|^2$
over the area where $\Psi^{(k)}(y,\bx,t)$
is not small, i.e., around $y_k(t)$ in $y$-space and all $\bx$-space,
and since all $\Psi^{(i)}(y,\bx,t)$ are normalized and don't overlap, this integral,
again in agreement with experiment, is equal to $|c_k|^2$.

The following features of von Neumann's measurement procedure deserve special mention:
\newline -- Although measurement statistics are determined by the wave function of the particle
alone, the result of every individual measurement (unless the particle was in the eigenstate of $O$
before it) is determined by the full \paqd\ states (i.e., positions and all momentums, or positions
and wave functions) of both particle and apparatus.
\newline -- If the paticle was not in the eigenstate of $O$, then the measurement's result $O_k$ is
completely unrelated to the value $O\psi(\bx)/\psi(\bx)$ (where $\bx$ is the particle's position)
of observable $O$ before the measurement. This and the previous note mean that unless the particle
was in a correspon\-ding eigenstate, $O$'s observed value is not really measured, but rather
created by the particle and apparatus jointly in the process of a measurement. What {\em is\,}
measured (by the corresponding relative frequencies of a series of measurements) is a set
of values of the squared amplitudes $|c_i|^2$.
\newline -- Unless the interaction Hamiltonian is proportional to $O$ and satisfies other
requirements
discussed above, after the measurement the particle doesn't have to be in a state with a
definite~$O$ value, let alone the state with $O$ equal to the measured eigenvalue $O_k$.
\newline -- To successfully perform a measurement, the apparatus doesn't have to be macroscopic.
The only necessary condition is that the packets $\Psi^{(i)}(y,\bx,t)$ with different $i$ do not
overlap after some time $\Delta t$ (the duration of the measurement). In a Stern-Gerlach
experiment, where the apparatus is the particle itself, the measurement ends and the wave function
collapses not when the particle is detected after passing the magnet and we learn the spin
measurement's result, but earlier, when the wave packets corresponding to the different spins cease
to overlap. See, however, the next note.
\newline -- Although we are discussing the wave function collapse,
the ``empty" packets $\Psi^{(i)}(y,\bx,t)$ with $i\neq k$ do not disappear, but just move away from
the ``active" packet $\Psi^{(k)}(y,\bx,t)$, so that their contribution to momentums and, therefore,
their influence on the dynamics of the system vanishes. If, in their future evolution, all or some
of the packets $\Psi^{(i)}$ have again overlapped with $\Psi^{(k)}$, then the measurement would be
``undone", the wave function would ``uncollapse", and the value of $O$ would again
become undetermined \big(in a sense that instead of being equal to $\Psi^{(k)}$, the wave function
would become equal to the linear combination of $\Psi^{(k)}$ and overlapping packets
$\Psi^{(i)}$\big). This overlap, however, should happen in an $\bx$-$y$ space of dimensionality
$\mathrm{dim}\,\bx + \mathrm{dim}\,y$. For the purpose of this argument, $y$ should include all
coordinates of the apparatus and its environment that are connected by a chain of nonnegligible
interactions. Consequently, while $\mathrm{dim}\,y$ is small (as before the particle is detected in
a Stern-Gerlach experiment) such reversion of the measurement can, in principle, be accomplished.
However, as soon as $\mathrm{dim}\,y$ becomes macroscopically large (as when the particle is
detected or observed by any macroscopic, conscious or not, observer) the reversion becomes
practically impossible, and its possibility may be neglected.
\newline -- The set of possible final states $\big\{\Psi^{(i)}(y,\bx,t)\big\}$ of the system is
predetermined by the measurement apparatus and does not depend on the initial wave function of the
particle. Consequently, after the wave function collapses into one of these states, all information
about the particle's initial state, and all influence of this state on the future history of the
system is lost. On a positive side, that means that von Neumann's measurements are convenient for
experiment preparation. Indeed, after the
obser\-vable $O$ is measured and found equal to $O_k$, say, we know that the system is prepared in
the state~$\Psi^{(k)}(y,\bx,t)$, regardless of the initial wave function of a particle.

\subsection{von Neumann's measurements with continuous spectrum}

An analysis, similar to that just presented, is also possible when the spectrum of a measured
observable $O$ is continuous. First consider the case when $O$ is not a particle's position.
Here it will be easier to use an explicit consideration, based on a particle-apparatus interaction
Hamiltonian $H_{int}$ proportional to $O$. Following \cite{dbb,dbb2,vNn,bohmqt}, choose it
in the form $H_{int} = g(t) O p_y$, where $p_y = -i\hbar\partial/\partial y$ is the momentum
conjugate to the apparatus position~$y$, and the factor $g(t)$ represents the switching of the
interaction on and off. Assume it has an impulsive character, so that $g(t) = g_0$ for $0<t<\tau$
and $g(t) = 0$ for $t<0$ and $t>\tau$, where~$\tau$ is the duration of the measurement. Consider
the limit of very small $\tau$ and large~$g_0$. The influence of the particle's and apparatus'
own Hamiltonians on the evolution of the wave function during the measurement may then be neglected
compared to the influence of $H_{int}$, so that between $t=0$ and $t=\tau$ the \Sch equation may be
approximated by
\begin{equation}\label{}
    i \hbar \,\dovd{\Psi}{t}\, = \, H_{int} \Psi \,=\, - i \hbar g_0 O \,\dovd{\Psi}{y} \, .
\end{equation}
Let $\psi_a(x)$ be eigenfunctions of $O$, $\,O\psi_a(x) = a\psi_a(x)$, normalized so that
\begin{equation}\label{}
    \int \psi^*_{a'}(x) \psi_{a''}(x) dx = \delta(a' - a'') ,
\end{equation}
and $c(a)$ be the coefficients of an expansion of the initial particle's wave function
$\psi(x,t=0)$ in an integral over them:
\begin{equation}\label{}
    \psi(x,0) = \int c(a) \psi_a(x)\, da \, .
\end{equation}
As before, assume the initial wave function $\varphi(y)$ of the apparatus to be centered at
$y=0$, have width $\Delta y$, and be normalized, $\int|\varphi(y)|^2 dy = 1$. The total wave
function of a system $\Psi(x,y,t)$ may be expanded as an integral over $\psi_a(x)$ as
\begin{equation}\label{}
    \Psi(x,y,t) = \int c_a(y,t) \psi_a(x) \, da \, .
\end{equation}
It is easy to see that the expansion coefficients $c_a(y,t)$ must satisfy the equation
\begin{equation}\label{}
    \dovd{c_a(y,t)}{t} = - g_0 a \, \dovd{c_a(y,t)}{y}
\end{equation}
with the initial condition $c_a(y,0) = c(a) \varphi(y)$. Then the solution for $c_a$ is
\begin{equation}\label{}
    c_a(y,t) = c_a(y - g_0 a t, 0) = c(a) \varphi(y - g_0 a t) ,
\end{equation}
so that at the moment $t=\tau$ at the end of the measurement, the system's wave function will be
\begin{equation}\label{}
    \Psi(x,y,\tau) \, = \, \int c(a) \psi_a(x) \varphi(y - g_0 \tau a) \, da \, .
\end{equation}

It is convenient to introduce a new apparatus coordinate $\wty = y / (g_0 \tau)$ and new function
$\wtp(z) = \varphi(g_0 \tau z)$, which becomes negligible when $|z| > \sigma$, where the half-width
$\sigma = \Delta y / (2 g_0 \tau)$. The system's final wave function can then be written as
\begin{equation}\label{psif}
    \Psi(x,y,\tau) = \int c(a) \psi_a(x) \wtp(\wty - a) \, da \, .  \dsplbl{psif}
\end{equation}
The directly observable coordinate $\wty$ plays now the role of a pointer for the measurement
of~$O$. Indeed, if the initial wave function of a particle $\psi(x,0)$ is an eigenstate of $O$, say
$\psi_{a_0}(x)$, then $c(a) = \delta(a - a_0)$ and $\Psi(x,y,\tau) = \psi_{a_0}(x) \wtp(\wty-a_0)$.
Since $\wtp(\wty - a_0)$ vanishes for $|\wty - a_0| > \sigma$, the value of $\wty$ at $t = \tau$
will be between $a_0 - \sigma$ and $a_0 + \sigma$, so that $\wty$ points to the correct value of
$O$ with precision $\sigma$. It is assumed, that parameters $\Delta y$, $g_0$, and $\tau$ may be
chosen at will, and so $\sigma$ can be made arbitrary small. Consequently, although never exact,
the measurement of $O$ can be made arbitrarily precise.

In a general situation, when $\psi(x,0)$ is not an eigenstate of $O$, the scaled position $\wty$
evolves during the measurement according to the equations of motion, and ends at $t = \tau$ at
some~$\wty_0$. Similar to the discrete spectrum case, because of
the properties of the function $\wtp$, the evolution of the system's coordinates and momentums
will then be the same as if the wave function at $t = \tau$ instead of being $\Psi(x,y,\tau)$,
Eq.~(\ref{psif}), was equal to
\begin{equation}\label{cwf}
    \Psi^{(\wty_0)}(x,y,\tau) = \int_{\wty_0 - \sigma}^{\wty_0 + \sigma} c(a) \psi_a(x)
    \wtp(\wty - a)\,da\, .      \dsplbl{cwf}
\end{equation}
Since $c(a)$ and $\psi_a(x)$ are smooth (analytic) functions of $a$, for sufficiently small
$\sigma$ the function $\Psi^{(\wty_0)}(x,y,\tau)$ may be approximated with arbitrary precision as
\begin{equation}\label{}
    \Psi^{(\wty_0)}(x,y,\tau) = c(\wty_0) \psi_{\wty_0}(x) f(\wty) \, , \quad\; f(\wty) =
                               \int_{\wty_0 - \sigma}^{\wty_0 + \sigma} \wtp(\wty - a)\,da\, ,
\end{equation}
so that the function $\Psi^{(\wty_0)}(x,y,\tau)$, to which the system appears to collapse, is an
eigenstate of $O$ with the eigenvalue $\wty_0$. The probability $p(a_0, da_0)$ to find the value of
$O$ (i.e., the value of~$\wty_0$) between $a_0$ and $a_0 + da_0$ is equal to $\rho(a_0) da_0$,
where $\rho(a_0)$ is the corresponding probability density. By the general rules of section~6 we
have for it
\begin{equation}\label{}
    \rho(a_0) = g_0 \tau \int \big|\Psi(x, g_0 \tau a_0, \tau)\big|^2 dx = g_0 \tau \int
                |c(a)|^2 |\wtp(a_0 - a)|^2 da \, ,
\end{equation}
where we used the normalization condition for the functions $\psi_a(x)$ and the factor $g_0 \tau$
appears because $dy = g_0 \tau \,d\wty$. For sufficiently small
$\sigma$, $|c(a)|^2$ in the integrand may be again approximated by $|c(a_0)|^2$ with negligible
error, and using normalization of the function $\wtp$, we obtain, in agreement with von Neumann's
postulate and experiment, the standard result $\rho(a_0) = |c(a_0)|^2$.

To summarize, von Neumann's measurement procedure of an observable with continuous spectrum that
is not a particle's position is similar to the one
with discrete spectrum, and has the same, listed above, properties. In particular, unless the
particle was initially in the eigenstate of $O$, the measurement's result is unrelated to the value
$O\psi(\bx)/\psi(\bx)$ of $O$ before the measurement, and with arbitrary precision the set of
possible final states of the particle-apparatus system is predetermined and does not depend on the
initial state of the particle. For the measurement procedure, considered above, the corresponding
set of possible particle final states is just the set $\{\psi_a(x)\}$ of eigenstates of~$O$.

We now consider von Neumann's measurement of a particle's position, and show that, in contrast to
other physical quantities, this measurement results in the true \paqd\ particle position. Indeed,
we have, obviously, for the eigenfunctions $\psi_a(x)$ of an operator $O = x$ and coefficients
$c(a)$ of expansion of the particle's wave function $\psi(x,0)$ in this case, $\psi_a(x) =
\delta(x-a)$ and $c(a) = \psi(a,0)$. Although $\delta(x-a)$ is not an analytic function of $x$,
and so cannot be considered a legitimate \paqd\ wave function, we still can use it in intermediate
mathematical transformations. The integral over $a$ in Eq.~(\ref{psif}) can then be immediately
calculated to give $\Psi(x,y,\tau) = \psi(x,0) \wtp(\wty - x)$. Let now the particle's position
before the measurement be $x_0$. Assume that the function $\varphi(y)$ is real. Then the motion of
the particle during the measurement may be neglected. The final value $\wty_0$ of $\wty$ will
now be such that $\Psi(x_0,g_0\tau\wty_0,\tau)$ does not vanish. Due to the properties of the
function $\wtp$, that means that $\wty_0$ should be in a $\sigma$-vicinity of~$x_0$, $|\wty_0 -
x_0| \leq \sigma$, as was asserted. The collapsed wave function $\Psi^{(\wty_0)}(x,y,\tau)$,
Eq.~(\ref{cwf}), is easily calculated to be equal to $\psi(x,0) \wtp(\wty - x)$ for $|x - \wty_0|
\leq \sigma$ and equal to zero for $|x - \wty_0| > \sigma$. It is, therefore, contained in a
$\sigma$-vicinity of $\wty_0$, and for $\sigma \rightarrow 0$, as in other cases of von Neumann's
measurements, loses the memory of the particle's initial state.

\subsection{Position measurements of the second kind and the double-slit experiment}

In practice, von Neumann's measurement is never used for a particle's position. The real position
measurement is carried out by such devices as a bubble chamber or photographic plate. As we will
see, wave function collapse and some other important features of this measurement are significantly
different from those of von Neumann's, which justifies calling it the measurement of the
second kind.

The measurement of a particle's position by a photographic plate or in a bubble chamber may be
described as follows. The physical state is filled with microscopic detectors (molecules of
photo-emulsion for photographic plate, or of overheated liquid for bubble chamber), which change
their state (chemical changes in emulsion, ionization in a liquid) if the measured particle passes
in close vicinity to them. Due to the
special physics of detectors, this changed microstate evolves then in such a way as to produce
directly observable macroscopic changes (dark spot on developed plate, bubble in a chamber). The
detectors with such changed state mark the position of a particle.

In our analysis of this procedure, we
again use the fact that the wave function in \paqd\ has its own dynamics, the same as in standard
quantum mechanics, and so its evolution may be analyzed without reference to a particular
particle's trajectory that is responsible for this wave function's existence, and which may be
included in the analysis later. Then the following crude model may be suggested for the description
of a position measurement. Consider first just one detector, fixed at a point with position $\tx$.
Let the detector itself be characterized by the parameter $y$, which in the initial state is close
to zero, so that the initial wave function of the detector
is, for example, $\exp(-y^2/4)$ (we will not worry about wave function normalization here).
Let the particle's wave function be $\psi(x)$, so that an initial wave function of the
particle-detector system is $\psi(x)\exp(-y^2/4)$.
Let the physics of the detector and its interaction with the
particle be such that within a short measurement time $\Delta t$, $y$ moves from the vicinity of
zero to the vicinity of some $Y\gg 1$, so that the detector's wave function becomes, for example,
$\exp\big[-(y-Y)^2/4\big]$, if during this time the particle's distance from the detector was less
than some characteristic distance $\sigma$. The evolution of a particle's wave function during the
measurement time due to its own Hamiltonian, i.e.,
without interaction with the detector, will be of no importance for us.
We can, therefore, consider the measurement to be instantaneous, i.e., $\Delta t$ to be so small
that the change of the particle's wave function during the measurement due to its own dynamics is
negligible. The wave function of the particle-detector system immediately after the measurement can
then be written in a general form as
\begin{equation}\label{}
    \widetilde{\Psi} = \psi(x) \left\{a(x - \tx,y) \exp\left[-\frac{1}{4}\,(y-Y)^2\right]
                   + b(x - \tx,y) \exp\left(-\frac{1}{4}\, y^2\right) \right\} \, .
\end{equation}
The functions $a$ and $b$ reflect the physics of the particle-detector interaction.
All we know about
them is that $a(x-\tx,y)$ vanishes and $b(x-\tx,y)$ converges to $1$ when the distance from $x$ to
$\tx$ becomes larger than $\sigma$, and that $b(x-\tx,y)$ vanishes when this distance is smaller
than~$\sigma$. The dependence on $y$ is included in $a$ and $b$ for generality, and is supposed to
leave the general character of $y$-dependence of the corresponding terms intact, i.e., the
probability density is concentrated near $y=Y$ in the first term, and near $y=0$ in the second. Now
if there are many detectors like that, then the initial wave function will be $\psi(x) \prod_i
\exp(-y_i^2/4)$, and the wave function after the measurement will be
\begin{equation}\label{amwf1}
    \Psi = \psi(x) \prod_i \left\{a(x - x_i,y_i) \exp\left[-\frac{1}{4}\,(y_i-Y)^2\right]
          + b(x - x_i,y_i) \exp\left(-\frac{1}{4}\, y_i^2\right) \right\} \, ,   \dsplbl{amwf1}
\end{equation}
where $y_i$ is the $y$-coordinate of the $i$-th detector, and $x_i$ is its position.
To avoid unnecessary
complications, we will make the simplifying assumption that $\sigma$-vicinities of different
detectors do not overlap and, at the same time, do not leave any places in the $x$-space uncovered.
The after-measurement wave function may then be rewritten as
\begin{equation}\label{amwf}
    \Psi = \psi(x) \sum_j a(x - x_j, y_j) \exp\left[-\frac{1}{4}\,(y_j-Y)^2\right] \prod_{i \neq j}
           \exp\left(-\frac{1}{4}\, y_i^2\right).   \dsplbl{amwf}
\end{equation}
Let now the particle's position $x_p$ at this moment happen to be in a $\sigma$-vicinity of the
$k$-th detector, $x_p\approx x_k$. Then obviously in the sum over $j$ in (\ref{amwf}), all terms
except the $k$-th, will give a negligible contribution to momentums (i.e., derivatives of
$(\hbar/i)\ln \Psi$ at that $x_p$ and proper $y_i$'s) and so the future motion of the particle and
detectors will proceed as if the wave function was equal to this $k$-th term, i.e., underwent a
collapse
\begin{equation}\label{fwf}
    \Psi \longrightarrow \psi(x)\, a(x - x_k, y_k) \exp\left[-\frac{1}{4}\,(y_k-Y)^2\right]
                         \prod_{i \neq k} \exp\left(-\frac{1}{4}\, y_i^2\right).   \dsplbl{fwf}
\end{equation}
In this state, the $y$-coordinate of the $k$-th detector will then be found near $Y$, and all
others will be near zero, and the particle's probability density,
although far from being a delta-function
centered at~$x_p$, will be concentrated in $x_p$'s and the $k$-th detector's $\sigma$-vicinity,
where $\sigma$ may be considered as a measurement precision. With this precision, therefore, a
position measurement of the second kind, like its von Neumann's counterpart, measures the true
\paqd\ position of a particle. We note, however, that although for both kinds of measurements the
final selection of a member of a linear superposition, to which the wave function would collapse,
is made by some variable which may have one, and only one value, for von Neumann's measurement this
variable is the apparatus position $y$, while for a position measurement of the second kind, it is
the measured particle's position $x_p$.
Also, for a position measurement of the second kind, the precision
$\sigma$ is fixed by the physics of detectors and so cannot be made arbitrary small. Consequently,
while for an arbitrarily precise von Neumann measurement, the final wave function becomes equal to
one possible function from the predetermined set of them, the final wave function after a position
measurement of the second kind
does depend on the initial wave function of the particle. Indeed, its $x$-dependence is essentially
given by the product $\psi(x)\, a(x - x_k, Y)$, i.e., is equal to the initial function $\psi(x)$
modulated by a factor $a(x - x_k, Y)$. If this factor is smooth enough, and if the characteristic
wavelengths in $\psi(x)$ are much smaller than $\sigma$, then the packet $\psi(x)\, a(x - x_k, Y)$
will keep propagating without spreading much along a trajectory that is close to the one the
particle would have by itself, i.e., if its position was not measured. In a bubble chamber, this
packet will then trigger other detectors, thus producing a track which approximates the particle's
unperturbed trajectory.

The same consideration may be also applied to the double-slit experiment discussed by
Feynman~\cite{fnmn}. In this case, variable $x$ in Eqs.~(\ref{amwf1})-(\ref{fwf}) will denote the
coordinate on the screen, and $x_i$, $i=1,2$ --- the position of the $i$-th slit. Without
detectors, the wave function of the particle immediately behind the screen would be equal to
$\psi(x) \sum_{j=1,2}a(x-x_j)$, where $\psi(x)$ is the wave function in front of the screen, and
the ``shadow function" $a(x-x_j)$ is nonzero only for $x$ inside the $j$-th slit.
Propagating away from the screen, the waves from the two slits would overlap
and create an interference pattern. On the other hand, in the presence of detectors the wave
function will be given by Eq.~(\ref{amwf}) with indices $i,j$ there taking the values of 1 and~2.
The condition that detectors work well and allow to determine
through which slit the particle have passed means then exactly that the packets from the two slits
remain well separated with respect to coordinates $y_1$ and~$y_2$, and so the interference between
them is impossible. As was explained above, if the particle have passed the slit $k$ and was
detected there, then its future motion and the
motion of detectors will be the same as in the state with the wave function~(\ref{fwf}), i.e., as
if the other slit was closed. In agreement with \cite{fnmn}, the observation of an interference is,
therefore, incompatible with the detection of the path chosen by the particle. These two operations
are just mutually exclusive: the interference happens when the packets overlap, while the detection
of the path requires them to be well separated. Note that this conclusion remains perfectly valid
even when the detectors are microscopic, like the one-bit detectors discussed in~\cite{sur}.

\section{Nonlocality, analyticity, and covariance}
\setcounter{equation}{0}

Although in \paqd, as in classical mechanics, particles move along well defined trajectories, the
equations of motion in these theories are fundamentally different. The only momentums that
contribute to the equations of classical mechanics are the first order momentums $p_{j_k}$, where
$k$ runs from 1 to the number of particles $n_p$, and for every $k$, $j_k$ runs from 1 to the
dimension of physical space~$n$. Every momentum $p_{j_k}$, therefore, is ``bound" to a
corresponding particle~$k$, and changes only due to the presence of forces, described by a
potential function~$U$. In nonrelativistic mechanics, the forces corresponding to this potential
normally vanish with distance, while in relativistic cases the potential propagates with finite
speed, which is not larger
than the speed of light $c$. Classical mechanics is, therefore, local: to predict the behavior of a
particle separated by a large distance from others during some time $\Delta t$, one doesn't have
to know what happens further than the distance of $c\Delta t$ from it. This locality, we see,
is a consequence of the fact that in classical mechanics particles influence each other only
through the action of the potential, which has the described properties.

In \paqd\ the situation is different. To avoid tedious manipulations with a multiparticle
Hamiltonian~(\ref{hmulti}), we may simply make all masses $m_k$ equal to each other and
denote them as~$m$. The Hamiltonian~(\ref{hmulti}) will then look exactly like the one-particle
Hamiltonian~(\ref{hp}), but in an $n_p n$, rather than in an $n$-dimensional space.
Correspondingly,
Eq.~(\ref{psigdot}) for the evolution of momentums will hold, with summation over repeating
indices $j$ there running from 1 to~$n_p n$. Now if particles are entangled, i.e., the system's
action function is not equal to the sum of separate particles' actions (or system's wave
function to the product of separate particles' wave functions) then there exist nonzero
momentums $p_\mu$ with ``mixed" multi-indices $\mu$, which include indices from different
particles. Eq.~(\ref{psigdot}) will then interconnect the time evolutions of all possible
momentums $p_\mu$, and with them of particles' velocities.
Since the momentum-dependent part of~(\ref{psigdot}) does not depend on the particle positions, and
all momentums are taken at the same time, they obtain the status of global variables: each momentum
affects the time evolution of all others at the same moment of time, independently of the particle
positions and the distances between them. Thus in this new (i.e., nonclassical, ``nonpotential")
way, the particles in \paqd\ influence each other on the whole hypersurface $t=\mbox{const}$
instantaneously, and over arbitrary distance. Clearly, the reason for this nonlocality is that an
analytic function is a fundamentally nonlocal object --- the set of its derivatives in any point of
space determines its behavior arbitrarily far from this point.

The nonlocal kind of behavior described above, is, according to Bell's theorem,
necessary for any theory that dynamically derives experimentally observed nonlocal correlations
between entangled particles, rather than just predicts them, as does standard quantum
me\-cha\-nics \cite{bell,norsen}. \paqd\ is built as an ``ODE side" of quantum mechanics,
which always agrees, of course, with its ``PDE side", i.e., the \Sch equation and the conventional
theory based on it. As such, \paqd\ {\em must} be nonlocal: if it
were local, so would the standard quantum mechanics. Note also, that \paqd\ does not conflict
with our intuition: indeed, our intuition is classical, but the classical limit of \paqd\ is just
the usual, completely local classical mechanics! \paqd\ thus has the desired feature of being a
fundamentally nonlocal theory with a local classical limit.

The described nonlocal behavior was first discovered in the framework of DBBT and discussed
extensively there \cite{dbb2}. It was soon realized that DBBT's nonlocality is in perfect accord
with the requirements of Bell's theorem and is, in this respect, welcomed \cite{bell}. There
remained, however, a difficult question about the theory's relativistic invariance. The influence,
propagating with infinite speed, seems to be in an obvious conflict with the requirements of
special relativity theory. This concern is addressed in \cite{dbb2}, where it is proved
that such influence
cannot be used for transmission of superluminal signals. Still, there is the other concern:
propagation of influence with infinite speed requires a selection of preferred reference frame, in
which this propagation happens along surfaces $t = \mbox{const}$, in contradiction with the spirit
of the theory of relativity, which demands that physical laws must be the same in every inertial
frame of reference. This is generally considered to be a serious problem for DBBT \cite{MYRVOLD}.
We will now show, that, thanks to the additional requirement of analyticity,
\paqd\ may be formulated in an arbitrary analytic foliation of space-time,
and will have the same form in each of
them. Our consideration will be nonrelativistic. It will be argued at the end of this section,
however, that its relativistic version, although still nonlocal, will be not only Lorentz
invariant, but can be also made generally covariant.

Indeed, consider an arbitrary analytic foliation of space-time, generated by a single-valued
analytic function~$f(\bx,\tau)$, i.e., a partition of space-time into 3-dimensional hypersurfaces
(leaves of foliation)
\begin{equation}\label{tftau}
    t = f(\bx,\tau) \, ,    \dsplbl{tftau}
\end{equation}
where $\tau$ parameterizes hypersurfaces (we can, for example, conveniently require $\tau =
f(0,\tau)$) monotonically, so that $\partial f(\bx,\tau)/\partial \tau > 0$ for all $\bx$ and
$\tau$, and such that the whole space-time is covered (so that every point $(\bx_0,t_0)$ belongs to
some hypersurface, i.e., $t_0 = f(\bx_0,\tau_0)$ with some $\tau_0$). In relativistic theory we
require the surfaces $\tau = \mbox{const}$ to be space-like. We will call this foliation
$f$-foliation. The standard partition of space-time into surfaces $t = \mbox{const}$ (``standard
foliation") corresponds to a function $f(\bx,\tau) = \tau$. We now want to introduce wave
functions, defined on surfaces $\tau = \mbox{const}$, rather than $t = \mbox{const}$. For a
one-particle case, the wave function $\psi(\bx,t)$ was introduced as a solution of a \Sch equation,
analytic with respect to $\bx$ for every $t$. It is then also analytic with respect to both $\bx$
and $t$, and so the function $\psi^{(f)}(\bx,\tau) = \psi\big(\bx,f(\bx,\tau)\big)$ is analytic
with respect to $\bx$ and $\tau$. To define a wave function on surfaces $\tau=\mbox{const}$ in a
multi-particle case, we will borrow from relativistic theory the multi-time formalism \cite{dfpw},
where each particle has its own individual time, and the multi-time wave function of $n_p$
particles $\psi(\bx_1,t_1,\ldots,\bx_{n_p},t_{n_p})$ depends on positions and times of all of them.
Detailed analysis of the physical meaning of this wave function and of corresponding analytical
quantum dynamics will be a subject of relativistic consideration. In nonrelativistic theory, where
the interaction between particles is mediated by an instantaneous potential function,
the multi-time formalism can be defined only for particles that do not interact with each other
(but can interact with an external potential). It will be sufficient for our purpose, however, to
consider such noninteracting particles, because here we are only interested in nonlocal
correlations, caused by entanglement, and not in correlations due to an interparticle interaction.
The multi-time wave function then satisfies the system of equations
\begin{equation}\label{mtse}
    \left(i\hbar\dovd{}{t_k}\, -\, \hat{H}_k\right)\psi(\bx_1,t_1,\ldots,\bx_{n_p},t_{n_p}) = 0 \,
    , \,\,\,\, k = 1,\ldots,n_p \, ,     \dsplbl{mtse}
\end{equation}
where
\begin{equation}\label{}
    \hat{H}_k = -\frac{\hbar^2}{2m_k} \, \Delta_k + U_k(\bx_k,t_k)
\end{equation}
is the Hamilton operator of the $k$-th particle in the external potential $U_k(\bx,t)$, $\Delta_k$
being a Laplace operator, acting on the coordinates of the $k$-th particle $\bx_k$. The wave
function $\psi(\bx_1,t_1,
\ldots,\bx_{n_p},t_{n_p})$ may be obtained by path integration over all the paths such that for
every $k = 1,\ldots,n_p$, the paths for the $k$-th particle terminate in a point $(\bx_k,t_k)$. The
action functions $S(\bx_1,t_1,\ldots,\bx_{n_p},t_{n_p})$ and $R(\bx_1,t_1,\ldots,\bx_{n_p},
t_{n_p})$, defined from $\psi(\bx_1,t_1,\ldots,\bx_{n_p},t_{n_p})$ as in Eq.~(\ref{psimp}), satisfy
a system of quantum Hamilton-Jacobi equations
\begin{equation}\label{mpqhje}
        \dovd{S}{t_k} + H^S_k = 0\, , \quad \quad \dovd{R}{t_k} + H^R_k = 0 \, ,\quad \,\,
        k = 1,\ldots,n_p \, ,     \dsplbl{mpqhje}
\end{equation}
similar to Eq.~(\ref{szetadot}), with $H^S_k$ and $H^R_k$ given by Eqs.~(\ref{hs}) and (\ref{hz}),
where the index $j$ in those equations means the derivative with respect to the $j$-th coordinate
of the $k$-th particle, and the potential $U$ is understood as~$U_k(\bx_k,t_k)$.
As will become clear soon, it is appropriate to postulate, in a straightforward generalization of
Eq.~(\ref{vlct}) and corresponding one-time theory, that if in a state described by a wave
function $\psi(\bx_1,t_1,\ldots,\bx_{n_p},t_{n_p})$, the particles have space-time positions
$(\bx_1,t_1),\ldots,(\bx_{n_p},t_{n_p})$, then their velocities are given by
\begin{equation}\label{mpvlct}
    \bv_k = \dovd{H^S_k}{S_{\bx_k}} = \dovd{H^R_k}{R_{\bx_k}} = \frac{1}{m_k} \,
    S_{\bx_k}(\bx_1,t_1,\ldots,\bx_{n_p},t_{n_p})\, ,  \dsplbl{mpvlct}
\end{equation}
where $S_{\bx_k}(\bx_1,t_1,\ldots,\bx_{n_p},t_{n_p}) = \partial S(\bx_1,t_1,\ldots,\bx_{n_p},
t_{n_p})/\partial \bx_k$, and similarly for $R_{\bx_k}$.

With so-defined multi-time wave function, the wave function on a hypersurface $\tau=\mbox{const}$
of any $f$-foliation is obtained by placing each particle on this hypersurface,
\begin{equation}\label{}
    \psi^{(f)}(\bx_1,\ldots,\bx_{n_p},\tau) = \psi\big(\bx_1,f(\bx_1,\tau),\ldots,\bx_{n_p},
    f(\bx_{n_p},\tau)\big) \, ,
\end{equation}
and it is an analytic function of all $\bx_k$ and $\tau$.
The whole theory developed above for a standard foliation may then be reproduced for an arbitrary
analytic $f$-foliation. The $\tau$-evolution of the wave function $\psi^{(f)}(\bx_1, \ldots,
\bx_{n_p},\tau)$ is governed by the equation
\begin{equation}\label{}
    i\hbar\dovd{\psi^{(f)}}{\tau} = \hat{H}^{(f)} \psi^{(f)} .
\end{equation}
The transformation of space coordinates does not affect our nonrelativistic analysis, and so we
will use the same coordinates in all foliations. Then
\begin{equation}\label{}
    \hat{H}^{(f)} = \sum_{k=1}^{n_p} \hat{H}^{(f)}_k \, , \quad\; \hat{H}^{(f)}_k = f_\tau(\bx_k,
    \tau) \hat{H}_k \, ,
\end{equation}
where $f_\tau$ denotes the derivative of $f$ with respect to its second argument, i.e.,
$f_\tau(\bx_k,\tau) = \partial f(\bx_k,\tau)/\partial\tau$. In what follows, we also need the
derivative $f_{\bx}$ of $f$ with respect to its first argument, $f_{\bx}(\bx_k,
\tau) = \partial f(\bx_k,\tau)/\partial\bx_k$. The action functions $p^{(f)}(\bx_1,\ldots,
\bx_{n_p},\tau)$, $S^{(f)}(\bx_1,\ldots,\bx_{n_p},\tau)$, and $R^{(f)}(\bx_1,\ldots,\bx_{n_p},
\tau)$ on $f$-foliation are again obtained from $\psi^{(f)}$ as in Eq.~(\ref{psimp}), and then the
quantum Hamilton-Jacobi equation, the momentums $p^{(f)}_\sigma$, $S^{(f)}_\sigma$, and
$R^{(f)}_\sigma$ and the equations of motion for them are introduced in the same way as for a
standard foliation. In particular, the quantum Hamilton-Jacobi equation for $S^{(f)}$ is
\begin{equation}\label{}
    \dovd{S^{(f)}}{\tau} + H^{(f)^S} = 0 \, ,
\end{equation}
where
\begin{equation}\label{hfs}
    H^{(f)^S} = \sum_{k=1}^{n_p} H^{(f)^S}_k \, , \quad\; H^{(f)^S}_k = f_\tau(\bx_k,
    \tau) H^S_k \, ,     \dsplbl{hfs}
\end{equation}
and particle velocities in the $f$-foliation, i.e., with respect to a new ``time" $\tau$, are given
by the usual relation $\bv^{(f)}_k = \partial H^{(f)^S} / \partial S^{(f)}_{\bx_k}$, where
$S^{(f)}_{\bx_k} = \partial S^{(f)}/\partial \bx_k$, or, using (\ref{hfs}),
\begin{equation}\label{vfk}
    \bv^{(f)}_k = f_\tau(\bx_k,\tau)\, \dovd{H^S_k}{S^{(f)}_{\bx_k}} \, .   \dsplbl{vfk}
\end{equation}
Note, that the derivative over $\bx_k$ in $S^{(f)}_{\bx_k}$ is taken along the leaf of the
$f$-foliation, i.e., for $\tau = \mbox{const}$, contrary to the derivative in $S_{\bx_k}$, which is
taken for $t = \mathrm{const}$.

We can now show that velocities $\bv^{(f)}_k$, Eq.~(\ref{vfk}), and $\bv_k$, Eq.~(\ref{mpvlct}),
correspond to the same motion of the $k$-th particle. We note first that if this particle moves
from point $\bx_k$ on leaf~$\tau$ to point $\bx_k + d\bx_k$ on leaf $\tau+d\tau$, so that its
$\tau$-velocity is $\bv^{(f)}_k = d\bx_k/d\tau$, then by Eq.~(\ref{tftau}) we have for a
corresponding time interval
\begin{equation}\label{}
    \begin{array}{ccl}
      dt_k & = & f_\tau(\bx_k,\tau)\, d\tau + f_{\bx}(\bx_k,\tau)\, d\bx_k \\[0.2cm]
      & = & \big( f_\tau(\bx_k,\tau) + f_{\bx}(\bx_k,\tau)\, \bv^{(f)}_k \big)\, d\tau \, .
    \end{array}
\end{equation}
Consequently, $t$-velocity $\bv_k = d\bx_k/dt_k$ should be equal to
$\bv^{(f)}_k/\big(f_\tau + f_{\bx}\, \bv^{(f)}_k\big)$, or
\begin{equation}\label{vfkvk}
    \bv^{(f)}_k = \bv_k\, \big(f_\tau + f_{\bx}\, \bv^{(f)}_k\big) \, .    \dsplbl{vfkvk}
\end{equation}
To demonstrate that this relation between $\bv^{(f)}_k$ and $\bv_k$ does indeed take place, we
need to express $S_{\bx_k}$ in Eq.~(\ref{mpvlct}) through $S^{(f)}_{\bx_k}$ in Eq.~(\ref{vfk}).
For the space-time of the $k$-th particle, consider the surface $\tau=\mbox{const}$, or $t_k =
f(\bx_k,\tau)$. We have for the derivatives of the action function $S(\bx_1,t_1,\ldots,\bx_{n_p},
t_{n_p})$ along this surface
\begin{equation}\label{}
    \begin{array}{ccl}
      \dsp\left.\dovd{S}{\bx_k}\, \right|_{\tau=\mathrm{const}} & = & \dsp \left.\dovd{S}{\bx_k}\,
      \right|_{t_k = \mathrm{const}}+\,\,\left.\dovd{S}{t_k}\, \right|_{\bx_k=\mathrm{const}} \cdot
      \,\,\left.\dovd{t_k}{\bx_k}\, \right|_{\tau = \mathrm{const}} \\[0.6cm]
      & = & S_{\bx_k} - H^S_k f_\bx (\bx_k,\tau) \, .
    \end{array}
\end{equation}
But $\partial S/\partial \bx_k|_{\tau=\mathrm{const}} = S^{(f)}_{\bx_k}$, and so $S_{\bx_k} =
S^{(f)}_{\bx_k} + f_{\bx}H^S_k$, and therefore
\begin{equation}\label{}
    \dovd{H^S_k}{S^{(f)}_{\bx_k}} = \dovd{H^S_k}{S_{\bx_k}} \left(1 + f_{\bx}
    \dovd{H^S_k}{S^{(f)}_{\bx_k}} \right) .
\end{equation}
Now multiplying this equation by $f_\tau(\bx_k,\tau)$ and using Eqs.~(\ref{mpvlct}), (\ref{vfk}),
we immediately obtain the desired relation (\ref{vfkvk}).

Thus, for any function $f$, the description provided by a foliation-based
wave function $\psi^{(f)}$ agrees with the one provided by a multi-time wave function $\psi$.
We have, therefore, the following
situation. In every foliation, the state of a system is described by particle positions and by all
momentums. Being the derivatives of the logarithm of a wave function along the leaves of the
foliation, momentums depend on the foliation chosen, and so in any given system's state,
there are different sets of momentums, corresponding to different possible foliations. As for a
standard foliation, by the equations of motion momentums, corresponding to any foliation,
are global variables --- they are bound to the leaves of their foliation, rather than to the points
on these leaves, and influence each other (and, consequently, the particle velocities) over the
whole leaves of this foliation instantaneously. \paqd, therefore, does not require a preferred
frame of reference: in every foliation, the theory, expressed
through the foliation's momentums and Hamiltonian function, looks the same. At the same time,
considerations, based on different foliations, agree with each other in terms of actual motion of
particles, because they all predict the same motion as the consideration based on a common object,
the multi-time wave function $\psi(\bx_1,t_1,\ldots,\bx_{n_p},t_{n_p})$, as was just discussed.
On the other hand, this multi-time wave function in all $R^{4n_p}$ may be
uniquely obtained, for any $f$ and~$\tau_0$, from a function
$\psi^{(f)}_{\tau_0}(\bx_1,\ldots,\bx_{n_p}) = \psi^{(f)}(\bx_1,\ldots,\bx_{n_p},\tau_0)$ by path
integration or by solving equations~(\ref{mtse}), and so every function $\psi^{(f)}_{\tau_0}$
contains the same information as the multi-time wave function $\psi$ in all $R^{4n_p}$.

In relativistic theory, the leaves of foliations corresponding to different Lorentz frames are
flat, and the angles between different foliations' leaves correspond to relative velocities of
respective frames. In addition to a time transformation, a Lorentz transformation of the space
coordinates inside the leaves should be done. It seems then reasonable to expect that in
relativistic theory the set of momentums, corresponding to
each Lorentz frame, will behave as described above, i.e., momentums will affect each other
instantaneously in this frame, and in each frame the theory will be the same. Also, if there
are two space-like separated entangled measurements, then neither of them can be considered as
causing the result of the other. Indeed, in different foliations their time order will be
different, and their results are unambiguously determined by the state (i.e., positions and
all momentums, or positions and wave function) on any leave of an arbitrary foliation. Foliations
with nonflat leaves (i.e., leaves that in different space-time points are angled differently with
respect to the leaves of ``inertial" foliations)
will appear, when the theory is formulated in reference frames with local
accelerations. Additional terms corresponding to inertial forces, proportional to particle
masses, will then appear in Hamiltonian functions, so that every foliation will come with its own
field of these forces. The equivalence of different foliations, i.e., the general covariance of
the theory, can then be restored in a standard
way by introducing a gauge field, which would adsorb the potential of inertial forces, in what
seems to be a natural route leading to a gauge theory of gravitation \cite{bla}.

\section[\,\,\,Analytical quantum dynamics of particles with spin]
{Analytical quantum dynamics of particles with spin}
\setcounter{equation}{0}

In this section we show how to describe in \paqd\ particles with spin. Since the wave
function of a particle with spin $s$ is a $2s+1$-component spinor,\footnote{We will denote
particle's spin by the small letter $s$,
to distinguish it from the real part $S$ of the action function $p$.}
there seem to be two possible ways to include spin in the theory. The first one is to somehow
define corresponding $2s+1$ complex or $2(2s+1)$ real action functions, satisfying evolutionary
equations of (\ref{evpde}) type with Hamiltonians that depend only on derivatives of the action and
satisfy Hamiltonian conditions. This, however, does not seem to be possible. Indeed, for the \Sch
equation we passed from the wave function to its logarithm, the action function,
in order to obtain an
evolutionary (namely, quantum Hamilton-Jacobi) equation with Hamiltonian that depends only on
derivatives of the action function, rather than on this function itself. For a multi-component wave
function, this simple trick will work only in a trivial case when every component satisfies its
own equation, independent of others. Moreover, in case of several, say $n_p$, particles, one would
have to find not $2s+1$ complex action functions, which satisfy equations of the required form, but
$(2s+1)^{n_p}$ of them! Clearly, this approach doesn't appear promising. The second possible
approach is to transform a system of $2s+1$ equations for spinor components into an equivalent
equation of the desired form for one new wave function. This can be done by using spin coherent
states, and this is the approach that we will employ here.

Spin coherent states are defined with the help of a spin rotating operator that rotates the spin
state through an angle $\alpha$ about direction $\bn$. The explicit form of this operator is
$\exp\left(-i\alpha\bn\bs\right)$, where $\bs$ is a spin operator in units of $\hbar$.
The rotation $\cR(\chi,\theta,\vphi)$, corresponding to Euler angles $\chi$, $\theta$, $\vphi$, is
obtained as a rotation through the angle $\chi$ about axis $Oz$, followed by rotation through angle
$\theta$ about axis $Oy$, followed by another rotation through angle $\vphi$ about $Oz$, and is
described by the product of the three corresponding operators: $\cR(\chi,\theta,\vphi) = e^{-i\vphi
s_z} e^{-i\theta s_y}e^{-i\chi s_z}$. Let $|s,m\rangle$ be a standard eigenstate of the spin
operators: $\bs^2 |s,m\rangle = s(s+1) |s,m\rangle$, $s_z |s,m\rangle = m |s,m\rangle$. Then the
spin coherent state is defined as the
maximally polarized state $|s,s\rangle$ rotated by the operator $\cR(\chi,\theta,\vphi)$: $|\chi,
\theta,\vphi\rangle = e^{-i\vphi s_z} e^{-i\theta s_y} e^{-i\chi s_z} |s,s\rangle$. The explicit
representation of the coherent state~is
\begin{equation}\label{scs}
    |\chi,\theta,\vphi\rangle = \sqrt{(2s)!} \sum_{m=-s}^s \frac{u^{s+m} v^{s-m}}{\sqrt{(s+m)!
    (s-m)!}}\, |s,m\rangle \, ,    \dsplbl{scs}
\end{equation}
where complex parameters $u$ and $v$ are defined as
\begin{equation}\label{uv}
    u = \cos \frac{\theta}{2}\,\, e^{-i(\vphi+\chi)/2} \, , \quad \, v = \sin \frac{\theta}{2}\,\,
    e^{i(\vphi-\chi)/2}.     \dsplbl{uv}
\end{equation}
Clearly, if $u = u_1 + iu_2$ and $v = v_1 + iv_2$, where $u_{1,2}$ and $v_{1,2}$ are real,
then $|u|^2 + |v|^2 = u_1^2 + u_2^2 + v_1^2 + v_2^2 = 1$,
so that parameters $u$ and $v$ (or $u_1$, $u_2$, $v_1$, $v_2$) live on the
three-dimensional unit sphere $S^3$ in the four-dimensional real space~$R^4$. Denote the set
$(\chi,\theta,\vphi)$, or corresponding sets $(u,v)$ or $(u_1,u_2,v_1,v_2)$, as~$\Omega$, and the
coherent state (\ref{scs}) as~$|\Omega\rangle$.
The expression (\ref{scs}) for it may be easily derived, for example, by
using the Schwinger bosons representation of spin operators \cite{Assa}. It is well known
\cite{Assa,Perelomov,Rdclf} that the system of spin coherent states is not orthogonal,
overcomplete, and allows a resolution of unity in a Hilbert space of states with spin~$s$:
\begin{equation}\label{ru}
    \frac{2s+1}{\pi^2} \int d\Omega |\Omega\rangle \langle\Omega| = \sum_m |s,m\rangle
    \langle s,m| \, ,         \dsplbl{ru}
\end{equation}
where
\begin{equation}\label{domg}
    d \Omega = \frac{1}{8} \, \sin \theta\, d \chi d \theta d \vphi    \dsplbl{domg}
\end{equation}
is the area element on $S^3$. If $w_1,w_2,w_3,w_4$ are Cartesian coordinates in $R^4$ that are
related to angular coordinates $\chi,\theta,\vphi$ and radial distance $r$ by
\begin{equation}\label{vt}
    w_{1,2} = r u_{1,2} \, , \quad \, w_{3,4} = r v_{1,2} \, ,   \dsplbl{vt}
\end{equation}
then the integration measure in $R^4$ is related to $d\Omega$ by
\begin{equation}\label{}
    d w_1 d w_2 d w_3 d w_4 = r^3 d r d\Omega \, ,
\end{equation}
which follows from the expression (most easily obtained by direct calculation with
{\em Ma\-the\-ma\-ti\-ca}) for the jacobian of the transformation (\ref{vt})
\begin{equation}\label{}
    \det \frac{\partial(w_1,w_2,w_3,w_4)}{\partial(\chi,\theta,\vphi,r)} =
    \frac{1}{8} \, r^3 \sin \theta \, .
\end{equation}
The action of the spin operators $s_\pm = s_x \pm i s_y$ and $s_z$ on spin coherent states is
described by equations \cite{Assa}
\begin{equation}\label{spins}
    \begin{array}{lcl}
      s_+ |\Omega\rangle & = & \dsp v \partial_u |\Omega\rangle , \\[0.15cm]
      s_- |\Omega\rangle & = & \dsp u \partial_v |\Omega\rangle , \\[0.1cm]
      s_z |\Omega\rangle & = & \dsp \half \, (u \partial_u - v \partial_v) |\Omega\rangle .
      \dsplbl{spins}
    \end{array}
\end{equation}

Using spin coherent states, the one-component wave function, corresponding to a spin state~
$|\psi\rangle$ with spin~$s$, is defined as a scalar product $\psi(\Omega) = \langle \Omega |\psi
\rangle$. It is clear from the resolution of unity (\ref{ru}) that using its wave function, the
state $|\psi\rangle$ may be expanded over spin coherent states as
\begin{equation}\label{}
    |\psi\rangle = \frac{2s+1}{\pi^2} \int d\Omega \, \psi(\Omega) |\Omega\rangle \, ,
\end{equation}
so that all information about the state is contained in its wave function and vice versa. The
argument $\Omega$ in the wave function describes a rotation with respect to Cartesian coordinates
in three-dimensional space, and so when the coordinate system itself is rotated the wave function
transforms accordingly. The rotations are elements of the three-dimensional rotation group SO(3),
where the spin wave function is defined. As is well known, SO(3) is not simply connected: its
fundamental group is cyclic group of order 2. As was discussed in section 2.5 and
section~3, this means that the spin wave function defined in SO(3) may be double-valued, as it
indeed is when spin $s$ is half-integer. The universal cover of SO(3) is the group SU(2) that
covers SO(3) two-to-one, and so the spin wave function is single-valued in SU(2). Elements of SO(3)
and SU(2) are parameterized by points on the sphere $S^3$ considered above \cite{Perelomov}, and
the area element $d\Omega$, Eq.~(\ref{domg}), is the Haar measure of these groups, so the spin wave
function may be considered as defined on $S^3$. Now if $|\psi\rangle = \sum_{m=-s}^s \psi_m(\bx,t)
|s,m\rangle$ is a state of a particle with spin~$s$, then the corresponding wave function is
\begin{equation}\label{wfs}
    \psi(\bx,\Omega,t) = \sqrt{(2s)!} \sum_{m=-s}^s \frac{\ubar^{s+m} \, \vbar^{s-m}}{\sqrt{(s+m)!
    (s-m)!}} \, \psi_m(\bx,t) \, .   \dsplbl{wfs}
\end{equation}
Thus a wave function of a particle with spin $s$ is defined in configuration space
$Q = R^3 \times S^3$ and is
an analytic function of $\ubar$ and $\vbar$, selected from arbitrary analytic functions of these
variables by the condition of being a homogeneous function of power $2s$, i.e., by condition
\begin{equation}\label{homcond}
    (\ubar\dubar + \vbar\dvbar) \psi(\bx,\Omega,t) = 2 s \psi(\bx,\Omega,t) \, .  \dsplbl{homcond}
\end{equation}
In the space of analytic functions of $\ubar$ and $\vbar$, an operator $(1/2)(\ubar\dubar +
\vbar\dvbar)$ plays, therefore, the role of a total spin operator. From Eq.~(\ref{spins}) we obtain
the action of spin operators on a wave function:
\begin{equation}\label{}
    \begin{array}{lcl}
      \langle\Omega|s_+|\psi\rangle & = & \ubar \dvbar \psi(x,\Omega,t) , \\[0.2cm]
      \langle\Omega|s_-|\psi\rangle & = & \vbar \dubar \psi(x,\Omega,t) , \\[0.1cm]
      \langle\Omega|s_z|\psi\rangle & = & \dsp \half\,(\ubar\dubar-\vbar\dvbar) \psi(x,\Omega,t) ,
    \end{array}
\end{equation}
where we used that operators $s_+$ and $s_-$ are hermitian conjugates of each other. Note, that
spin operators make a complete set of non-trivial first order differential operators that leave
a wave function in the form~(\ref{wfs}). The only remaining operator, $\ubar\dubar + \vbar\dvbar$,
gives $2s$ by Eq.~(\ref{homcond}).

The behavior of a particle of charge $e$ in an electric field with scalar potential $A_0$ and
magnetic field $\bB$ with vector potential $\bA$ is described by
\begin{equation}\label{ses}
    i\hbar\dovd{\psi}{t} = \left[\frac{1}{2m} \left(\hi\nabla - \frac{e}{c}\, \bA \right)^2 + eA_0
    - \gamma \bB \bs \right] \psi \, ,   \dsplbl{ses}
\end{equation}
where $c$ is the speed of light.
For a particle with spin $1/2$, Dirac theory gives for a constant $\gamma$ the value of
$e\hbar/mc$. If $\psi$ is the just-defined wave function in a spin coherent state representation,
then the term $\bB\bs$ expands~as
\begin{equation}\label{bsin}
    \begin{array}{ccl}
      \bB\bs & = & \dsp \half \, \left(B_+ s_- + B_- s_+\right) + B_z s_z \\[0.3cm]
      & = & \dsp \half \, \big[B_+ \vbar\dubar + B_- \ubar\dvbar + B_z (\ubar\dubar-\vbar\dvbar)
      \big] \, ,    \dsplbl{bsin}
    \end{array}
\end{equation}
where $B_\pm = B_x \pm i B_y$. Using matrix notations and standard Pauli matrices, we have then
\begin{equation}\label{bs}
    \begin{array}{ccl}
      \bB\bs & = & \dsp \half \, (\ubar, \vbar)
                    \left(\begin{array}{cc}
                             B_z & B_- \\
                             B_+ & -B_z
                    \end{array} \right)
                    \left(\begin{array}{c}
                             \dubar \\
                             \dvbar
                   \end{array} \right)
                 \\[0.35cm]
      & = & \dsp \half \, (\ubar, \vbar) \, \bB \bsigma
            \left(\begin{array}{c}
                      \dubar \\
                      \dvbar
            \end{array} \right).    \dsplbl{bs}
    \end{array}
\end{equation}
As in the spinless case, expressing the wave function as
\begin{equation}\label{psis}
    \psi(\bx,\Omega,t) = \dsp \exp \left(\ih\,p(\bx,\Omega,t) \right) , \quad \quad p(\bx,\Omega,t)
                       = S(\bx,\Omega,t) + \hi\,R(\bx,\Omega,t)  \dsplbl{psis}
\end{equation}
introduce the action function $p(\bx,\Omega,t)$ and its real and imaginary parts $S(\bx,\Omega,t)$
and $-\hbar R(\bx,\Omega,t)$. It is convenient to use a gauge $\mbox{div} \bA = 0$. Momentums that
correspond to spin variables, such as $\ubar$, $u_1$, or $\chi$ (i.e., partial derivatives of the
action with respect to these variables) will be denoted by the corresponding indices. The
derivatives with respect to complex variables are understood as in Eq.~(\ref{cderivs}). Then
sub\-sti\-tu\-ting (\ref{psis}) into (\ref{ses}), obtain for a particle with spin a quantum
Hamilton-Jacobi equation (\ref{qhje}) with Hamiltonian function
\begin{equation}\label{hspin}
    H = \frac{1}{2m} \left(p_j - \frac{e}{c} \, A_j\right)^2 + eA_0 + \frac{\hbar}{2im} \,
        p_{jj} - \frac{i\gamma}{2\hbar} \, (\ubar, \vbar) \, \bB \bsigma
            \left(\begin{array}{c}
                      p_{\ubar} \\
                      p_{\vbar}
            \end{array} \right).     \dsplbl{hspin}
\end{equation}
The Hamiltonian (\ref{hspin}) is of the first order with respect to the spin variables.
Consequently, HC1 is satisfied for it automatically, while HC2 is satisfied due to analyticity, as
was discussed in section 2.6. The whole theory of section 2, therefore, is applicable, but this 
time in
configuration space $Q = R^3 \times S^3$, so that at any time the particle has its space position
in $R^3$, ``internal" SU(2) position on $S^3$, and all corresponding momentums. To guarantee that
the particle's spin is equal to $s$, $S^3$ positions and momentums should satisfy
\begin{equation}\label{homcondp}
    \ih (\ubar p_{\ubar} + \vbar p_{\vbar}) = 2s \, ,   \dsplbl{homcondp}
\end{equation}
which follows from Eq.~(\ref{homcond}). Since spin operators change only the projections of spin,
and not its value, for any Hamiltonian that, as in Eq.~(\ref{ses}), depends only on spin
operators, it is sufficient if this condition is satisfied at the initial moment of time. According
to the general theory of section 2, particle velocity in physical space $R^3$ is given by
Eq.~(\ref{cqdot}), i.e.,
\begin{equation}\label{vwithA}
    v^j = \frac{1}{2m}\, (p_j + \bar{p}_j) - \frac{e}{mc}\, A_j \, ,   \dsplbl{vwithA}
\end{equation}
while SU(2) variables evolve (see section~2.6) according to
\begin{equation}\label{ubd}
    (\dot{\ubar}, \dot{\vbar}) = \left(\dovd{H}{p_{\ubar}}\, , \dovd{H}{p_{\vbar}}\right) =
    - \frac{i\gamma}{2\hbar} \, (\ubar, \vbar) \, \bB \bsigma     \dsplbl{ubd}
\end{equation}
or, after hermitian conjugation,
\begin{equation}\label{ubdot}
    \left(\begin{array}{c}
            \dot{u} \\
            \dot{v}
          \end{array}\right) = \frac{i\gamma}{2\hbar} \, \bB \bsigma
          \left(\begin{array}{c}
            u \\
            v
          \end{array}\right) .    \dsplbl{ubdot}
\end{equation}
This is an equation of spinor rotation with angular velocity $\boldsymbol{\omega} = - (\gamma/
\hbar)\bB$. The time evolution of spinor $(u,v)^T$ (where $T$ indicates transposition), composed
of SU(2) coordinates of a particle, is, therefore, very simple: at any moment it rotates with this
angular velocity, $\bB$ being the magnetic field at the current particle's position. Using
Eqs.~(\ref{ubd}), (\ref{ubdot}), it is easy to demonstrate that the value $|u|^2 + |v|^2$ is
conserved along the spinor's trajectory, and so remains equal to one, if it was equal to it
initially.

Equations of motion for momentums couple all kinds of them: ``space" momentums, with multi-indices
composed of $x$, $y$, and $z$, ``spin" momentums with multi-indices composed of $\ubar$ and
$\vbar$, and ``mixed" momentums, with multi-indices composed of both kinds of variables. These
equations decouple if the magnetic field $\bB$ is spatially uniform, and the initial wave function
factorizes in the form $\psi(\bx,\Omega,0) = \psi^{(x)}(\bx,0) \psi^{(s)}(\Omega,0)$ or
$p(\bx,\Omega,0) = p^{(x)}(\bx,0) + p^{(s)}(\Omega,0)$, where the indices $x$ and $s$ mark the
space and spin parts. The mixed momentums then remain equal to zero and the wave function remains
factorized at all times. The space part of the wave/action function satisfies the equations for a
spinless particle, and so the particle moves in the physical space as if it didn't have any spin.
Using Eq.~(\ref{ubd}), the equation
\begin{equation}\label{}
    \dovd{p^{(s)}}{t} - \frac{i\gamma}{2\hbar} \, (\ubar, \vbar) \, \bB \bsigma
            \left(\begin{array}{c}
                      p^{(s)}_{\ubar} \\
                      p^{(s)}_{\vbar}
            \end{array} \right) = 0
\end{equation}
for a spin part of an action function may be written in the form $\dot{p}^{(s)} = 0$, where
$\dot{p}^{(s)} = \partial p^{(s)}/\partial t + \dot{\ubar} \dubar p^{(s)} + \dot{\vbar} \dvbar
p^{(s)}$. As for every homogeneous PDE of the first order \cite{arnld2}, the solution $p^{(s)}$,
therefore, remains constant along the equation's characteristic curve, i.e., along the trajectory
$\big(u(t),v(t)\big)$ in $S^3$ described by Eq.~(\ref{ubdot}). Along with $p^{(s)}$, the wave
function $\psi^{(s)}$ also remains constant, i.e., $\langle u(t),v(t)|\psi^{(s)}(t)\rangle =
\mbox{const}$.
Consequently, like a spinor $(u,v)^T$, the spin part $\psi^{(s)}$ of the wave function rotates with
angular velocity $\boldsymbol{\omega} = - (\gamma/\hbar)\bB$, exhibiting the well-known spin
precession in a spatially uniform magnetic field.

Like a spinless \Sch equation, Eq.~(\ref{ses}) may be obtained from a variational principle
\begin{equation}\label{vps}
    \delta \int \cL \, d\Omega^{(u,v)} \prod_{j=1}^3 dx^j dt = 0    \dsplbl{vps}
\end{equation}
with Lagrangian density
\begin{equation}\label{lds}
    \cL = \frac{\hbar}{2i} \left(\psi^*\partial_t\psi - \psi\partial_t\psi^*\right) +
    \frac{\hbar^2}{2m}\, \left(\partial_j\psi^* + \frac{ie}{\hbar c} A_j \psi^*\right) \left(
    \partial_j \psi - \frac{ie}{\hbar c} A_j \psi\right) + e A_0 \psi^* \psi - \gamma \psi^*
    \bB\bs\psi \, .       \dsplbl{lds}
\end{equation}
The spin part of the integration measure in Eq.~(\ref{vps}) is $d\Omega^{(u,v)} = \rmd u_1 \wedge
\rmd u_2 \wedge \rmd v_1 \wedge \rmd v_2 = - (1/4) \rmd u \wedge \rmd \ubar \wedge \rmd v \wedge
\rmd \vbar$, and integration over $u_1$, $u_2$, $v_1$, $v_2$ runs over the whole space $R^4$, so
that this variational principle defines Eq.~(\ref{ses}) in the whole space $R^4$, and not only on
the unit sphere $S^3$. The phase invariance of the Lagrangian density $\cL$ leads, by Noether's
theorem, to the corresponding conservation law, which now has the form
\begin{equation}\label{cls}
    \partial_0 j^0 + \mbox{div} \bj + \dubar j^{\ubar} + \dvbar j^{\vbar} = 0 \, ,   \dsplbl{cls}
\end{equation}
where the components of the current are given by Eq.~(\ref{crnt}), but with $i=0, \ldots,
3,\ubar,\vbar$ this time. Substituting there $\Delta = (i/\hbar)\psi$, $\Delta^* =
-(i/\hbar)\psi^*$, $\Lambda^i=0$, and using Eq.~(\ref{bs}) for $\bB\bs$, one gets for the current
\begin{equation}\label{}
    \begin{array}{ccl}
      \left(j^0,\bj,j^{\ubar},j^{\vbar}\right) &=&\dsp\left(|\psi|^2, \,\frac{\hbar}{2im}\, (\psi^*
      \nabla\psi - \psi\nabla \psi^*) - \frac{e}{mc} \, \bA |\psi|^2, \,- \frac{i\gamma}{2\hbar} \,
      |\psi|^2 (\ubar, \vbar) \, \bB \bsigma \right) \\[0.4cm]
      & = & |\psi|^2 (1,\bv,\dot{\ubar},\dot{\vbar}) \, .
    \end{array}
\end{equation}
Note, that since this current doesn't have a radial component in $R^4$, the conservation law
(\ref{cls}) is satisfied on every sphere with the center in the origin there, including a unit
sphere~$S^3$, where we need it. The results of the previous sections can now be immediately
generalized to the case of particles with spin.
Most importantly, $|\psi|^2$ becomes the probability
density in configuration space $R^3 \times S^3$ with respect to a measure $d\Omega\prod_{j=1}^3
dx_j$, where $d\Omega$ is a measure (\ref{domg}) on $S^3$, and the measurement of spin-related
physical quantities is described by the same theory of section~8 as for space-related quantities.

Although, as was discussed in section~2.6, the above derivation in complex coordinates
$\ubar, \vbar$ is equivalent to the one that uses coordinates $u_1, u_2, v_1, v_2$, it
may be instructive to present a direct derivation in these real coordinates. For that, it is
convenient to present the last term in Eq.~(\ref{ses}) in the form
\begin{equation}\label{}
    - \gamma\bB\bs\psi = \hi\big(U^*\dubar + V^*\dvbar\big)\psi \, ,
\end{equation}
where we introduced $U^* = U_1 - iU_2$ and $V^* = V_1 - iV_2$ for which, by Eq.~(\ref{bsin}), we
have
\begin{equation}\label{U12V12}
    \begin{array}{ccrcr}
      U_1 & = & \dsp -\frac{\gamma}{2\hbar}\,\Re\big[i(B_+\vbar + B_z\ubar)\big] & = & \dsp
      -\frac{\gamma}{2\hbar}\,(B_x v_2 - B_y v_1 + B_z u_2) \, , \\[0.4cm]
      U_2 & = & \dsp \frac{\gamma}{2\hbar}\,\Im\big[i(B_+\vbar + B_z\ubar)\big] & = & \dsp
      \frac{\gamma}{2\hbar}\,(B_x v_1 + B_y v_2 + B_z u_1) \, , \\[0.4cm]
      V_1 & = & \dsp -\frac{\gamma}{2\hbar}\,\Re\big[i(B_-\ubar - B_z\vbar)\big] & = & \dsp
      -\frac{\gamma}{2\hbar}\,(B_x u_2 + B_y u_1 - B_z v_2) \, , \\[0.4cm]
      V_2 & = & \dsp \frac{\gamma}{2\hbar}\,\Im\big[i(B_-\ubar - B_z\vbar)\big] & = & \dsp
      \frac{\gamma}{2\hbar}\,(B_x u_1 - B_y u_2 - B_z v_1) \, .
    \end{array}
\end{equation}
Since $\psi$ is an analytic function of $\ubar$ and $\vbar$, we have, using the Cauchy-Riemann
equations, $\dubar\psi = \partial_{u_1}\psi = i\partial_{u_2}\psi$, and so $U^*\dubar\psi = U_1
\partial_{u_1}\psi - iU_2\, i\partial_{u_2}\psi = U_1 \partial_{u_1}\psi + U_2 \partial_{u_2}\psi$
and also $V^*\dvbar\psi = V_1 \partial_{v_1}\psi + V_2 \partial_{v_2}\psi$. Equation (\ref{ses})
can now be written in the form
\begin{equation}\label{sesr}
    i\hbar\dovd{\psi}{t} = \left[\frac{1}{2m} \left(\hi\nabla - \frac{e}{c}\, \bA \right)^2 + eA_0
    +\hi\big(U_1 \partial_{u_1} + U_2 \partial_{u_2} + V_1 \partial_{v_1} + V_2 \partial_{v_2}\big)
    \right] \psi \, ,   \dsplbl{sesr}
\end{equation}
and after substituting (\ref{psis}), we obtain for the action function a quantum Hamilton-Jacobi
equation (\ref{qhje}) with Hamiltonian function
\begin{equation}\label{}
    H = \frac{1}{2m} \left(p_j - \frac{e}{c} \, A_j\right)^2 + eA_0 + \frac{\hbar}{2im} \,
        p_{jj} + U_1 p_{u_1} + U_2 p_{u_2} + V_1 p_{v_1} + V_2 p_{v_2} \, ,
\end{equation}
so that the spin coordinates have velocities $\dot{u}_{1,2} = U_{1,2}$, $\dot{v}_{1,2} =
V_{1,2}$, which agrees with Eq.~(\ref{ubdot}). Equation (\ref{sesr}) may be obtained from the
variational principle (\ref{vps}) with the same Lagrangian density $\cL$ as in Eq.~(\ref{lds}),
but with the spin term $-\gamma\psi^*\bB\bs\psi$ there presented as $(\hbar/i)\psi^*(U_1
\partial_{u_1} + U_2 \partial_{u_2} + V_1 \partial_{v_1} + V_2 \partial_{v_2})\psi$. The phase
invariance of $\cL$ leads then to the conservation law
\begin{equation}\label{}
    \partial_0 j^0 + \mbox{div} \bj + \partial_{u_1} j^{u_1} + \partial_{u_2} j^{u_2} +
    \partial_{v_1} j^{v_1} + \partial_{v_2} j^{v_2} = 0
\end{equation}
with current
\begin{equation}\label{}
    \begin{array}{ccl}
      \left(j^0,\bj,j^{u_1},j^{u_2},j^{v_1},j^{v_2}\right) & = & \dsp\left(|\psi|^2,\,
      \frac{\hbar}{2im}\, (\psi^*\nabla \psi - \psi \nabla \psi^*) - \frac{e}{mc} \, \bA |\psi|^2,
      \, |\psi|^2 \big(U_1,U_2,V_1,V_2\big) \right) \\[0.4cm]
      & = & |\psi|^2 (1,\bv,\dot{u}_1,\dot{u}_2,\dot{v}_1,\dot{v}_2)
    \end{array}
\end{equation}
and, therefore, to the probabilistic interpretation of $|\psi|^2$ and the measurement theory of
section~8.

Finally, we present the theory in ``natural" coordinates $\chi, \theta, \vphi$ on $S^3$. For
that, note that $u_1, u_2, v_1, v_2$ in Eq.~(\ref{sesr}) are just Cartesian coordinates in
$R^4$, running from $-\infty$ to $\infty$. To avoid confusion, rename them as $w_1,w_2,w_3,w_4$
and make the transformation (\ref{vt}), where now $u_{1,2}$ and $v_{1,2}$ are real and
imaginary parts of complex coordinates $u$ and $v$, Eq.~(\ref{uv}), on~$S^3$. Let $J =
\partial(w_1,w_2,w_3,w_4)/\partial(\chi,\theta,\vphi,r)$ be the jacobian matrix of this
transformation. The spin term in the \Sch equation (\ref{ses}) may then be presented, using its
form in (\ref{sesr}), as
\begin{equation}\label{}
    - \gamma\bB\bs\psi = \hi\big(U_\chi\partial_\chi + U_\theta\partial_\theta +
    U_{\vphi}\partial_{\vphi} + U_r\partial_r\big)\psi \, ,
\end{equation}
where
\begin{equation}\label{}
    \big(U_\chi,U_\theta,U_{\vphi},U_r\big) = \big(U_1,U_2,V_1,V_2\big) \left(J^{-1}\right)^T
    \big|_{r=1} \, .
\end{equation}
Since we only need this transformation on $S^3$, i.e., for $r=1$, $U_{1,2}$ and $V_{1,2}$ here are
given by Eq.~(\ref{U12V12}) where $u_{1,2}$ and $v_{1,2}$ are real and imaginary parts of $u$ and
$v$, Eq.~(\ref{uv}). Direct calculation using {\em Mathematica} then gives
\begin{equation}\label{}
    \begin{array}{ccl}
      U_\chi & = & \dsp -\frac{\gamma}{\hbar}\,\frac{1}{\sin\theta}\,(B_x \cos\vphi +
      B_y \sin\vphi) ,\\[0.4cm]
      U_\theta & = & \dsp \frac{\gamma}{\hbar}\, (B_x \sin\vphi - B_y \cos\vphi) ,\\[0.4cm]
      U_{\vphi} & = & \dsp \frac{\gamma}{\hbar} \left[\frac{\cos\theta}{\sin\theta}\, (B_x
      \cos\vphi + B_y \sin\vphi) - B_z\right] ,
    \end{array}
\end{equation}
and, as expected, $U_r=0$. The \Sch equation now has the form
\begin{equation}\label{sesa}
    i\hbar\dovd{\psi}{t} = \left[\frac{1}{2m} \left(\hi\nabla - \frac{e}{c}\, \bA \right)^2 + eA_0
    + \hi \big(U_\chi \partial_\chi + U_\theta \partial_\theta + U_{\vphi} \partial_{\vphi}\big)
    \right] \psi \, ,    \dsplbl{sesa}
\end{equation}
and the Hamiltonian function in a quantum Hamilton-Jacobi equation (\ref{qhje}) will become
\begin{equation}\label{hspinchi}
    H = \frac{1}{2m} \left(p_j - \frac{e}{c} \, A_j\right)^2 + eA_0 + \frac{\hbar}{2im} \,
        p_{jj} + U_\chi p_\chi + U_\theta p_\theta + U_{\vphi} p_{\vphi} \, ,  \dsplbl{hspinchi}
\end{equation}
so that $\dot{\chi} = U_\chi$, $\dot{\theta} = U_\theta$, $\dot{\vphi} = U_{\vphi}$.

Equation (\ref{sesa}) may be obtained from a variational principle $\delta\int\cL\,d\Omega
\prod_{j=1}^3dx^jdt = 0$, where the measure $d\Omega$ is given by Eq.~(\ref{domg}) and the
Lagrangian density $\cL$ by Eq.~(\ref{lds}) with spin term $-\gamma\psi^*\bB\bs\psi$ there
presented as $(\hbar/i)\psi^*(U_\chi \partial_\chi + U_\theta \partial_\theta + U_{\vphi}
\partial_{\vphi})\psi$. The measure $d\Omega$ is not homogeneous --- it is equal to the product of
differentials of independent variables times the function $f=\sin\theta$. In such cases, to ensure
the possibility of all necessary integrations by parts, the derivation of the equations of motion
and conservation laws from the variational principle differs by using instead of the usual
derivatives $\partial_k$ the operator $\partial_k^{(f)} = (1/f)\partial_kf$, which acts on any
function $p$ as $\partial_k^{(f)}p = (1/f)\partial_k(fp)$ \cite{sch}.
In our case, when $f$ is a function of only one variable~$\theta$, all derivatives except
$\partial_\theta$ remain unchanged. The correct form of a
conservation law that follows from the phase invariance of $\cL$ in a space $R^3\times S^3$ with
an inhomogeneous integration measure $d\Omega\prod_{j=1}^3dx^j$, and has the usual meaning and
consequences there, is
\begin{equation}\label{lastconslaw}
    \partial_0 j^0 + \mbox{div} \bj + \partial_\chi j^\chi + \frac{1}{\sin\theta}\,\partial_\theta
    \big(\sin\theta j^\theta\big) + \partial_{\vphi} j^{\vphi} = 0
\end{equation}
where
\begin{equation}\label{lastcurrent}
    \begin{array}{ccl}
      \left(j^0,\bj,j^\chi,j^\theta,j^{\vphi}\right) & = & \dsp\left(|\psi|^2,\,
      \frac{\hbar}{2im}\, (\psi^*\nabla \psi - \psi \nabla \psi^*) - \frac{e}{mc}\, \bA |\psi|^2,\,
      |\psi|^2 \big(U_\chi,U_\theta,U_{\vphi}\big) \right) \\[0.4cm]
      & = & |\psi|^2 (1,\bv,\dot{\chi},\dot{\theta},\dot{\vphi}) .
    \end{array}
\end{equation}
The conservation law (\ref{lastconslaw}) may also be obtained directly by substituting there
expressions (\ref{lastcurrent}) for components of the current, noticing that $|\psi|^2 = e^{2R}$,
and using quantum Hamilton-Jacobi equation with Hamiltonian function~(\ref{hspinchi}).

We thus demonstrated that the analytical quantum dynamics of particles with spin can be developed
by an extension of particle configuration space from $R^3$ to $R^3\times S^3$.
The de~Broglie\,-\,Bohm\,-\,style theory of spin in $R^3\times S^3$,
in which particles are considered as a point limit of
extended rigid objects, is developed in chapter~10 of Holland's book \cite{dbb2}. Our spin theory
uses an infinite phase space over $R^3\times S^3$, and all the theory of the previous sections is
applicable to it. In particular, particles move in $R^3\times S^3$ along trajectories that are well
defined by the equations of motion, $|\psi|^2$ is the probability density in $R^3 \times S^3$, and
the measurement of a spin component in a Stern-Gerlach experiment is a typical example of von
Neumann's measurement procedure with discrete spectrum.

\section[\,\,\,Conclusion]{Conclusion}
\setcounter{equation}{0}

Let us summarize the main points of \paqd. It is straightforward to verify that for a sum $p^s$ of
a Taylor series~(\ref{us}) to satisfy PDE~(\ref{evpde}), it is necessary and sufficient if the
Taylor coefficients, or momentums, $p^r_\sigma$ satisfy the ODEs~(\ref{tottd}), where $\dot{\bq}$
is the velocity of an expansion point moving in the configuration space. Consequently, we have a
simple and universal connection between evolutionary PDE~(\ref{evpde}) and the dynamical
ODEs~(\ref{tottd}), where the
velocity $\dot{\bq}$ is still arbitrary. If, further, the Hamiltonian function $H^r$ in
Eq.~(\ref{evpde}) satisfies the Hamiltonian conditions of section~2.3, then there exists a
special velocity, given by Eq.~(\ref{preheq}), which leads to Eq.~(\ref{hamcond}), the variational
principles of section~2.4, and the hierarchical Hamiltonian structure of the whole
theory. Thus, there is a general ODE/PDE Hamiltonian formalism that may be filled with different
physical contents, depending on the form of a Hamiltonian function. In classical mechanics, the
Hamiltonian function is of the first order. As was explained in section~2.7, the ODE part
of the theory in this case simplifies into an ordinary Hamiltonian mechanics in the usual phase
space. Quantum theory utilizes the second available option, with a Hamiltonian of a higher order 
and ODEs residing in an infinite phase space. More specifically, it appears that in a 
nonrelativistic domain for spinless particles, nature builds
quantum theories by the recipe of section~4, so that any theory of this kind is defined by
Eq.~(\ref{psigser}) with some Lagrangian function, quadratic in velocity and such that the
corresponding Hamiltonian function satisfies Hamiltonian condition (\ref{hcond}).
The theory will then automatically exhibit the superposition principle, path-integral 
representation, wave-particle duality (which is shown to be possible only in the infinite phase
space), and the classical limit, described by Hamilton-Jacobi PDE and Hamilton ODEs with
a Hamiltonian, corresponding to the Lagrangian function in Eq.~(\ref{psigser}). Also, the resulting
\Sch equation will be automatically obtainable from a variational principle, so its symmetries
will lead to corresponding conservation laws. Since Hamiltonian functions in our theory depend
only on derivatives of unknown functions, they are automatically invariant with respect to shifts
of these functions by arbitrary constants. This symmetry leads to a current conservation, a current
being defined with the correct velocity~(\ref{preheq}), and to invariance of a measure $|\psi|^2dV$
with respect to equations of motion. This invariance leads then to the probability density
$\rho=|\psi|^2$ in the same way as invariance of the Liouville measure leads to the microcanonical
distribution in classical statistics, the difference in probability densities resulting from 
different forms of equations of motion. The probabilistic interpretation of the wave function is,
therefore, deduced in \paqd\ rather than being postulated.
The multiparticle generalization of the theory leads to the
standard picture of quantum particles in a classical macroscopic environment, and being applied to
specially constructed apparatuses, to the quantum theory of measurement. The measurements of
classical quantities that may be used as parts of particle-apparatus interaction Hamiltonians,
appear to have the desired features in this theory.
On the other hand, the measurement and observation
of nonlocal momentums, such as $p_{x_iy_j}$, where $i\neq j$ are particle indices, is impossible,
because Hamiltonians built by the rules of section 4 cannot contain such terms. The presence of
such momentums makes the whole multiparticle theory nonlocal, and explains the mechanism of
nonlocal correlations. On the other hand, their nonobservability prevents using them for the
transmission of superluminal signals. As was discussed in section 9, in spite of being nonlocal,
the relativistic version of the theory seems to be presentable in a Lorentz invariant and even
generally covariant way. Finally, the theory of particles with nonzero spin resides in
configuration space that includes, along with particle's space coordinates, its internal SU(2)
degrees of freedom.

\begin{table}[htb]
\begin{center}
\begin{picture}(460,220)
\put(230,200){\makebox(120,20){\paqd}}
\put(350,200){\makebox(110,20){QM}}
\put(0,180){\makebox(230,20){Relation of mathematical theory to experiment}}
\put(230,180){\makebox(120,20){Straightforward}}
\put(350,180){\makebox(110,20){Needs interpretation}}
\put(0,160){\makebox(230,20){Laws of nature}}
\put(230,160){\makebox(120,20){Deterministic}}
\put(350,160){\makebox(110,20){Indeterministic}}
\put(0,140){\makebox(230,20){Particles' behavior}}
\put(230,140){\makebox(120,20){Always particle-like}}
\put(350,140){\makebox(110,20){Complementary}}
\put(0,120){\makebox(230,20){Wave-particle duality}}
\put(230,120){\makebox(120,20){Mathematically derived}}
\put(350,120){\makebox(110,20){Verbally described}}
\put(0,100){\makebox(230,20){Non-statistical interpretation of wave function}}
\put(230,100){\makebox(120,20){Exists}}
\put(350,100){\makebox(110,20){Does not exist}}
\put(0,80){\makebox(230,20){Statistical interpretation of wave function}}
\put(230,80){\makebox(120,20){Mathematically derived}}
\put(350,80){\makebox(110,20){Postulated}}
\put(0,60){\makebox(230,20){Measurement problem}}
\put(230,60){\makebox(120,20){Does not exist}}
\put(350,60){\makebox(110,20){Unsolved}}
\put(0,40){\makebox(230,20){Nonlocality vs. Lorentz invariance conflict}}
\put(230,40){\makebox(120,20){Cleared}}
\put(350,40){\makebox(110,20){Unresolved}}
\put(0,20){\makebox(230,20){Classical limit}}
\put(230,20){\makebox(120,20){Direct}}
\put(350,20){\makebox(110,20){Indirect}}
\put(0,0){\makebox(230,20){Geometric picture of classical mechanics}}
\put(230,0){\makebox(120,20){Generalized}}
\put(350,0){\makebox(110,20){Lost}}
\multiput(0,0)(0,20){11}{\line(460,0){460}}
\put(230,220){\line(230,0){230}}
\put(0,0){\line(0,200){200}}
\put(230,0){\line(0,220){220}}
\put(350,0){\line(0,220){220}}
\put(460,0){\line(0,220){220}}
\end{picture}
\caption{The comparison of \paqd\ and QM.}
\end{center}
\end{table}

In Table 1, different aspects of \paqd\ and QM are compared in a self-explanatory form. The
comparison clearly demonstrates that in spite of experimental agreement, the two theories are
evidently different and draw different pictures of the physical reality. Several additional
remarks may be useful. First, as was discussed in sections~2.5 and 3, the actual
solution of the equations of motion may be obtained by the generalized Jacobi method from a known
wave function. In the spinless case, the particles will then move along Bohmian trajectories. The
equations of motion, however, determine the motion completely and unambiguously themselves, and so
a technique of their direct solution, without any use of a wave function, should be possible to
develop if desirable. Second, the theory is formulated in terms of momentums, and as was just
discussed not all of them are observable. We note, however, that all coordinates and momentums
that {\em are} observable (for example in such devices as bubble chambers) are reflected in the
theory, and nonobservable momentums are nonobservable not because they are postulated to be such,
but because this is a property of an observation/measurement procedure that follows from the
basic equations of the theory. The situation here should be compared with the one in QM, which is
formulated entirely in terms of nonobservable wave functions, and brings in the observable 
quantities (coordinates and momentums) only through the interpretational part of the theory. Third,
the theory of von Neumann's measurements, presented in section 8, exhibits all the properties
attributed to the measurement procedure in standard QM, in particular such measurements must
satisfy the uncertainty relations. However, contrary to QM, \paqd\ gives detailed description of
the behavior of both observed system and observing apparatus before, during, and after their 
interaction. In this situation, it is not unthinkable to speculate that new ``measurement-like" 
procedures may be found, which will provide more information than is permitted by the uncertainty
principle, or will generate experimental situations (especially when mesoscopic objects are 
involved) in which QM (but not \paqd) fails to give unambiguous predictions. The investigation of
such possibilities lies, however, outside of the scope of the present work.

To summarize, the mathematical theory developed in section 2 allows to give a simple description
of quantum phenomena as resulting from generalized Hamiltonian motion of particles. The present
theory does not suffer from the shortcomings discussed in the Introduction. It appears especially
important, that although the specific form of the theory is completely determined by the \Sch
equation, it nevertheless allows to simultaneously deduce the statistical interpretation, which in
existing quantum theory is described by the separate and independent axioms. We conclude, that it
seems not unreasonable to believe that \paqd\ may indeed provide the true and fundamental 
description of nature.

\section*{Acknowledgements}
\addcontentsline{toc}{section}{Acknowledgements}

I am grateful to Joseph Krasil'shchik for critical reading of section 2 of this work and valuable
comments.

\renewcommand{\thesection}{Appendix:}
\renewcommand{\theequation}{A\arabic{equation}}

\section[\hspace{1.6cm}Statistical distribution and entropy growth in classical statistics]
{Statistical distribution and entropy growth in classical statistics}
\setcounter{equation}{0}

To facilitate the comparison in section 6 of statistical distribution in \paqd\ and in the
classical theory, here we review the basics of classical equilibrium and nonequilibrium statistics
in a convenient for this comparison form. A classical system is represented by a point in a phase
space $P$ of dimension $2n_pn_s$, where $n_p$ is the number of particles and $n_s$ is the dimension
of the physical space. Denote a general point
of $P$ by $\gamma$. Invariant measure in $P$ is a Liouville measure $\prod_{i=1}^{n_p}d^{n_s}r_i
d^{n_s}p_i$, and we denote it by $d\Gamma$. Besides the description in a phase space $P$, or the
space of microstates that we will also call ``full description space" (FDS), classical statistics
uses a much cruder description of systems by sets of kinetic or thermodynamic parameters forming
``reduced description space" (RDS) or the space of macrostates of dimension $n_R \ll 2n_pn_s$. The
coordinates $R_i$ of this space are the values of some functions $r_i(\gamma)$, $i=1,\ldots,n_R$ of
the microstate, and the macrostate is considered fully specified by the known values of all the
$R_i$, i.e., by the set $R = \{R_i, i = 1,\ldots,n_R\}$. In other words, phase space $P$ is broken
into subspaces $P_R$ that correspond to small cells in RDS: $P_R = \{\gamma\in P: \, R_i \leq
r_i(\gamma) \leq R_i+\delta_i, i=1,\ldots,n_R\}$, and a crude description of a system at any time
$t$ is given by the corresponding $R(t)$, i.e., by specifying a subspace that the point $\gamma$
resides in at
this moment. Different points of $P_R$ represent then different microstates, compatible with the
same macrostate $R$, so that every time the system is prepared in this macrostate, its microstate
will be represented by some random point $\gamma\in P_R$. As it was with a quantum particle, in
the mathematical limit of an infinite number of such preparations, these points will form an
ensemble $A_R$ with probability density $\rho_R$ that maximizes the corresponding Gibbs entropy
$S_G^{(R)}(\rho)$. Repeating the steps that led from~(\ref{ge}) to~(\ref{sgr}), but using the
Liouville measure this time, it is easy to show that
\begin{equation}\label{sgrc}
    S_G^{(R)}(\rho) = - \int_{P_R}\!\! \rho \ln\rho \,d\Gamma \, .   \dsplbl{sgrc}
\end{equation}
We call the density~$\rho_R$ that maximizes this $S_G^{(R)}(\rho)$ a microcanonical
density, and corresponding ensemble $A_R$ a microcanonical ensemble. Let us show that the
microcanonical density is constant:
\begin{equation}\label{mcd}
    \rho_R = \mbox{const} = \frac{1}{\Gamma(P_R)} \, ,   \dsplbl{mcd}
\end{equation}
where $\Gamma(P_R) = \int_{P_R} d\Gamma$ is a phase volume of $P_R$. Indeed, from $\rho_R =
\mbox{const}$ and $\ln x \leq x-1$, we have for any other normalized probability density $\rho$ in
$P_R$:
\begin{equation}\label{}
    \begin{array}{ccl}
      \dsp - \int_{P_R}\!\! \rho_R \ln\rho_R \,d\Gamma & = & \dsp - \int_{P_R}\!\! \rho \ln\rho
      \,d\Gamma  - \int_{P_R}\!\! \rho \ln\frac{\rho_R}{\rho}\, d\Gamma \\[0.5cm]
      & \geq & \dsp - \int_{P_R}\!\! \rho \ln\rho \,d\Gamma + \int_{P_R}\!\! \rho \left(1 -
      \frac{\rho_R}{\rho}\right) d\Gamma \\[0.5cm]
      & = & \dsp - \int_{P_R}\!\! \rho \ln\rho \,d\Gamma \, .
    \end{array}
\end{equation}
Due to the constancy of $\rho_R$, the averaging over $A_R$, i.e., over a set of systems randomly
created in a macrostate~$R$, reduces to the averaging over $P_R$ with the measure $d\Gamma$.
Consequently, the need for introducing ensembles $A_R$ disappears; instead, we will use as
ensembles the corresponding subsets $P_R$ of FDS.

The maximum possible value of the Gibbs entropy on $P_R$ will be called the Boltzmann entropy of
$R$ and denoted as $S_B(R)$. We have, obviously,
\begin{equation}\label{sbr}
    S_B(R) = \max_{\{\rho\}} S_G^{(R)}(\rho) = S_G^{(R)}(\rho_R) = -\ln \rho_R
             = \ln \Gamma (P_R) \, .   \dsplbl{sbr}
\end{equation}
While Gibbs entropy is a function of the probability density $\rho$ in FDS,
Boltzmann entropy is a function of a set $R$ that belongs to a RDS and describes a macroscopic
state of a system. The corresponding functions $r_i$ may be, for example, the particle, energy,
and momentum densities in small cells, covering the volume of a system or, for another example,
the densities of particles in the cells that cover a six-dimensional one-particle phase space
(i.e., essentially, the values of the Boltzmann distribution function in different points of this
space). The fact that the Boltzmann entropy $S_B$ depends on the macroscopic state $R$ of a system,
rather than on its microscopic state $\gamma$, is a manifestation of its ``anthropomorphic" nature,
as was discussed by Jaynes \cite{jayns}: by its very thermodynamic definition, the difference of a
system's entropy between different states depends on which parameters were held fixed and which
were allowed to change during the transition from one state to the other.

The relations (\ref{mcd}) and (\ref{sbr}) are based only on the $N\rightarrow\infty$ limit \big(see
Eq.~(\ref{rf0})\big). They are, therefore, universally applicable to any system and any functions
$r_i$, even if they don't have a macroscopic character.
In macroscopic systems, however, functions $r_i$ may be selected in a way that
allows them to possess additional important properties \cite{lbvtz}. Namely, functions
$r_i$ give a crude description of a system; as was just discussed, usually they represent the
properties of particles in small cells in coordinate or phase spaces. Then in macroscopic systems,
with their enormous quantity of particles, the size of these cells may be chosen in such a way
that on one hand they are sufficiently small,
so that the parts of the system inside them appear homogeneous
and further division does not lead to more refined description, while on the other hand they are
large enough to still contain a macroscopic number of particles. The behavior of all physical
values that are additive with respect to contributions of separate particles
or small groups of particles, will then be regulated by a central limit theorem: in a typical
microstate $\gamma \in P_R$, these values, as well as their time derivatives, will be close to the
corresponding averages over $P_R$ with negligible dispersion. The averaging over $P_R$
serves, therefore, as a method of calculation of typical, i.e., observable, magnitudes of
thermodynamic values, with the microcanonical ensemble often being replaced by a canonical one for
calculational convenience.

The existence of such intermediate scale of description refinement, and of corresponding typical
behavior, practically identical to average one, is the first important property of macroscopic
systems. Another property is related to the character of their time evolution. Namely, let at time
$t_0$ the system be described by parameter set $R(t_0)$, so that its microstate belongs to an
ensemble (i.e., subspace of FDS) $P_{R(t_0)}$. We let all states of this ensemble evolve until the
time $t_1>t_0$ and denote the resulting ensemble $P_{R(t_0)}(t_1)$. As was just discussed, typical
values of the functions $r_i$ in this ensemble will be close, with negligible dispersion, to
corresponding averages over it, which we will denote by $R_i(t_1)$. The typical, i.e., observable,
macrostate $R(t_1)$ for $t_1>t_0$ is obtained, therefore, by direct averaging of the equations of
motion over $P_{R(t_0)}$. We refer to the corresponding averaged equations as generalized kinetic
equations. Let us now compare the ensemble $P_{R(t_0)}(t_1)$ with a microcanonical ensemble
$P_{R(t_1)}$ that corresponds to the values of the parameters $R$, observed at time $t_1$.
Neglecting extremely rare nontypical microstates, we can say that the values of $R$ in all states
$\gamma\in P_{R(t_0)}(t_1)$ are equal to $R(t_1)$, so that $P_{R(t_0)}(t_1)$ is a subset of
$P_{R(t_1)}$, $P_{R(t_0)}(t_1)\subset P_{R(t_1)}$. We need to elaborate in what sense the
microstates of $P_{R(t_0)}(t_1)$ that do not belong to $P_{R(t_1)}$ are ``extremely rare."
Note that phase volumes that are essential for our conclusions are such that their logarithms
are extensive, i.e., scale proportionally to the number of particles $n_p$ in a system.
Consequently, we are only interested in the logarithms of these volumes per particle in the limit
as $n_p\rightarrow\infty$. Then the inclusion $P_{R(t_0)}(t_1)\subset P_{R(t_1)}$ should be
understood as
\begin{equation}\label{}
    \lim_{n_p\rightarrow\infty} \frac{1}{n_p} \Big( \ln \Gamma\left(P_{R(t_0)}(t_1)\right) -
    \ln \Gamma\left(P_{R(t_0)}(t_1){\textstyle \bigcap} P_{R(t_1)}\right) \Big) = 0 \, ,
\end{equation}
so that the part of $P_{R(t_0)}(t_1)$ that falls outside of $P_{R(t_1)}$ is inessential in this
sense of logarithm per particle. This consideration also explains why we should not care about the
exact values of the $\delta_i$ in the definition of $P_R$ --- they are inessential in the same
sense.

We are especially interested in the situation
when at time $t_0$ the system was in a nonequilibrium state. The states of the ensemble
$P_{R(t_0)}(t_1)$ will not be typical for $P_{R(t_1)}$ in this case, for they will have nontypical
for $P_{R(t_1)}$ correlations. These correlations will manifest themselves under time inversion:
the states of $P_{R(t_0)}(t_1)$ will return back to the macrostate $R(t_0)$, which is further
from equilibrium than $R(t_1)$, while the typical states of $P_{R(t_1)}$ will approach equilibrium
(modulus tiny thermodynamic fluctuations) with deviation of time from $t_1$ in both directions.
Also, if we wait for the whole Poincar\'e cycle to pass, then we will observe another
manifestation: the states of $P_{R(t_0)}(t_1)$ will deviate off equilibrium as far as to $R(t_0)$,
while the typical states of $P_{R(t_1)}$ will deviate less --- up to $R(t_1)$. It happens,
however, and this is the second important property of macroscopic systems, that the correlations,
which are different in $P_{R(t_0)}(t_1)$ and $P_{R(t_1)}$, are $n$-particle correlations with
macroscopically large $n$ (``large-$n$ correlations"), while ``small-$n$ correlations" (i.e.,
one-particle densities and correlations between a small number of particles) in $P_{R(t_0)}(t_1)$
and $P_{R(t_1)}$ are practically the same. At the end of the Poincar\'e cycle nontypical large-$n$
correlations will conspire to coherently affect
small-$n$ ones and drive the system into an abnormally (for
$P_{R(t_1)}$) nonequilibrium state $R(t_0)$, but before that the influence of large-$n$
correlations on small-$n$ ones may be neglected. But it is only these small-$n$ correlations, and
not large-$n$ ones, that contribute to observable and measurable physical values of interest.
Consequently, in the normal physical experiment, when time goes only forward, but not as
far forward as for the length of the Poincar\'e cycle, nontypical for $P_{R(t_1)}$ large-$n$
correlations in $P_{R(t_0)}(t_1)$ do not manifest themselves, and we can regard the states of
$P_{R(t_0)}(t_1)$ as typical states of $P_{R(t_1)}$. In other words, the origin of the current
microstate $\gamma(t_1)$ of the system, reflected in the ensemble $P_{R(t_0)}(t_1)$, becomes
irrelevant, the only important question being in which subspace $P_R$ of $P$ point $\gamma$
resides now. This means that for every time $t>t_0$, ensemble
$P_{R(t_0)}(t)$, representing the system, can (and must, for that makes the calculations much
simpler) be replaced by a microcanonical ensemble $P_{R(t)}$, where $R(t)$ is obtained from initial
$R(t_0)$ by solving the generalized kinetic equations. Now from the inclusion $P_{R(t_0)}(t)\subset
P_{R(t)}$ and the invariance of the Liouville measure, we have for $t>t_0$
\begin{equation}\label{}
    \Gamma\big(P_{R(t_0)}\big) = \Gamma\big(P_{R(t_0)}(t)\big) \leq \Gamma\big(P_{R(t)}\big)
\end{equation}
and so $S_B\big(R(t)\big) \geq S_B\big(R(t_0)\big)$, i.e., Boltzmann entropy never decreases and
achieves its maximum in equilibrium, when the system's macrostate $R$ ceases to change.

\vfill\eject

\end{document}